\documentclass[a4paper,12pt,twoside,titlepage]{article}
\pdfoutput=1
\RequirePackage[OT1]{fontenc}
\RequirePackage{amsthm,amsmath,amssymb}
\RequirePackage{natbib}

\usepackage{tikz,soul}                     
\usetikzlibrary{arrows.meta}     
\usepackage[ruled,norelsize,vlined]{algorithm2e}
\usepackage[noend]{algpseudocode}      
\usepackage{dsfont}                 
\usepackage{gensymb}             
\usepackage{multirow}  
\usepackage{makecell}
\usepackage{mathrsfs}

\newcommand{\Mu}{\boldsymbol{\mu}}
\newcommand{\x}{\mathbf{x}}

\newcommand{\smixture}{\mathcal{S}(\cdot|\cdot)}
\newcommand{\bmixture}{\mathcal{B}(\cdot|\cdot)}
\newcommand*\Bell{\ensuremath{\boldsymbol\ell}}
\newcommand{\lon}{\Bell}
\newcommand{\lat}{\textbf{\textit{b}}}

\newcommand{\iter}{^{(t)}}
\newcommand{\Z}{\textrm{Z}}
\newcommand{\Hvec}{\textrm{H}}
\newcommand{\PSF}{\hbox{PSF}}
\newcommand{\imagemodel}{Spatial Model}
\newcommand{\jointmodel}{Joint Spatial-Spectral Model}

\usepackage{fancyhdr}
\title{\bf Identification of high-energy astrophysical point sources via hierarchical Bayesian nonparametric clustering\vspace{2cm}}

\author{{A. Sottosanti$^1$, M. Bernardi$^1$, A.R. Brazzale$^1$,} \\
{A. Geringer-Sameth$^2$, D.C. Stenning$^3$, R. Trotta$^{2,4}$} \\ {and D.A. van Dyk$^2$}
\\[6ex]
{$^1$University of Padova} \\
{$^2$Imperial College London} \\
{$^3$Simon Fraser University} \\
{$^4$International School for Advanced Study (SISSA) }\vspace{2cm}
}

\date{\today \\ \vspace{3cm}
Address for correspondence: \texttt{sottosanti@stat.unipd.it}}

\begin{document}

\maketitle

\begin{abstract}
The light we receive from distant astrophysical objects carries information about their origins and the physical mechanisms that power them. The study of these signals, however, is complicated by the fact that observations are often a mixture of the light emitted by multiple localized sources situated in a spatially-varying background. A general algorithm to achieve robust and accurate source identification in this case remains an open question in astrophysics.

This paper focuses on high-energy light (such as X-rays and $\gamma$-rays), for which observatories can detect individual photons (quanta of light), measuring their incoming direction, arrival time, and energy. Our proposed Bayesian methodology uses both the spatial and energy information to identify point sources, that is, separate them from the spatially-varying background, to estimate their number, and  to compute the posterior probabilities that each photon originated from each identified source.  This is accomplished via a Dirichlet process mixture while the background is simultaneously reconstructed via a flexible Bayesian nonparametric model based on B-splines. Our proposed method is validated with a suite of simulation studies and illustrated with an application to a complex region of the sky observed by the \emph{Fermi} Gamma-ray Space Telescope. 
\end{abstract}

\section{Introduction}
\label{sec:intro} 
	
Astronomy aims to extend our knowledge of the physical processes that underlie the wide variety of phenomena that exist in the Universe. Electromagnetic radiation, that is, light, is the primary carrier of astronomical information from the Universe to us. The nature of cosmic objects is imprinted in the light that they emit and that we subsequently detect. However, photons (that is, quanta of light) do not bear labels telling us what object they originated from. In fact, when it reaches Earth, light is characterized by only 6 numbers: its time of arrival and energy (measured in eV\footnote{An eV (electronvolt) is a unit of energy commonly used in particle physics. It gives the kinetic energy acquired by an electron when accelerated from rest across a potential difference of 1 volt. The symbol GeV denotes an energy of $10^9$ eV.}), and its two dimensional directions of provenance and polarization (a unit vector perpendicular to the direction of travel indicating the oscillation plane of the electromagnetic wave). This article focuses on high-energy light (e.g., X-rays and $\gamma$-rays), where we may hope to measure a subset of these properties for individually detected photons. At lower energies (e.g., optical, infrared, radio waves), photon properties are averaged over an area of the sky. In both cases, the resulting astronomical dataset is generally the realization of a mixture model in which light generated by multiple sources is blended together in a single image or collection of detected photons.
   
The light received from a region of the sky is the result of the superposition of different physical sources. A basic distinction, and an important one for astronomers, is between localized and diffuse sources: localized sources are those whose angular size, as viewed from the Earth, is much smaller than the angular resolution of the detector. A localized source appears as a sharply defined region of high intensity. The prototypical example is a star, which nearly always appears as a so-called ``point source'' in an astronomical image. In contrast, diffuse sources appear as extended regions of variable intensity across a portion of the sky. An important example in $\gamma$-ray astronomy are the clouds of gas in the Milky Way, which emit $\gamma$-ray photons as cosmic rays collide with the hydrogen and helium that make up the clouds.  

The necessary first step in many physics analyses is the the identification of distinct astrophysical sources --- a procedure called ``source extraction''. The goal of source extraction is to detect new astronomical objects and infer basic properties such as their location and the energy distribution of the light they emit. Once sources are identified, they can be further studied by restricting the analysis to the region or collection of photons that are attributed to them. Conversely, separating point sources from a diffuse background is essential to study the physical processes which give rise to the diffuse background itself. 
   
From a statistical perspective, identifying a source means quantifying the evidence of its presence in an observed region of the sky and placing constraints on its location and properties. In X-ray and $\gamma$-ray astronomy, the data consist of an event list giving the measured sky coordinates, energy, and arrival time of each photon detected by the instrument. The distribution of the sky coordinates of the events forms an image and the distribution of energies forms a spectrum.
   
\begin{figure}[t]
	\centering
	\includegraphics[width=0.65\linewidth]{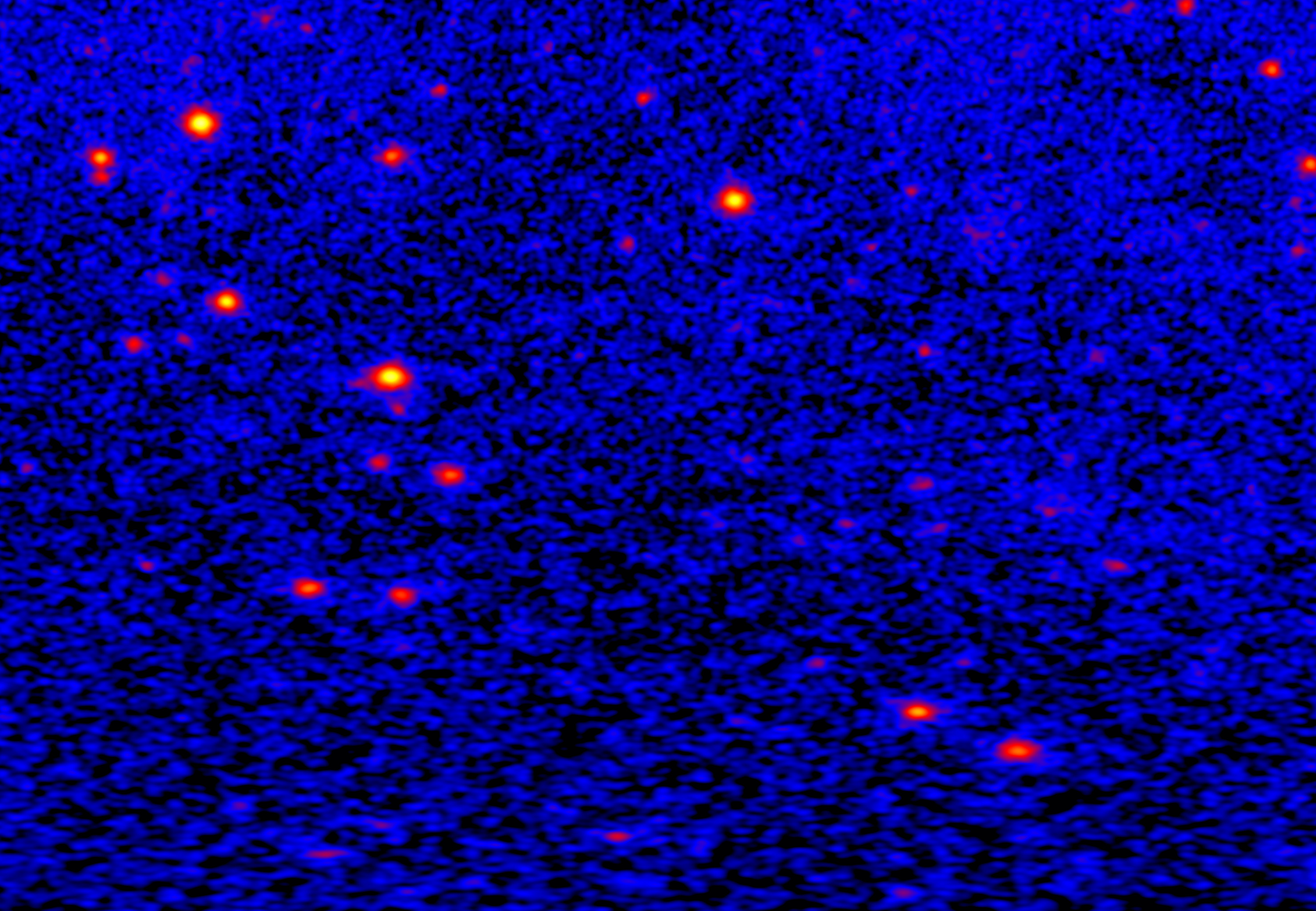}
	\caption{Image map of the $\gamma$-ray counts at energies larger than 1~GeV accumulated by the {Fermi} Large Area Telescope over five years of operation.  The panel shows a small region in the South-Eastern celestial hemisphere of size $65\degree\times 45\degree$, which is sufficiently far away from the Galactic plane so as not to be influenced by the Galactic interstellar emission.  The brighter the color appears, the brighter is the corresponding emitting source.  \emph{Image Credit: NASA/DOE/Fermi LAT Collaboration}}
	\label{fig:backgroundandsources}
\end{figure}

Identifying point sources is challenging for several reasons: the sky coordinates are noisy estimates, interesting sources may be faint and emit only a few photons, and point sources are embedded in a diffuse background, generally a significant component of the data. The intensity of the background and measurement error in the recorded sky coordinates of the photons both vary with location and energy, that is, across the joint domain of the image and the spectrum. Directions in the sky near the Galactic plane, for example, have a high level of emission from diffuse gas clouds, whereas emission at high Galactic latitudes is dominated by a more or less isotropic $\gamma$-ray background component due to the integrated emission of all the faint sources along each line of sight through the Universe, as shown in Figure~\ref{fig:backgroundandsources}. The process of source extraction is generally easier in the latter region where the background can be more simply modeled as a spatially uniform process.

\subsection{Astrophysical source extraction\label{sec:introsourceextraction}}
	
Source extraction has attracted a growing interest both in the astronomical and the statistical literature. In the last twenty years, new technologies have massively increased the precision of detectors and the size of the data sets they generate. Astronomical images now contain multitudes of localized sources over vast regions of the sky. Previously manual or {\em ad hoc} analyses are now impractical and introduce uncontrolled biases into the statistical characterization of populations of sources. The resulting rich new data resources pose significant statistical challenges and have stimulated the development of new statistical techniques and advanced computational methods.
	
\cite{hobson_etal.2010} differentiate methods according to whether they are designed to extract a single source or multiple sources simultaneously. While the former subject has received considerable attention since the early 1990s \citep{kraft_etal.1991, mattox_etal.1996, vandyk_etal.2001, protassov_etal.2002, park_etal.2006, weisskopf_etal.2007, knoetig.2014}, the latter has attracted a growing interest only recently, partially because of its higher computational cost. For example, see \cite{savage_oliver.2007}, \cite{guglielmetti_etal.2009}, \cite{primini_kashyap.2014} and \cite{jones_etal.2015} for X-ray data and \cite{acero_etal.2015} and \cite{selig_etal.2015} for $\gamma$-ray data. Simultaneous multiple source extraction is preferable over single-source extraction in cases where sources overlap spatially in images, a case in which single-source extraction would yield misleading results.  
	 
All source extraction methods must account for the presence of a diffuse background emission in the data.  When the background is low or relatively constant over the region of interest, its spatial and energy components can be modelled as uniform \citep[e.g.][]{jones_etal.2015,meyer_etal.2021}; alternatively, one can consider the Bayesian aperture-photometry approach of \cite{primini_kashyap.2014} for low-counts images. However, for multiple sources extraction methods the modeling of the background is more challenging, since the intensity of the background can vary across the region of interest, as shown in Figure~\ref{fig:backgroundandsources}. For example, \cite{guglielmetti_etal.2009} use a Poisson-based mixture model with thin-plate splines for the background of X-ray count images with intense and prominent background contamination. 

In contrast to the data-driven methods above, an alternative approach is to adopt a physics-based model for the diffuse background emission. For instance, \cite{acero_etal.2016} define a physical model of the Milky Way based on approximate knowledge of the distribution of gas in the Galactic disk, the locations and properties of sources of cosmic rays, dust maps, etc. Combined with physics models for cosmic ray diffusion and  high-energy particle physics, they generate ``templates'' for the approximate $\gamma$-ray emission expected from various components. A final background model is then found by fitting these templates to the full-sky $\gamma$-ray data collected by the \emph{Fermi} Large Area Telescope (LAT).  The method has been widely used to analyse the data collected by the \emph{Fermi} LAT to build detailed catalogues of sources, the latest of which is presented in \cite{abdollahi_etal.2020}. Additionally, the physics-based templates can be used, for example, in conjunction with the method of \cite{stein_etal.2015} to detect unknown structures added to a known image (e.g., a known background). Nonetheless, because the physics model is incomplete, the resulting templates are statistically inconsistent with the observed all-sky data and residuals between model and data are incorporated into the background model according to an {\em ad hoc} procedure. Therefore, the \emph{Fermi} model may inadvertently subsume some point sources into its background estimate, masking their signal in subsequent analyses. In addition, this tool provides only a point estimate of the background morphology, without any measure of uncertainty.
	
As a completely data-driven approach to characterize the diffuse background in photon observations, \cite{selig_ensslin.2015} propose their own empirical background reconstruction based on \emph{Information Field Theory}, an alternative approach to the one of \cite{acero_etal.2016}.  This tool is exploited by \cite{selig_etal.2015} to analyse the \emph{Fermi} LAT sky.  The most recent background fit of the Galactic center in $\gamma$-rays, provided by \cite{abazajian_etal.2020}, shows that a proper treatment of the \emph{Fermi} LAT data at their highest energies is fundamental to improve the understanding of the diffuse background component. The resulting inferences on physical process of interest (for example, the presence of a diffuse emission due to dark matter) can strongly depend on the accuracy of the source extraction procedure. Most available simultaneous sources extraction methods require specification of the number of sources in advance \citep{savage_oliver.2007,ray_etal.2011} or are computationally limited to a maximum number of sources detectable \citep{primini_kashyap.2014}.  An alternative approach, which also does not require the specification of the number of sources in advance, but only applies to spatially well-separated sources, is \cite{feroz_hobson.2008}.  This is a nested sampling method which exploits Bayesian model comparison to detect and characterise multiple modes in the marginal distribution of the data.  The recent work of \cite{jones_etal.2015}, which inspired this paper, considers a Bayesian extraction method based on a finite mixture model where the number of sources is inferred; see also \cite{daylan_etal.2017}.  Computationally, the model is fit using reversible-jump Markov chain Monte Carlo (MCMC) \citep{richardson_green.1997}.

\subsection{Main goals and outline} 

In this work we propose a novel, fully data-driven approach to simultaneously extract the signal of high-energy point sources and reconstruct the diffuse background emission that extends over a region of the sky. The method exploits both the spatial coordinates and the energy of the photons to probabilistically allocate them to sources. These probabilities can then be used in secondary analyses in place of an absolute classification. We utilize Bayesian nonparametric modelling to overcome the limitations of the existent approaches to locating sources in a map that is highly contaminated by background. In particular, our proposed method: i) automatically determines the unknown number of point sources in the map; ii) probabilistically clusters the photons into these sources according to their sky coordinates and energy; and iii) flexibly estimates the underlying intensity of the background emission without relying on either a previous physical understanding or empirical reconstructions of the background map.

Our approach exploits the advantages of mixture modelling and extends the model of \cite{jones_etal.2015} by using an infinite mixture induced by a Dirichlet process (DP) prior \citep{ferguson.1973}. In addition, we use Gibbs sampling algorithms for Bayesian nonparametric methods \citep{muller_etal.2015} which deliver better scaling properties with the mixture size than Green's reversible-jump MCMC, thus providing practical advantages in model fitting. Along with source extraction, our method simultaneously reconstructs the background using a flexible model -- a central advantage, as a poor background choice may lead to misleading results such as the identification of spurious sources, or the incorrect absorption of sources into the background. \cite{jones_etal.2015} employ a uniform  model both for the background map and spectrum. However, this assumption is inappropriate for heavily contaminated regions such as those close to the Galactic plane.  \cite{costantin_etal.2020} extend the model of \citet{jones_etal.2015} using a bivariate exponential distribution for the background map. However, their method is designed for the analysis of a particular field and cannot be directly applied more generally. Additionally, their method does not account for irregularities in the background and thus may result in a large number of false detections. Our innovative solution is to model the background component by combining Bayesian nonparametric techniques with B-splines to fully reconstruct the background morphology over the map. 
Previous comparable approaches  \citep{denison_etal.1998, biller.2000, dimatteo_etal.2001, sharef_etal.2010} adopt reversible-jump MCMC to select the active spline functions. Other competitive models based on splines require solving minimization problems, which in practice might be unfeasible \citep{guglielmetti_etal.2009, schellhase_kauermann.2012}. By contrast, we adopt faster Gibbs sampling algorithms for fitting our Bayesian nonparametric model.

This paper is structured in six sections. Section~2 presents our novel Bayesian nonparametric mixture model for signal extraction with high background contamination, first addressing spatial data only, that is, ignoring energy, and then extended to jointly model spatial data and spectra. Section~3 recasts the model as a mixture of DP mixtures and presents our Gibbs sampler. The model is validated in Section~4 with a suite of simulation studies, first illustrating the method and then demonstrating the advantage of including spectral data. An application to a region of the sky using \emph{Fermi} LAT data is presented in Section~5 and  discussion appears in Section~6. Technical details of the Gibbs sampling algorithm are presented in Appendix~A, and additional results related to our simulation studies appear in Appendix~B.

\section{The statistical model}
\label{sec:model}
Let $i=1,\dots,n$ index a collection of photons (that is, events) with sky coordinates $\x_i =(x_i,y_i)\in\mathcal{X} = (x_{\rm min},x_{\rm max}) \times (y_{\rm min},y_{\rm max})$ and energy $E_i\in \mathcal{E} = (E_{\rm min},E_{\rm max})$.  
	
\subsection{The spatial model}
\label{subsec:image}
\noindent
As each event may originate either from a point source or from background, we formalize a mixture model for the sky coordinates,
\begin{equation}
f(\mathbf{x}_i|\boldsymbol{\varTheta})=\delta \mathcal{S}(\x_i|\boldsymbol{\vartheta}_s)+(1-\delta) \mathcal{B}(\x_i|\boldsymbol{\vartheta}_b),\label{formula:3.1_mixturemodel}
\end{equation}
where $\mathcal{S}$ and $\mathcal{B}$ are the unknown density functions for the sky coordinates of source events and background events, with parameters $\boldsymbol{\vartheta}_s$ and $\boldsymbol{\vartheta}_b$, respectively. Finally, $\delta\in (0,1)$ is a mixing parameter, which follows a $\hbox{Beta}(\lambda,\lambda)$ distribution, where $\lambda>0$ is a known hyperparameter. We collect the unknown parameters in $\boldsymbol{\varTheta} = \{\boldsymbol{\vartheta}_s, \boldsymbol{\vartheta}_b, \delta\}$.

\subsection{The spatial source model}
\label{subsec:model_signal} 

\begin{figure}
	\centering
	\includegraphics[width=1\linewidth]{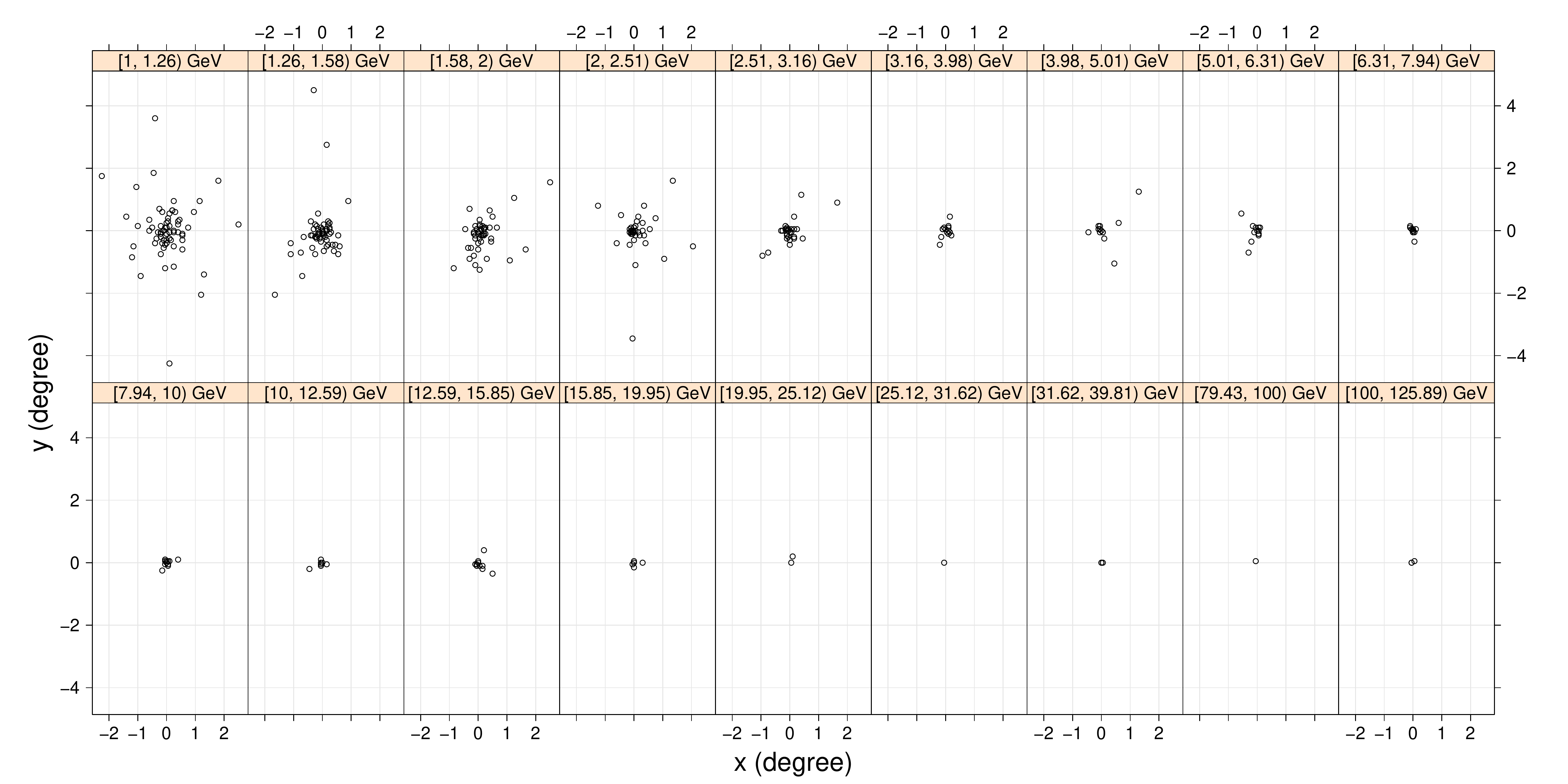}
	\caption{Simulated sky coordinates $\x$ of photons of different energies $E$ detected by the {Fermi} LAT $\PSF(\x|\Mu,E)$. The source is located at $\Mu=(0,0)$. As energy $E$ increases (from left to right and from row 1 to row 2) according to the energy ranges given in the figure headers, the events become rarer and more concentrated around the true source location.}
	\label{fig:sourcemodelsourcespread}
\end{figure}
	
\noindent
We begin with a model for the recorded sky coordinates of photons from source $j$, with location $\Mu_j = (\mu_{jx},\mu_{jy})$. Although we only consider point-like sources, due to measurement noise the observed sky coordinates form a distribution around the true source location. This distribution is called the \emph{Point Spread Function} (PSF). The shape of the PSF varies with photon energy: qualitatively, the \emph{Fermi} LAT instrument measures more accurately the arrival direction of high-energy photons (i.e., in this case the PSF concentrates more tightly around the true source location), while the PSF of low-energy photons has a larger spread. Additionally, high-energy photons are generically rarer than low-energy photons, as the flux from astrophysical sources generally follows a negative power law with energy (sometimes with exponential cut-offs past a certain energy threshold, though the details are unimportant here). The \emph{Fermi} LAT PSF \citep{ackermann_etal.2012} is tabulated so that $\PSF(\x|\Mu,E)$ gives the probability that a photon with energy $E$ and true sky coordinates $\Mu$ is recorded in the instrumental pixel containing $\x$. Here, the instrumental pixel is the smallest addressable unit, measured in degrees, into which the area of the whole sky map is subdivided. Figure \ref{fig:sourcemodelsourcespread} uses simulated data to illustrate how the distribution of the recorded sky coordinates for a point source observed with the \emph{Fermi} LAT varies with photon energy. 

Formulating a model based on the PSF for multiple sources requires care because (a) the total number of sources is unknown, (b) the sources have different intensities, and (c) no \textit{a priori} information on the sources' locations is available. To address these issues, we propose the following model:
\begin{equation}
\mathcal{S}(\x_i|E_i,\mathcal{F})=\int \PSF(\mathbf{x}_i|\Mu, E_i) \mathcal{F}(d\boldsymbol{\mu}),\hspace{1cm}\mathcal{F}\sim \mathcal{DP}(\alpha_s, \mathcal{F}_0),\label{formula:3.3_DPsources}
\end{equation}
where $\mathcal{DP}(r,\mathcal{P}_0)$ is a DP prior with concentration parameter $r$ and base measure $\mathcal{P}_0$. Model \eqref{formula:3.3_DPsources} is known as a DP mixture and formulates $\smixture$ by mixing the kernel $\PSF$ with respect to the unknown measure $\mathcal{F}$. $\mathcal{F}_0$ is the prior distribution for $\Mu$ and is assumed to be uniform over the map $\mathcal{X}$. (For a review of Bayesian nonparametrics, see \cite{muller_etal.2015}). 
	
Although \eqref{formula:3.3_DPsources} allows for any number of sources, in practice the observed number of photons originating from the point sources, $n_s$, serves as an upper bound. Thus, \eqref{formula:3.3_DPsources} is effectively a sum of at most $n_s$ components, although in practice the actual number of point sources is generally $\ll n_s$.

\subsection{The spatial background model}
\label{subsec:model_back} 

\noindent
In \eqref{formula:3.1_mixturemodel}, $\bmixture$ in the density of background photons across the image. As mentioned in Section~\ref{sec:intro}, high-energy astrophysics images can be heavily contaminated by a non-uniform background (both in spatial location and energy). Because the background is the result of several astronomical phenomena that cannot be predicted from first principles, no analytical distribution is available. We therefore propose a  flexible nonparametric solution based on B-splines.

The B-spline basis function of order $m\in \mathbb{N}^+$ is defined on a vector of $m+1$ knots $\boldsymbol{\tau} = (\tau_1,\dots,\tau_{m+1})$,
\begin{equation}
\mathscr{B}_{m}(x|\boldsymbol{\tau})=\frac{x-\tau_1}{\tau_{m}-\tau_1}\mathscr{B}_{m-1}(x|\boldsymbol{\tau}_{-(m+1)})+\frac{\tau_{m+1}-x}{\tau_{m+1}-\tau_{2}}\mathscr{B}_{m-1}(x|\boldsymbol{\tau}_{-1}),\label{formula:3.4_bspline}
\end{equation}
where $\boldsymbol{\tau}_{-j}$ is the vector $\boldsymbol{\tau}$ with element $j$ removed. Starting from  $\mathscr{B}_1(x|\textbf{\emph{a}}=(a_1,a_2)) = \mathds{1}_{[a_1,a_2)}(x)$, a basis of order $m$ is defined recursively from the bases of smaller orders. By construction, \eqref{formula:3.4_bspline} is always positive between $\tau_1$ and $\tau_{m+1}$ and zero elsewhere.  Although $\mathscr{B}_{m}(x|\boldsymbol{\tau})$ is unimodal for $m>1$, it can assume many different shapes depending on the location of the knots  and can be normalised to the density function $\tilde{\mathscr{B}}_m(x|\boldsymbol{\tau})=m\mathscr{B}_m(x|\boldsymbol{\tau})/(\tau_{m+1}-\tau_1)$; see \cite{deboor.2001} for a full review. We define the bivariate density 
\begin{equation*}
\varphi(\mathbf{x}_i|\lon,\textbf{\emph{b}})= \tilde{\mathscr{B}}_4(x_i|\lon)\tilde{\mathscr{B}}_4(y_i|\textbf{\emph{b}}), \label{formula:3.5_kernelbackground}
\end{equation*}
where $\lon=(\ell_1,\dots,\ell_5)$ and $\textbf{\emph{b}}=(b_1,\dots,b_5)$ denote the knots of the longitude and of the latitude B-splines, respectively; thus, $\ell_j\in(x_{\min},x_{\max})$ and $b_j\in(y_{\min},y_{\max})$, for $j = 1,\dots,5$. Here we use splines of order 4 as they ensure sufficient flexibility without requiring too many knots. We then model the background as
\begin{equation}
\label{formula:3.6_DPbackground}
\mathcal{B}(\x_i|\mathcal{G})=\int \varphi(\mathbf{x}_i|\lon,\textbf{\emph{b}}) \mathcal{G}(d\lon,d\textbf{\emph{b}}),\hspace{1cm}\mathcal{G}\sim \mathcal{DP}(\alpha_b,\mathcal{G}_0).
\end{equation}
This is an infinite mixture of normalised B-spline functions induced by a DP prior, but in practice a sample of $n_b$ photons from the background leads to a finite mixture with at most $n_b$ components. The base measure, $\mathcal{G}_0$, is the  prior distribution for the sets of longitude and latitude knots. The ordering of the knots is maintained by assuming the middle knot is uniformly distributed over the limits of the map, e.g., $\ell_{3}\sim \mathcal{U}(x_{\min},x_{\max})$, and that the remaining knots are conditionally uniformly distributed over their appropriate ranges. Specifically, given $\ell_{3}$, we assume $\ell_{2}\sim \mathcal{U}(x_{\min},\ell_{3})$ and $\ell_{4}\sim \mathcal{U}({\ell}_{3},x_{\max})$; given $\ell_{2}$, we assume $\ell_{1}\sim \mathcal{U}(x_{\min},\ell_{2})$ and given $\ell_{4}$ we assume $\ell_{5}\sim \mathcal{U}({\ell}_{4},x_{\max})$; likewise for $\boldsymbol{b}$, but replacing $x_{\min}$ and $x_{\max}$ with $y_{\min}$ and $y_{\max}$.

\subsection{The joint spatial-spectral model}
\label{subsec:spatialspectral}

\noindent
We wish to extend Model \eqref{formula:3.1_mixturemodel} to jointly model both the sky coordinates and energies of the photons. Astrophysical sources typically have power-law spectra \citep[e.g.,][]{acero_etal.2015}, where the number of emitted photons decreases with energy according to $E^{-\eta}$, where $\eta>0$. Therefore, we propose Pareto distributions for both the source and background energies, that is, 
\begin{equation}
\begin{aligned}
f_s(\x_i,E_i|\boldsymbol{\varTheta}^{\rm s})=\delta  \mathcal{S}(\x_i|\boldsymbol{\vartheta}_s)h(E_i|&E_{\min},\eta_s)+\\
& (1-\delta)\mathcal{B}(\x_i|\boldsymbol{\vartheta}_b)h(E_i|E_{\min},\eta_b),
\label{formula:extendedMixture}
\end{aligned}
\end{equation}
where $\boldsymbol{\varTheta}^{\rm s} = \{\boldsymbol{\varTheta},\eta_s,\eta_b \}$ and $h(\cdot|v, \eta)$ is a Pareto density with shape parameter $\eta$, scale parameter $v$, and support $[v, \infty)$; we set $v=E_{\rm min}$, the lower limit of the instrumental energy range. We specify $\eta_s\sim \hbox{Gamma}(a_{\eta_s},b_{\eta_s})$ and $\eta_b\sim \hbox{Gamma}(a_{\eta_b},b_{\eta_b})$ to exploit the conjugacy between the Gamma and the Pareto distributions.

Equation~\eqref{formula:extendedMixture} ignores two effects. First, not all photons that arrive at the telescope from a fixed direction are detected with the same probability. Very often this detection probability depends on energy and is quantified by the \emph{effective area} of the detector (which is a function of energy). Second, as with direction $\x_i$, photon energies $E_i$ are measured with uncertainty. Ideally, $h(E_i|E_{\min},\eta)$ in \eqref{formula:extendedMixture} should be replaced with 
\begin{equation}
h(E_i | E_{\min},\eta) = \int\limits p(E_i|E^{\rm true}) \frac{\epsilon(E^{\rm true})h_0(E^{\rm true}|E_{\min},\eta)}{\int \epsilon(E') h_0(E'|E_{\min},\eta) dE' } dE^{\rm true},
\label{formula:EAconvolution}
\end{equation}
where $E^{\rm true}$ is the true photon energy, $h_0(\cdot|E_{\min},\eta)$ is the intrinsic energy spectrum of the source or background component, $\epsilon(E^{\rm true})$ is the exposure (effective area integrated over the observation time), and $p(\cdot|E^{\rm true})$ is the probability density function of the observed energy given the true energy. This is straightforward to implement but introduces a performance penalty. In our application to the \emph{Fermi}~LAT in Sections~\ref{sec:simulationapplication} and~\ref{sec:Fermiapplication}, we use an energy range where the effective area is approximately constant and in which the energy uncertainty is relatively small, around 10\%. Thus, the approximation implicit in \eqref{formula:extendedMixture} is a good one. Alternatively, \eqref{formula:extendedMixture} can be interpreted as modelling the \emph{observed} energy spectrum, rather than the true spectrum, as a power-law. The parameter $\eta_s$ then potentially loses its physical meaning. However, for broad energy ranges, small energy uncertainty, and intrinsic power-law spectra, the observed power-law index equals the intrinsic index. 
	
Model~\eqref{formula:extendedMixture} specifies a common Pareto distribution for the spectra of all the sources. In practice, however, we expect the source spectra to differ. As we illustrate in Section~\ref{subsec:model_comparison}, even with this simplifying assumption, the {\jointmodel} defined in Formula~\eqref{formula:extendedMixture} out-performs the {\imagemodel} defined in \eqref{formula:3.1_mixturemodel} in classification accuracy.  

In practice, the {\jointmodel} can be used for a preliminary analysis to attribute photons to the several sources and background. The photons associated with a particular detected source can then be analyzed in a source-specific secondary analysis using a more sophisticated spectral model. 
Another practical advantage of the {\jointmodel} is that it has only two additional free parameters and therefore does not excessively increase the size of the parameter space, relative to the {\imagemodel}. 

\begin{figure}[t]
	\centering
	\scalebox{.95}{
		\begin{tikzpicture}
		\node (n)[below] at (11.35,-2.65) {$n$};
		\node (ns)[below] at (7.1,-2.3) {$n_s$};
		\node (ns)[below] at (11,-2.3) {$n_b$};
		\begin{scope}[every node/.style={circle,thick,draw}]
		\node (G0s)[shape=rectangle] at (5.5,2.5) {$\mathcal{F}_0$};
		\node (As)[shape=rectangle] at (3.925,1) {$\alpha_s$};
		\node (Gs) at (5.5,1) {$\mathcal{F}$};
		\node (mus) at (5.5,-0.5) {$\boldsymbol{\mu}_i$};
		\node [text width = 0.35cm,fill=gray!30] (Ys) at (5.5,-1.975) {$\hspace{0.03cm}\mathbf{x}_i$};
		\node [text width = 0.32cm,fill=gray!30] (Es) at (6.675,-1.975) {$\hspace{-.05cm}E_i$};
		\node (Etas)[dashed] at (6.675,-3.5) {$\eta_s$};
		
		\node (G0b)[shape=rectangle] at (10.575,2.5) {$\mathcal{G}_0$};
		\node (Ab)[shape=rectangle] at (12.15,1) {$\alpha_b$};
		\node (Gb) at (10.575,1) {$\mathcal{G}$};
		\node [text width = 0.4cm] (mub) at (10.575,-0.5) {$\hspace{-.1cm}\lon_i,\hspace{-.05cm}\textbf{\emph{b}}_i$};
		\node [text width = 0.35cm,fill=gray!30] (Y) at (10.575,-1.975) {$\hspace{0.03cm}\mathbf{x}_i$};
		\node [text width = 0.32cm,fill=gray!30,dashed] (Eb) at (9.325,-1.975) {$\hspace{-.05cm}{E}_i$};
		\node (Etab)[dashed] at (9.325,-3.5) {$\eta_b$};
		\node (Etasprior) [dashed,shape=rectangle] at (5,-3.5) {$a_{\eta_s}, b_{\eta_s}$};
		\node (Etabprior) [dashed,shape=rectangle] at (11,-3.5) {$a_{\eta_b}, b_{\eta_b}$};
			
		\node (a0)[shape=rectangle] at (8,2.85) {$\lambda$}; 
		\node (delta) at (8,1.3) {$\delta$}; 
		\node (Z) at (8,-0.25) {$\Z_i$};
			
		\draw (8.7,-2.7) -- (11.2,-2.7) -- (11.2,.2) -- (9.95,.2) -- (9.955,-1.25) -- (8.7,-1.25) -- (8.7,-2.7);
		\draw (7.3,-2.7) -- (7.3,-1.25) -- (6.05,-1.25) -- (6.05,.2) -- (4.8,.2) -- (4.8,-2.7) -- (7.3,-2.7);
		\draw (4.5,-3) -- (11.5,-3) -- (11.5,0.5) -- (4.5,0.5) -- (4.5,-3);
		\end{scope}
			
		\begin{scope}[>={Stealth[black]},]
		\path [->] (a0) edge (delta);
		\path [->] (delta) edge (Z);
		\path [->] (Z) edge (Ys);
		\path [->] (Z) edge (Y);          
			
		\path [->] (G0s) edge (Gs);
		\path [->] (As) edge (Gs);
		\path [->] (Gs) edge (mus);
		\path [->] (mus) edge (Ys);
		\path [->] (Es) edge (Ys);

		\path [->] (G0b) edge (Gb);
		\path [->] (Ab) edge (Gb);
		\path [->] (Gb) edge (mub);
		\path [->] (mub) edge (Y);
		
		\path [->,dashed] (Z) edge (Es);
		\path [->,dashed] (Etas) edge (Es);
		\path [->,dashed] (Z) edge (Eb);
		\path [->,dashed] (Etab) edge (Eb);
		\path [->,dashed] (Etabprior) edge (Etab);
		\path [->,dashed] (Etasprior) edge (Etas);
		
		\end{scope}
			
		\end{tikzpicture}
    }
\caption{Directed Acyclic Graph (DAG) of the \imagemodel\ in \eqref{formula:3.1_mixturemodel} and the \jointmodel\ in \eqref{formula:extendedMixture}. The arrows represent the direction of data generation under the models. Data are represented by grey circles, parameters by open circles, and hyperparameters by open rectangles. The \imagemodel~is represented by solid arrows, while the \jointmodel~is represented by both solid and dashed arrows.}
\label{figure4.1:graph_model}
\end{figure}
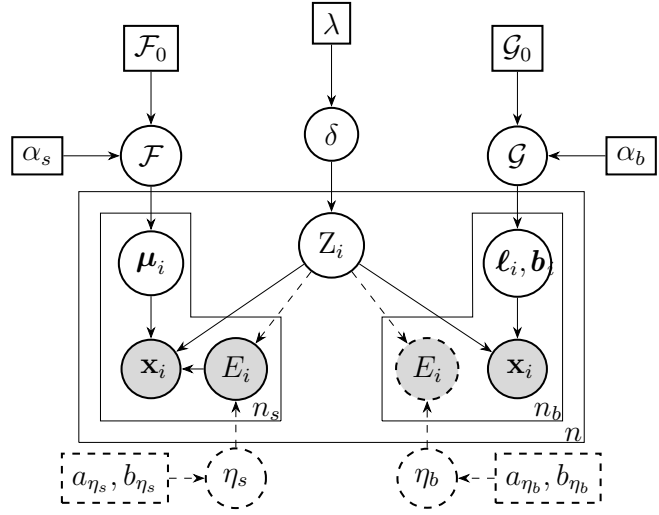
	
Figure~\ref{figure4.1:graph_model} represents the Spatial and the Joint Spatial-Spectral Models  with a direct acyclic graph \citep[DAG,][]{lauritzen_spiegelhalter.1988}. The indicator variable, $\Z_i$, in  Figure~\ref{figure4.1:graph_model} specifies whether photon $i$ originates from a source or from the background. The spatial source model, $\smixture$, is represented on the left of Figure~\ref{figure4.1:graph_model}, where the location parameter $\{\Mu_1,\dots,\Mu_{n_s}\}$ is sampled from a random probability measure $\mathcal{F} \sim \mathcal{DP}(\alpha_s,\mathcal{F}_0)$. Finally, the true sky coordinates of photon $i$ is convolved with the $\PSF(\cdot|\Mu_i,E_i)$ to obtain their observed location  $\{\x_1,\ldots, \x_{n_s}\}$. The background model, $\bmixture$, is represented on the right of Figure~\ref{figure4.1:graph_model}. The  parameters $\{(\lon_1,\lat_1),\dots,(\lon_{n_b},\lat_{n_b})\}$ are sampled from $\mathcal{G} \sim \mathcal{DP}(\alpha_b, \mathcal{G}_0)$. The sky coordinates of background photons, $\{\x_1,\ldots, \x_{n_b}\}$, are generated according to $\varphi(\cdot|\lon_i,\lat_i)$. For the \jointmodel, $E_i$, follows a Pareto distribution with scale $\eta_s$ for source photons and with scale $\eta_b$ for background photons.
	
The discreteness of random probability measures sampled from DP priors \citep{ferguson.1973, blackwell.1973} means we observe only $k_s$ distinct values among the $\{\Mu_1,\dots,\Mu_{n_s}\}$, and only $k_b$ values among the  $\{(\lon_1,\lat_1),\dots,\allowbreak (\lon_{n_b},\lat_{n_b})\}$, with $k_s\leq n_s$ and $k_b\leq n_b$. Thus, source photons with the same value of the model parameters naturally divide into clusters that correspond to sources. In this way, both the \imagemodel\ and the \jointmodel\ produce two levels of clustering: first photons are split between sources and background as quantified by $\Z_i$, and second source photons are split among the sources, as determined by the model parameters.

\subsection{Smoothing the background model}
\label{subsec:misclassification} 

\noindent
We are concerned with two opposing types of potential errors: the first occurs when groups of background photons are clustered spatially so as to mimic the signal from a point source; the second error occurs when the signal from a point source is attributed to the background. While the former cannot be eliminated, the latter might be incurred by excessive flexibility in the B-spline functions when combined with the DP: an excessively flexible background model may absorb sources, treating them as sharp background irregularities. In practice and based on physics arguments, we expect the diffuse background to be spatially much smoother than the sources, and aim for sharp spatial variations in the image to be attributed to a source, rather than to a background feature. To enforce a degree of spatial smoothness in the fitted background, we impose restrictions on the variance of its kernel density, given in \eqref{formula:3.6_DPbackground}. 

Using results in \cite{carlson.1991}, the variance of a random variable, $X$, with density $\tilde{\mathscr{B}}_m(\cdot|\boldsymbol{\tau})$ is 
\begin{equation}
Var(X)=\mathcal{V}(\boldsymbol{\tau}):=\frac{\sum_{p=1}^{m}\sum_{q=p+1}^{m+1}(\tau_p-\tau_q)^2}{(m+1)^2(m+2)};
\label{formula:3.7_bsplinevariance}
\end{equation} 
recall that $\boldsymbol{\tau}$ denotes the knots. We require this variance to be greater than a certain threshold in order to constrain the knots, $\lon$ and $\lat$, 	and discourage overly sharp spatial variability in $\bmixture$. Specifically, we require 
\begin{equation}
\label{formula:constraint}
\sqrt{\mathcal{V}(\lon)}>c_\ell, \ \hbox{ and } \ \sqrt{\mathcal{V}(\lat)}>c_b,
\end{equation}
for each component, where $c_\ell$ and $c_b$ have units of degrees and are tuned to control the smoothness of the background. If $c_\ell$ and $c_b$ are too small, the background model can easily mimic point sources, which are then missed. If instead they are too large, the model is excessively smooth and cannot capture real variation in the background, leading to spurious point source detections. 
	
When an approximate background model is available from a previous analysis, we propose to tune $c_\ell$ and $c_b$ by simulating a background image from this model and fitting it with the background-only model, $\bmixture$ in \eqref{formula:3.6_DPbackground}. This fit should yield a reasonable range for $\mathcal{V}(\lon)$ and $\mathcal{V}(\lat)$ and thus the lower bounds stipulated by \eqref{formula:constraint}. For example, $c_\ell$ and $c_b$ could be set to the twentieth percentile of the  posterior distribution of $\sqrt{\mathcal{V}(\lon)}$ and $\sqrt{\mathcal{V}(\lat)}$, respectively. In Section~\ref{sec:simulationapplication}, we use the background model of \citet{acero_etal.2016}, which yields a common value of $c_\ell$ and $c_b $ equal to $1\degree$. An alternative approach is to set the two lower limits $c_\ell$ and $c_b$ using the known PSF of the instrument. The rationale is that the density of observed photons is given by the convolution of the true density distribution of incoming photons with the PSF.  The typical angular uncertainty on the direction of an incoming photon is quantified by the 68\%-containment angle $c$ (about 1\degree\ for the \emph{Fermi} LAT at 1~GeV energy, though this value is generally energy-dependent). Therefore, we expect any cluster of \emph{observed} photons to spread out at least to size $c$.  This yields a principled choice of $c_\ell=c_b = c$.
This approach has the advantage that it does not require an approximate background model and is purely based on the known characteristics of the instrument.

\subsection{A mixture of DP mixtures}
\label{subsec:mixture_of_dp_mixtures}
\noindent
We generalize Model~\eqref{formula:3.1_mixturemodel} to a mixture of DP mixtures with more than two components. This enables us to present more general algorithms for posterior sampling in Section~\ref{sec:model_algo}. Specifically, let $\{ \x_i\}_{i=1}^{n}$ be a collection of independent data from the mixture of DP mixtures,  
\begin{equation}
\label{generalMixture.a}
f(\x_i|\boldsymbol{\delta},\{\mathcal{K}_j\})=\sum_{j=1}^{J} \delta_j g_j(\x_i|\mathcal{K}_j),
\end{equation}
\begin{equation}
\label{generalMixture.b}		
g_j(\x_i|\mathcal{K}_j)=\int q_j(\x_i|{\theta}_j)\mathcal{K}_j(d{\theta}_j),\hspace{.9cm}\mathcal{K}_j\sim \mathcal{DP}(\alpha_j, \mathcal{K}_{0,j}),
\end{equation}
where $\boldsymbol{\delta}\sim Dir(\boldsymbol{\lambda})$ and $J$ is a fixed value. Model \eqref{formula:3.1_mixturemodel} derives directly from \eqref{generalMixture.a} by setting $J=2$, $g_1(\cdot|\mathcal{K}_1)=\mathcal{S}(\cdot|E_i,\mathcal{F})$, $g_2(\cdot|\mathcal{K}_2)=\mathcal{B}(\cdot|\mathcal{G})$, $q_1(\cdot|\theta_1)=\PSF(\cdot|\Mu,E_i)$, $q_2(\cdot|\theta_2)=\varphi(\cdot|\lon,\lat)$, $\alpha_1 = \alpha_s$, $\alpha_2 = \alpha_b$, $\mathcal{K}_{0,1}=\mathcal{F}_0$ and $\mathcal{K}_{0,2}=\mathcal{G}_0$. A special case of Model \eqref{generalMixture.a}-\eqref{generalMixture.b} has been considered by \cite{do_etal.2005} when both $q_1(\cdot|\cdot)$ and $q_2(\cdot|\cdot)$ are Gaussian densities.

\section{Posterior inference and computation}
\label{sec:model_algo}

\subsection{Posterior analysis} 
\label{subsec:posterior_analysis}
	
We present MCMC techniques that exploit the models in Section~\ref{sec:model} to (a) locate regions of the map with substantial evidence for point sources, and (b) estimate the number of sources in each region, and the sky coordinates and intensity of each source. We also develop a post-processing routine to handle the multimodal nature of the posterior distributions. The multimodality, however, is not merely a technical hurdle, but rather, different modes may correspond to unique scientific interpretations of the data. We will treat this aspect in Section~\ref{subsec:post-processing}.

\subsection{A Collapsed Gibbs sampler for the mixture of DP mixtures}
\label{subsec:mcmc_simulation}
	
When $J=1$,  \eqref{generalMixture.a}-\eqref{generalMixture.b} simplifies to a standard DP mixture and we can obtain a sample from its posterior distribution using  the sampler of \cite{maceachern_muller.1998}. This method is based on the Blackwell-MacQueen scheme and incorporates earlier samplers proposed by \cite{escobar.1994}, \cite{escobar_west.1995}, and \citet{bush_maceachern.1996}. See also \citet[][Section 3.3.1]{ muller_rodriguez.2013} for a review of the method. MacEachern and M{\"u}ller's algorithm is a \emph{collapsed} Gibbs sampler \citep{liu.1994} in that it targets the posterior distribution marginalized over the mixture measure $\mathcal{K}_1$. We deploy the same strategy in our extension to the case where $J>1$, marginalizing over both mixture weights, $\boldsymbol{\delta}$, in the first level mixture in \eqref{generalMixture.a} and the mixture measures, $\mathcal{K}_j$, in the second level mixture in \eqref{generalMixture.b}; see the pseudo-code in Algorithm~\ref{algorithm:nestedPolya}. 
     
The more general case requires us to account for the two levels of clustering  described, for $J=2$, in Section~\ref{subsec:spatialspectral}. Specifically, the first level of clustering is formalized by the finite mixture in \eqref{generalMixture.a} and indexed by the $n\times 1$ vector $\mathbf{Z}$, where $\Z_i=j$ if event $i$ is associated with $g_j(\cdot|\cdot)$. The second level is formalized by the continuous mixture in \eqref{generalMixture.b} and indexed by the $n\times 1$ vectors $\mathbf{H}_j$, where $\Hvec_{ji}=l$ if event $i$ is associated with component $l$ of the finite representation of the mixture for $g_j(\cdot|\cdot)$ in \eqref{generalMixture.b}. Finally, $\Hvec_{ji}=0$ if $\Z_i\neq j$. In Algorithm~1, we use a superscript $(t)$ to denote quantities sampled during iteration $t$. 

\begin{algorithm}[th!]
\caption{Collapsed Gibbs sampler for model in \eqref{generalMixture.a} -- \eqref{generalMixture.b}}
\label{algorithm:nestedPolya}
\flushleft 
This pseudo-code shows how quantities are updated in iteration $t$, as random functions of quantities from iteration $t-1$, where iterations are indexed as parenthetical superscripts.
		
	\KwIn{Level~1 classification indicator,  $\mathbf{Z}^{(t-1)}$; for $j=1,\ldots, J$: level~2 classification indicator $\textbf{H}^{(t-1)}_j$; mixture sizes, $k^{(t-1)}_j$; model parameters $\boldsymbol{\theta}^{(t-1)}_j=(\theta_{j1},\dots,\theta_{jk_j})^{(t-1)}$; an integer $h_j$. (We suggest $h_j=5$ or 10 for computational efficiency.)} 
	\Begin{
		For $j=1,\dots,J$, draw the additional values $\tilde{\boldsymbol{\theta}}_j=(\tilde{\theta}_{j1},\dots,\tilde{\theta}_{jh_j})$ from $\mathcal{K}_{0,j}$ and set $k^{(t)}_j = k^{(t-1)}_j$.
		\newline
		\For{$i = 1,\dots,n$}{
			{\bf Step 1} Draw $\Z_i^{(t)}$ from
			\begin{equation*}
			p(\Z_i=j|\textbf{\Z}_{-i}^{(t-1)},
			\mathbf{x}_i, \textbf{\Hvec}_j^{(t-1)},\boldsymbol{\theta}_j^{(t-1)})\propto\frac{n^{-i}_{j}+\lambda}{n+J\lambda-1} g_j(\mathbf{x}_i|\textbf{\Hvec}_j^{(t-1)},\boldsymbol{\theta}_j^{(t-1)})
			\end{equation*}
			for $j=1,\ldots, J$, where $n^{-i}_{j} = \sum_{i'\neq i} \mathds{1}(\mathrm{Z}_{i'}^{(t-1)} = j)$.
				
			{\bf Step 2} Let assume $\Z^{(t)}_i=j'\in\{1,\dots,J\}$. Draw $\Hvec^{(t)}_{j'i}$ from
			\begin{equation*}
			p(\Hvec_{j'i}=l|\textbf{H}^{(t-1)}_{j',-i},\x_i, \boldsymbol{\theta}^{(t-1)}_{j'},\tilde{\boldsymbol{\theta}}_{j'},\alpha_{j'})\propto \left\{
			\begin{array}{@{}lr@{}}
			n^{-i}_{j'l} q_j(\mathbf{x}_i|\theta^{(t-1)}_{j'l}) & l=1,\dots,k^{(t)}_{j'},\\
			\vspace{-.15cm} & \\
			\frac{\alpha_{j'}}{h_{j'}}q_{j'}(\mathbf{x}_i|\tilde{\theta}_{j'h}) &
			\begin{array}{c@{}}
			l=k^{(t)}_{j'}+h,\\ 
			h=1,\dots,h_{j'},
			\end{array}\\
			\end{array}\right.
			\end{equation*}
			where $n^{-i}_{j'l} = \sum_{i'\neq i} \mathds{1}(\Hvec^{(t-1)}_{j'{i'}}=l)$. 
			If $l>k^{(t)}_{j'}$, then $\Hvec^{(t)}_{j'i}=k^{(t)}_{j'}+1$ and increase $k^{(t)}_{j'}$ by one. Draw a value  from the posterior distribution of $\theta_{j'(k_{j'}+1)}| \mathbf{y}_i$ and add it to $\boldsymbol{\theta}^{(t-1)}_{j'}$.
				
			{\bf Step 3} For each $j\neq j'$, set $\Hvec^{(t)}_{ji} = 0$. If a cluster is empty, reduces $k^{(t)}_{j'}$ by one and discard the corresponding value from $\boldsymbol{\theta}^{(t-1)}_j$.}
			{\bf Step 4} For $j = 1,\dots,J$ and $l = 1,\dots,k^{(t)}_j$, draw $\theta^{(t)}_{jl}$ from the conditional posterior distribution  
			\begin{equation}
			\label{formula:limitDistribution_algorithm}
			p(\theta_{jl}|\{\x_i\}, \mathbf{H}^{(t)}_j)\propto \mathcal{K}_{0,j}(\theta_{jl})\prod_{i: \Hvec^{(t)}_{ji}=l} q(\x_i|\theta_{jl}).   
			\end{equation}
		}
	\KwOut{$\textbf{\Z}\iter$ and $k\iter_j$, $\mathbf{H}\iter_j$, $\boldsymbol{\theta}\iter_j$, for $j=1,\dots,J$.}
\end{algorithm}	
	
The first three steps of Algorithm~1 are run for each of the $n$ observations. Step~1 updates each $\Z_i$ by assigning each event $i$ to one of the $J$ components of the mixture in \eqref{generalMixture.a}. Step~2 updates each $\Hvec_{ji}$ by assigning each event $i$ to one of the components of the (finite version) of mixture $\Z_i$ in \eqref{generalMixture.b}, or adds a new component to the model; this uses Algorithm~8 of \cite{neal.2000}. Step~3 sets $\textrm{H}\iter_{ji} = 0$ for $j\neq \Z_i$, and removes any empty clusters. Finally, Step~4 updates each set of model parameters according to a kernel with stationary distribution equal to \eqref{formula:limitDistribution_algorithm} given in Step 4.

Although Algorithm~1 is a \emph{collapsed} Gibbs sampler and does not provide a posterior sample of $\boldsymbol{\delta}$ or $\mathcal{K}_j$, we can easily obtain their sample after running the algorithm. Conditional on the other unknowns,  $\boldsymbol{\delta}$ follows a Dirichlet distribution \citep{richardson_green.1997} and we can sample the second-level mixture weights of the existing components with sequences of Beta draws using the stick-breaking updating formula \citep[][Section 3.4]{muller_rodriguez.2013}.

\subsection{A collapsed Gibbs sampler for the \jointmodel} 
\label{subsec:mcmc_fullmodel}

Because the spectral components of the joint model in \eqref{formula:extendedMixture} do not fall under the general framework of \eqref{generalMixture.a}-\eqref{generalMixture.b}, Algorithm~\ref{algorithm:nestedPolya} must be adapted to fit our \jointmodel.  To do this, Algorithm~\ref{algorithm:joint} replaces Step 4 of Algorithm 1 with Steps 4 and 5: the former to update the source locations, $\Mu_l$, and the knots, $(\lon_l, \lat_l)$, the latter to update the Pareto parameters, $(\eta_s, \eta_b)$, of the spectral models. Each parameter is sampled in turn from its conditional posterior distribution.

\begin{algorithm}[t!]
	\caption{Collapsed Gibbs sampler for \jointmodel~\eqref{formula:extendedMixture} and for \imagemodel~\eqref{formula:3.1_mixturemodel}}
	\label{algorithm:joint}
	\flushleft 
		
	This pseudo-code shows how to adapt Algorithm~\ref{algorithm:nestedPolya} to perform posterior inference on Model \eqref{formula:extendedMixture}. The algorithm works also for Model \eqref{formula:3.1_mixturemodel}, just avoiding Step 5.
		
	\KwIn{Level~1 classification indicator,  $\mathbf{Z}^{(t-1)}$; level~2 classification indicators, $\textbf{H}^{(t-1)}_s$ and $\textbf{H}^{(t-1)}_b$; mixture sizes, $k^{(t-1)}_s$ and $ k^{(t-1)}_b$; spatial model parameters $\boldsymbol{\theta}^{(t-1)}_s=(\Mu_{1},\dots,\Mu_{k_s})^{(t-1)}$ and  $\boldsymbol{\theta}^{(t-1)}_b=((\lon_1,\lat_1),\dots,(\lon_{k_b},\lat_{k_b}))^{(t-1)}$; spectral model parameters $(\eta_s,\eta_b)^{(t-1)}$; integers $h_s = h_b = 5$; covariance matrix $\tilde{\Sigma} = \tilde{\sigma}^2\textnormal{I}_2$.}
		
    \Begin{
    Draw the additional values $\tilde{\boldsymbol{\theta}}_s=(\tilde{\Mu}_{1},\dots,\tilde{\Mu}_{h_s})$ from $\mathcal{F}_{0}$ and $\tilde{\boldsymbol{\theta}}_b=((\tilde{\lon}_{1},\tilde{\lat}_{1}),\dots,(\tilde{\lon}_{h_b},\tilde{\lat}_{h_b}))$ from $\mathcal{G}_{0}$; set $k\iter_s = k^{(t-1)}_s$ and $k\iter_b = k^{(t-1)}_b$.
        
    \For{$i = 1,\dots,n$}{
        Run \textbf{Steps 1 -- 3} of Algorithm~\ref{algorithm:nestedPolya}, where the subscript $s$ corresponds to $j = 1$ and the subscript $b$ to $j = 2$.
    }
    {\bf Step 4} For $l = 1,\dots,k\iter_s$, draw $\Mu^*_l\sim \mathcal{N}(\Mu^{(t-1)}_l,\tilde{\Sigma})$. Then $\Mu\iter_l = \Mu^*_l$ with probability $\min\{1,A_l\}$, where
    $$
    A_l = \prod_{i:\mathrm{H}\iter_{si} = l} \frac{\PSF(\x_i|\Mu^*_l,E_i)}{\PSF(\x_i|\Mu^{(t-1)}_l,E_i)},
    $$
    otherwise $\Mu\iter_l = \Mu^{(t-1)}_l$.\newline
    For $l = 1,\dots,k\iter_b$, draw $(\lon_l,\lat_l)\iter$ using the rejection sampler given in Appendix~\ref{app.A}, conditioned to $\{\x_i:\mathrm{H}\iter_{bi}\neq 0\}$.
        
    {\bf Step 5 (only for \eqref{formula:extendedMixture})} Draw $\eta\iter_s$ from the posterior distribution of $\eta_s|\{E_i:\Z\iter_i = s\},\alpha_{\eta_s},\beta_{\eta_s}$ and $\eta\iter_b$ from $\eta_b|\{E_i:\Z\iter_i = b\},\alpha_{\eta_b},\beta_{\eta_b}$.
    }
    \KwOut{$\textbf{\Z}\iter$ and  $k\iter_j$, $\mathbf{H}\iter_j$, $\boldsymbol{\theta}\iter_j$, $\eta\iter_j$, for $j=s,b$.}
\end{algorithm}	

Specifically, Step~4 updates each set of parameters from the spatial source model and the spatial background model. Each source location $\Mu_{l}$ is updated using a Metropolis algorithm with a Gaussian jumping rule centered at the current iteration and with variance-covariance matrix $\tilde{\Sigma} = \tilde{\sigma}^2\textnormal{I}_2$. In our simulations, $ \tilde{\sigma}^2 = 0.001$ yields an acceptance rate of around 0.4. Each knot of the background model in $(\lon_l,\lat_l)$ is updated by drawing from its conditional posterior distribution given the other knots using a rejection sampler with a uniform proposal distribution. Before accepting the knots, we check that condition in \eqref{formula:constraint} holds; details appear in Appendix \ref{app.A}.  
    
Finally, Step~5 takes advantage of the conjugacy between the Gamma and the Pareto distributions to update $\eta_s$ and $\eta_b$ from independent Gamma distributions, given the other unknown quantities. By removing Step 5, Algorithm \ref{algorithm:joint} can be used for posterior inference on the \imagemodel~\eqref{formula:3.1_mixturemodel}.

\subsection{Post-processing} 
\label{subsec:post-processing}

\begin{algorithm}[t]
	\caption{Posterior analysis algorithm}
	\label{algorithm:labelswitching}
	\KwIn{The posterior draws $\{\Mu^{(t)}_l,t = 1,\dots,T\textnormal{ and }l=1,\dots,k^{(t)}_s\}$. Values $p^*=0.95$ and $k^*= \arg\max_k \sum_t \mathds{1}(k\iter_s = k)$. An odd integer $d_\mathcal{R}$.}
		
	\Begin{
		{\bf Step 1} Pool the draws $\{\Mu^{(t)}_l\}$ into a grid of pixels of given size, counting the number of draws in each pixel. 
			
		{\bf Step 2} Select the $k^*$ pixels which correspond to the highest local maxima (a local maximum is a pixel which has more counts than in any of its 8 neighbors), and label these pixels as the regions $\mathcal{R}_1,\dots,\mathcal{R}_{k^*}$.
			
		{\bf Step 3} Compute $\Pr(K_{\mathcal{R}_m} \geq 1|\dots)$, the posterior probability that there is at least one source in region $\mathcal{R}_{m}$, using \eqref{formula_algorithm:probability_postprocessing}.
			
		\For{$m=1,\dots,k^*$}{
			\While{$\Pr(K_{\mathcal{R}_m} \geq 1|\dots)<p^*$}{
		    {\bf Step 4} extend $\mathcal{R}_m$ by including the adjacent pixel which would increase $\Pr(K_{\mathcal{R}_m} \geq 1|\dots)$ by the largest amount among the $d_\mathcal{R}\times d_\mathcal{R}$ square of pixels around the local maximum pixel. Recompute $\Pr(K_{\mathcal{R}_m} \geq 1|\dots)$.
				
			{\bf Step 5} If the size of $\mathcal{R}_m$ is equal to a $d_\mathcal{R}\times d_\mathcal{R}$ square of pixels, then {\bf break} the {\bf while} cycle.
			}
		}

		{\bf Step 6} 
		For every $t$ and $l$, if $\Mu^{(t)}_l\in \mathcal{R}_m$, then set $\phi\iter_l=m$. If $\Mu^{(t)}_l\notin \{\bigcup_m \mathcal{R}_m\}$, put $\phi\iter_l=0$.
		}
	\KwOut{$\{\Pr(K_{\mathcal{R}_1} \geq 1|\dots),\dots,\Pr(K_{\mathcal{R}_{k^*}} \geq 1|\dots)\}$ and $\{\phi\iter_l,t = 1,\dots,T\textnormal{ and }l=1,\dots,k^{(t)}_s\}$.}
\end{algorithm}
	
MCMC samplers for mixture models are prone to label switching, that is, the swapping of mixture component indices in the evolving Markov chain  \citep{fruhwirth-schnatter.2011}. Without correction, this leads to multiple artificial modes in the posterior distribution of the model parameters. In our numerical studies, label switching arises in both the DP mixtures $\smixture$ and $\bmixture$. Multimodality can also arise when the first level mixture labels $\Z_i$ switch repetitively, causing DP mixture components to be added or removed (in Steps~2 and 3 of Algorithm~\ref{algorithm:nestedPolya}, respectively).

To enable mixture-component-specific parameter inference, we propose Algorithm~\ref{algorithm:labelswitching} for post-processing. Algorithm~\ref{algorithm:labelswitching} determines regions of the map that are likely to contain sources by relabelling the location parameters $\Mu_l$ in the Markov chain. As a by-product, the algorithm also addresses the label switching issue. 

Step~1 of Algorithm~\ref{algorithm:labelswitching} pools the chains of posterior draws of all the source locations, $\{\Mu\iter_l \}_{t,l}$, and tabulates them together in a grid of pixels of given size. Step~2 collects the $k^*$ highest local maxima (a local maximum is a pixel which has more counts than in any of its 8 neighbors)  and labels them as the regions $\mathcal{R}_1,\dots,\mathcal{R}_{k^*}$. Let $K_{\mathcal{R}_m}$ be the number of sources in the $m$-th region. Step~3 computes the posterior probability that at least one source is in $\mathcal{R}_m$ as
\begin{equation}
\label{formula_algorithm:probability_postprocessing}
	\Pr(K_{\mathcal{R}_m}\geq 1|\dots)=\sum_t \mathds{1}\left\{\sum_l \mathds{1}\left(\Mu\iter_l\in \mathcal{R}_{m}\right) >0  \right\}/T,
\end{equation}
where \textquoteleft$\dots$\textquoteright~ denotes the conditioning on the data. If $\Pr(K_{\mathcal{R}_m}\geq 1|\dots)< p^*$, for some threshold $p^*$ (e.g., $p^* = 0.95$), Steps~4 - 5 enlarge $\mathcal{R}_m$ by adding adjacent pixels contained in the $d_\mathcal{R}\times d_\mathcal{R}$ square of pixels around the local maximum pixel until $\Pr(K_{\mathcal{R}_m} \geq 1|\dots)\geq p^*$; alternatively, if the size of $\mathcal{R}_{m}$ grows to a certain size (e.g., a $d_\mathcal{R}\times d_\mathcal{R}$ square of pixels), no further pixels are added. In Sections 4 and 5, which are based on the \emph{Fermi} LAT data, we will set $d_\mathcal{R} = 3$ because of the narrow \PSF. Even for wider \PSF, we advice against a large value for $d_\mathcal{R}$ as excessively large regions are  of  little use. Finally, Step~6 sets $\phi\iter_l=m$ if $\Mu\iter_l\in\mathcal{R}_m$, and $\phi\iter_l=0$ if $\Mu\iter_l$ is not contained in any of the regions $(\mathcal{R}_1,\dots,\mathcal{R}_{k^*})$. 

In some cases, $\phi\iter_l = \phi\iter_{l'}$ for some $l\neq l'$, meaning two source locations drawn in the same MCMC iteration are assigned to the same region. 
The conditional posterior probability of the number of sources in $\mathcal{R}_m$, given that there is at least one source, is 
$$
\Pr(K_{\mathcal{R}_m} = k|K_{\mathcal{R}_m}>0, \dots) = \frac{\sum_t \mathds{1}\left\{\sum_l \mathds{1}\left(\phi\iter_l = m\right) = k  \right\}}{\sum_t \mathds{1}\left\{\sum_l \mathds{1}\left(\phi\iter_l = m\right) > 0 \right\}}.
$$
For instance, $\Pr(K_{\mathcal{R}_m}>1|K_{\mathcal{R}_m}>0,\dots)$ is the probability that $\mathcal{R}_m$ contains the signal of multiple overlapping sources, assuming the presence of at least one source in $\mathcal{R}_m$.
 
Finally, we consider MCMC iterations with the same value of $\phi$ when conducting mixture-component-specific parameter inference. For example, conditional on $K_{\mathcal{R}_m} = 1$, the set of draws $\{\Mu\iter_l: \phi\iter_l = m \hbox{ and }\phi\iter_{l'}\neq m \hbox{ for } l'\neq l\}$ is a posterior sample of the location of the only source contemplated in region $\mathcal{R}_m$.

\section{Simulation Studies}
\label{sec:simulationapplication}

\noindent
We demonstrate our methods on simulation-based experiments of increasing complexity. We simulate the $10\degree\times10\degree$ region of the $\gamma$-ray sky, as observed by the {\em Fermi}~LAT, surrounding the dwarf spheroidal galaxy \emph{Hydrus I} \citep{koposov_etal.2018}. The center of the region is $37\degree$ south of the Galactic Plane, relatively far from the complex diffuse emission arising from the Galactic Disk.

\subsection{Simulation model}
\label{subsection:SimulationModel}

\noindent
Although our inferential model uses the continuous measured positions and energies of the photons, it is most convenient to simulate data in discrete spatial and energy bins. The spatial region is divided into $200\times 200$ pixels of size $0.05\degree \times 0.05\degree$. The energy range (1~GeV to 316~GeV) is divided into 25 $\log_{10}$-equispaced bins of size $\Delta \log_{10} E/\text{GeV} = 0.1$. We generate simulated photon counts in a 3-dimensional array, $Y_{uvz}$, where the first two indices correspond to spatial pixel and the third corresponds to energy. The centroid of spatial bin $(u,v)$ (that is, the arithmetic mean of the bin limits) is denoted $(x_u,y_v)$, while $E_z$ is the centroid of energy bin $z$ (geometric mean of the bin limits). After generating photon counts in each bin, the individual photons are assigned positions and energies equal to the centroid $(x_u,y_v,E_z)$ of the bin they occupy. The spatial and energy bins are fine enough that this discretization has a minimal impact compared to a fully continuous sampling of photon positions and energies.
	 
Let $S$ be the number of simulated sources, with locations $(\boldsymbol{\psi}_1,\dots,\boldsymbol{\psi}_\varsigma)$. The photon counts in array element $(u,v,z)$ are simulated as
\begin{equation}
Y_{uvz}\sim \hbox{Poisson}\left(\sum_{\varsigma=1}^{S}\varLambda_{uvz}^{\boldsymbol{\psi}_\varsigma}+\varLambda_{uvz}^b\right), \hspace{.5cm}
\begin{array}{l@{}}
u,v = 1,\dots,200,\\ 
z = 1,\dots,25,
\end{array}
\label{formula:data_generator_poisson}
\end{equation}
where $\varLambda_{uvz}^{\boldsymbol{\psi}_\varsigma}$ is the expected simulated photon count from the source located at $\boldsymbol{\psi}_\varsigma$ and $\varLambda_{ijk}^{b}$ is the expected photon count from the background. In order to conduct realistic experiments, we simulate the source emissions using the \emph{Fermi} LAT \PSF~and a power-law spectral model,
\begin{equation}
\label{formula:data_generator_sources}
\varLambda^{\boldsymbol{\psi}_\varsigma}_{uvz} = F_{0,\varsigma}\left(\frac{E_z}{1\;\text{GeV}}\right)^{-\varrho_\varsigma} \cdot \text{PSF}(x_u,y_v|\boldsymbol{\psi}_\varsigma,E_z)\cdot \epsilon(E_z),
\end{equation}
where 
$F_{0,\varsigma}$ denotes the \emph{amplitude} of source $s$ and controls its brightness (that is, the total flux of photons reaching the detector), $\varrho_\varsigma$ denotes the \emph{spectral shape} of source $\varsigma$ and controls the distribution of its photon energies, and $\epsilon(E_z)$ denotes the \emph{exposure} of energy bin $z$ and combines the duration of the observation and the instrument's \emph{effective area}.  

Here, the point spread function, $\PSF(x_u,y_v|\boldsymbol{\psi}_\varsigma, E_z)$  denotes the probability that a photon with location $\boldsymbol{\psi}_\varsigma$ and energy $E_z$ is recorded in the spatial pixel centered at $(x_u,y_v)$. The background expectation $\varLambda_{uvz}^b$ is given by the model developed by \cite{acero_etal.2016} and shown in Appendix \ref{app.B}, Figure \ref{fig:appendix_simulationdatasets}(a). Both the PSF and background expectation map are prepared as described in Section~5.4 of~\cite{koposov_etal.2018}. 

In our experiments we consider the energy range above 1~GeV, where the effective area of the \emph{Fermi} LAT is essentially constant in energy and the measurement uncertainty for photon energies is small. Thus, we need not incorporate the refinement to the spectral model given in \eqref{formula:EAconvolution}. However, when we fit Model~\eqref{formula:3.1_mixturemodel}, which ignores photon energies, the simulation and inference model are mismatched. Still, we may expect reasonable performance because, for power-law spectra ($\varrho_\varsigma >0$), the data are dominated by lower-energy photons, around 1~GeV in our case. Therefore, to a first approximation, the resulting inference on the spatial location of a source can be expected to be similar to the case in which all photons have the same energy. We explore in Section~\ref{subsec:model_comparison} the performance improvement when we include the additional energy information in our model.

\subsection{An illustrative demonstration}
\label{subsection:illustrative_example}

\noindent
As a first illustrative example, we simulate 9 equally bright sources, all with the same amplitude, $F_{0,\varsigma}= 10^{-9}$, common spectral shape, $\varrho_\varsigma=2$ ($\varsigma=1,\dots,9$), and with a moderate level of background emission generated from the model of \cite{acero_etal.2016}. 

\begin{figure}[t]
	\centering
	\includegraphics[width=6.2cm, height=5.5cm]{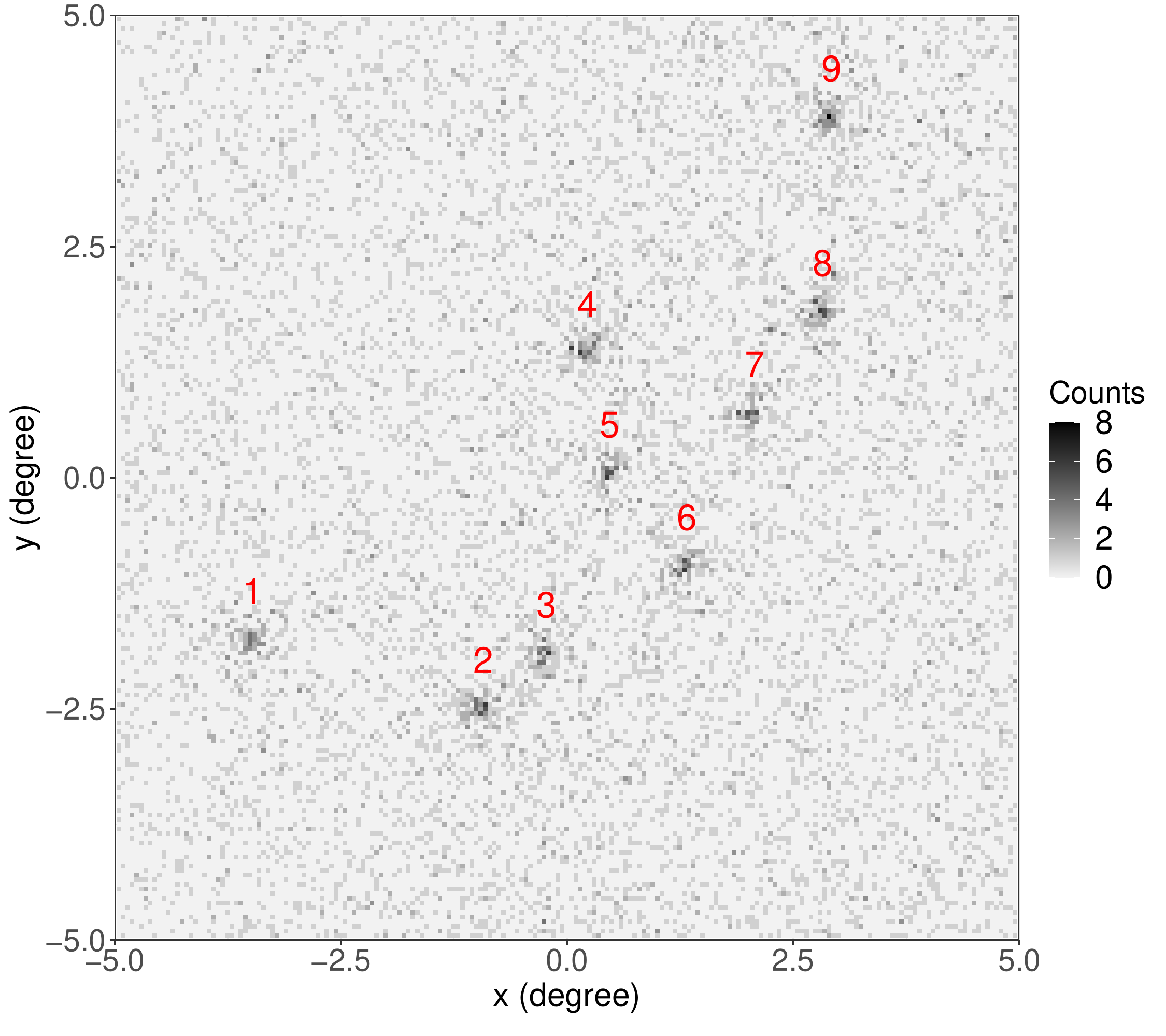}
	\raisebox{-.015\height}{\includegraphics[width=6.1cm, height=5.5cm]{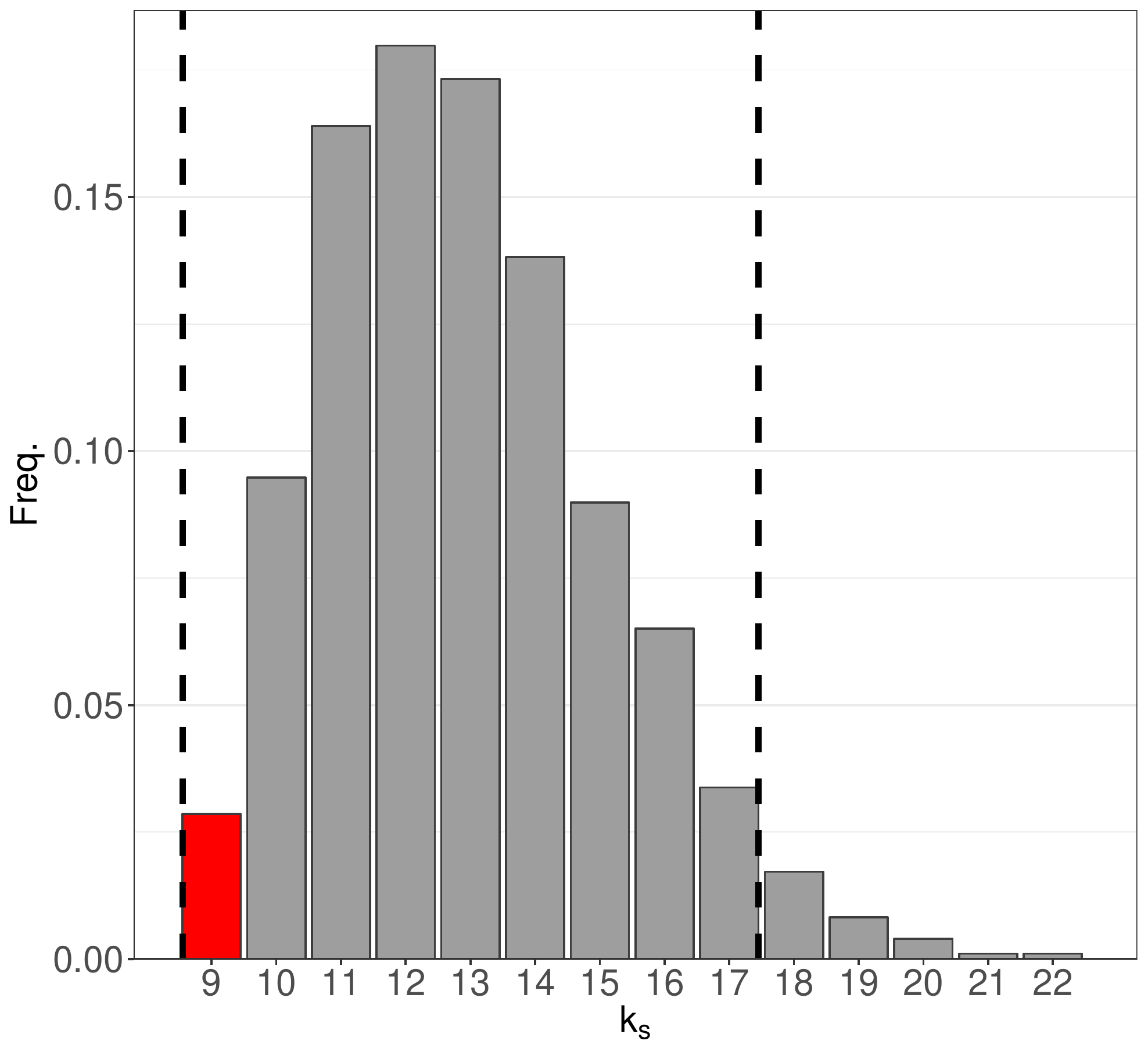}}
	\caption{Left: binned map of 10,000 simulated photons from 9 identical sources ($F_{0,\varsigma} =  10^{-9}$, $\varrho_\varsigma = 2$ for $\varsigma = 1,\dots,9$) and a moderate background contamination generated by the background model of \cite{acero_etal.2016} (see Figure \ref{fig:appendix_simulationdatasets} (a) of Appendix \ref{app.B}). Red numbers label the 9 sources, which in this example are readily distinguished by eye. Right: posterior distribution of $k_s$. The number of sources used for the simulation is indicated in red. Dashed vertical lines delimit the 95\% HPD interval. Notice that there is no MCMC iteration where the source mixture has size $k_s <9$.}
	\label{fig:applied1}
\end{figure}

Simulation under \eqref{formula:data_generator_poisson} yielded 25,140 photons, corresponding approximately to an observation period of 9.4~years.  To reduce computation time, we work with a dataset of 10,000 randomly selected counts from this simulation, which would be observed over a period of approximately 3.7~years. The final binned map of photons is shown in left panel of Figure~\ref{fig:applied1}; the sources are labelled from 1 to 9 in longitude ordering. 

In this section, we conduct inference only under the \imagemodel, postponing comparisons with the \jointmodel\ to Section~\ref{subsec:model_comparison}. To fit the \imagemodel, we must set the hyperparameters of the DPs. The  concentration parameters $\alpha_s$ and $\alpha_b$ control the number of components of the mixtures $\smixture$ and $\bmixture$, respectively: the larger the concentration parameter, the more components the DP tends to add to the fitted model. We want a prior set-up that favours the inclusion of new sources, allowing for a reasonably large $k_s$, which therefore limits the increase of the background mixture size $k_b$.   \citet[][Section~3.1.2]{muller_rodriguez.2013} specifies that $\mathbb{E}(k_s|n_s)=\sum_{i=1}^{n_s} \alpha_s/(\alpha_s+i-1)$, and, from our model, $n_s|\delta\sim \hbox{Binomial}(n,\delta)$ and $\delta\sim \hbox{Beta}(\lambda,\lambda)$. By setting $\lambda = 1$, our inference models assumes that all photon have equal chances \textit{a priori} of coming from the sources or from the background. Additionally, we encourage $k_s$ to be larger than $k_b$ setting $\alpha_s=2$ and $\alpha_b = 1.5$. With this set-up, the \textit{a priori} expectation of the number of sources is approximately 16, and the expectation of the number of background components is approximately 12. Additional simulation experiments show that, due to the  large size of the simulated dataset, the final conclusions do not change appreciably if we vary the prior set-up.

\begin{figure}
	\centering
	\scalebox{.95}{
		\hspace{-.4cm}
		\includegraphics[height=4.8cm, width=4.8cm]{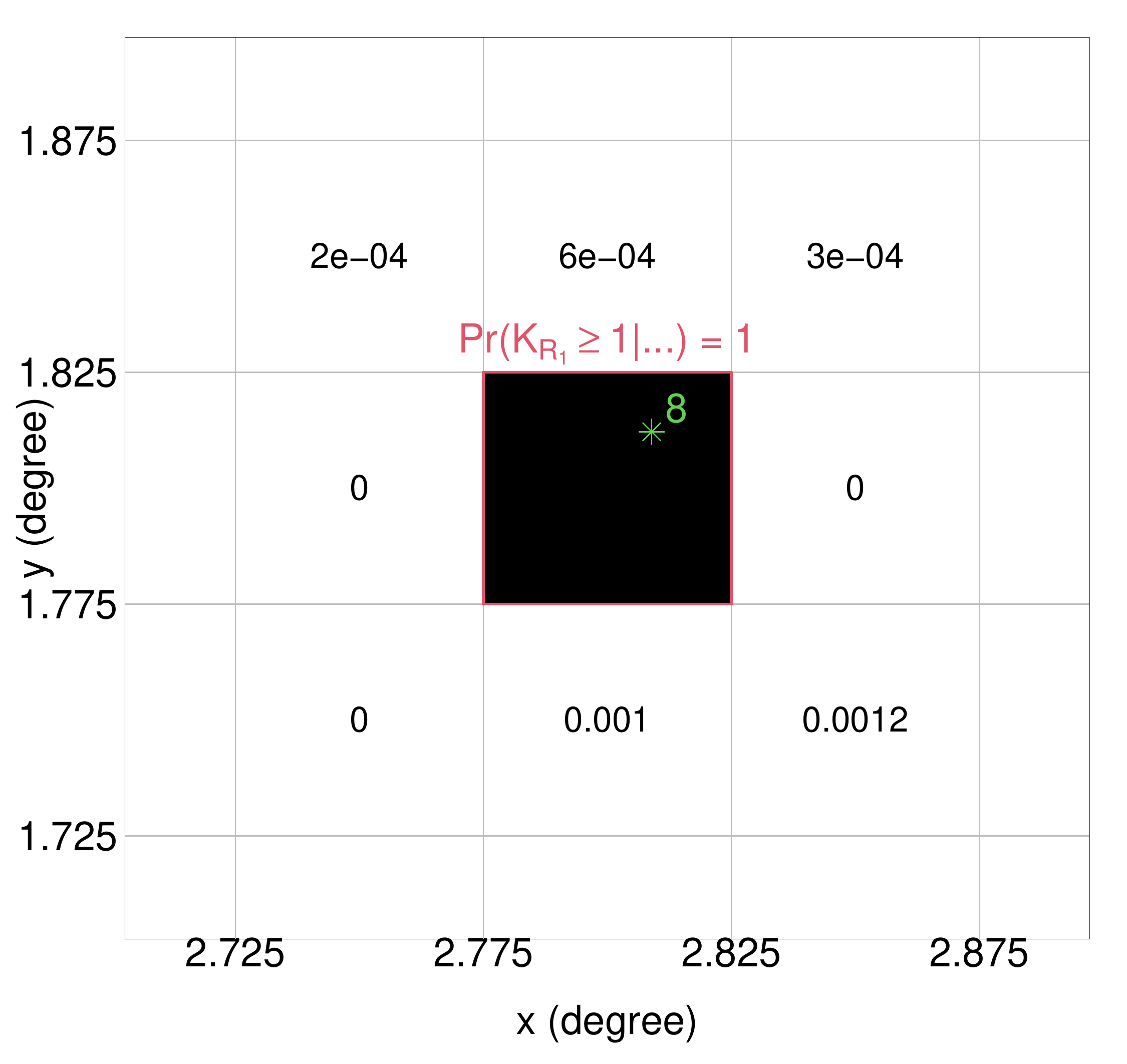}\hspace{-.2cm}
		\includegraphics[height=4.8cm, width=4.8cm]{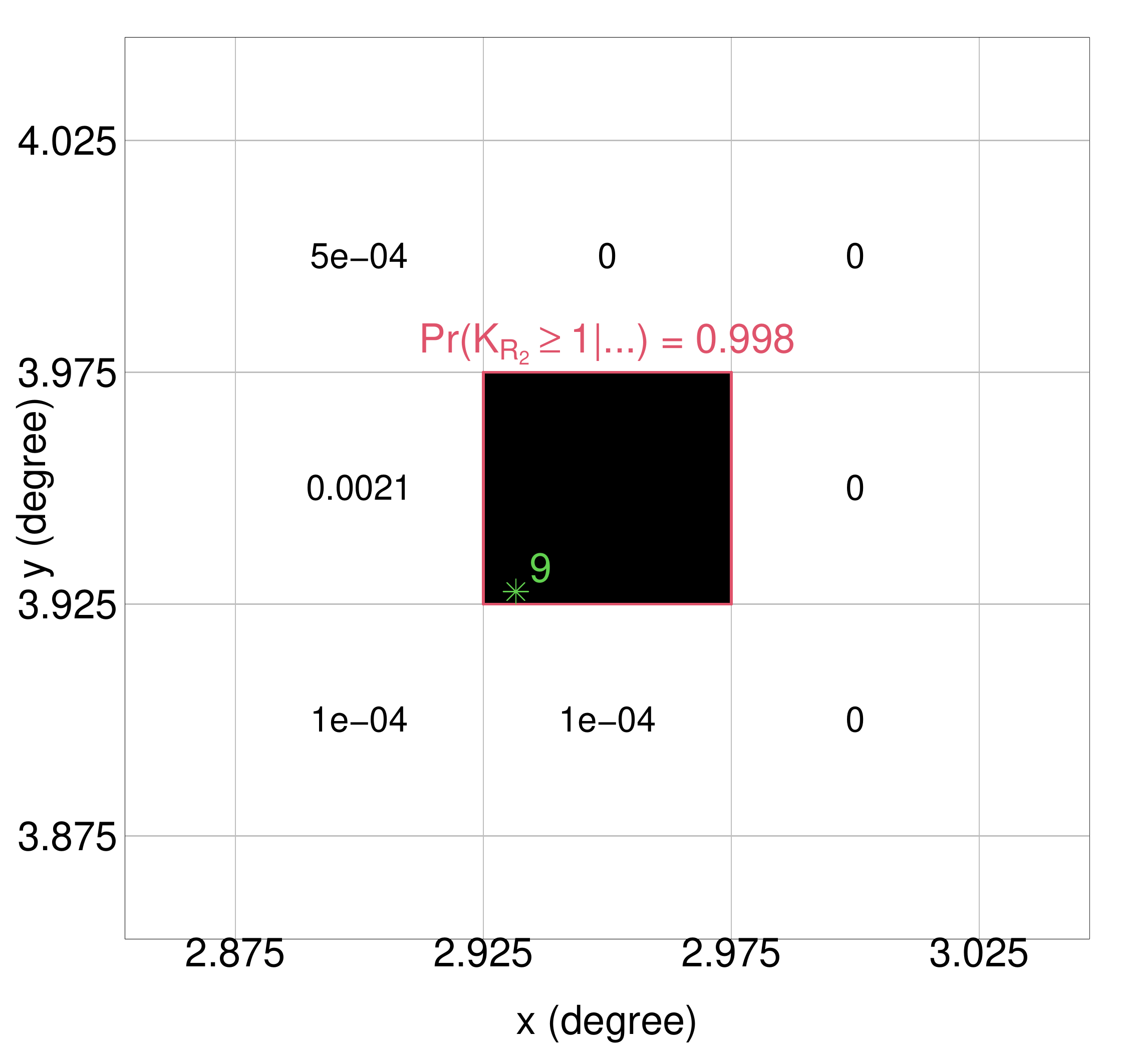}\hspace{-.2cm}
		\includegraphics[height=4.8cm, width=4.8cm]{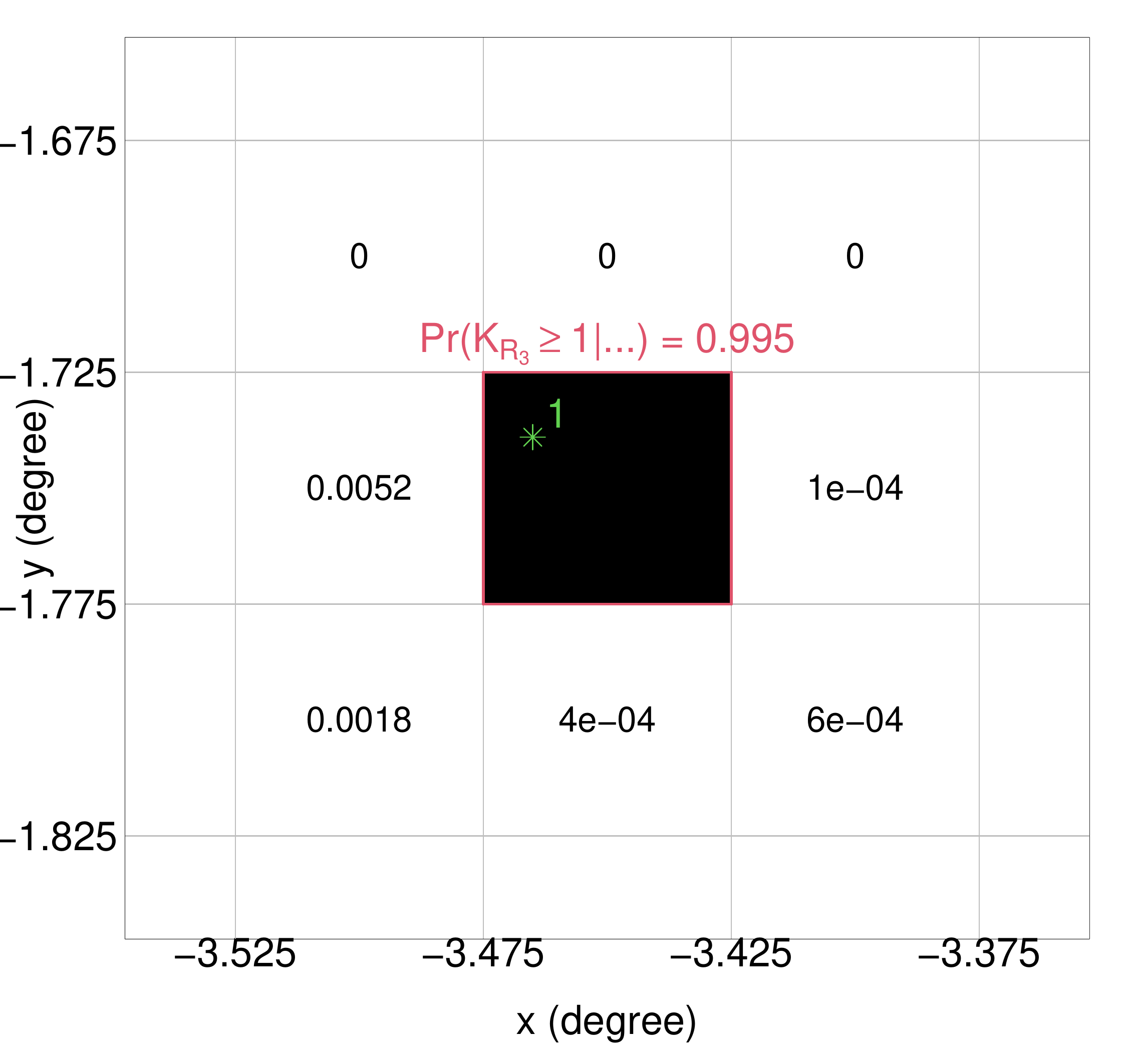}}\\
	\vspace{-.1cm}
	\scalebox{.95}{
		\hspace{-.4cm}
		\includegraphics[height=4.8cm, width=4.8cm]{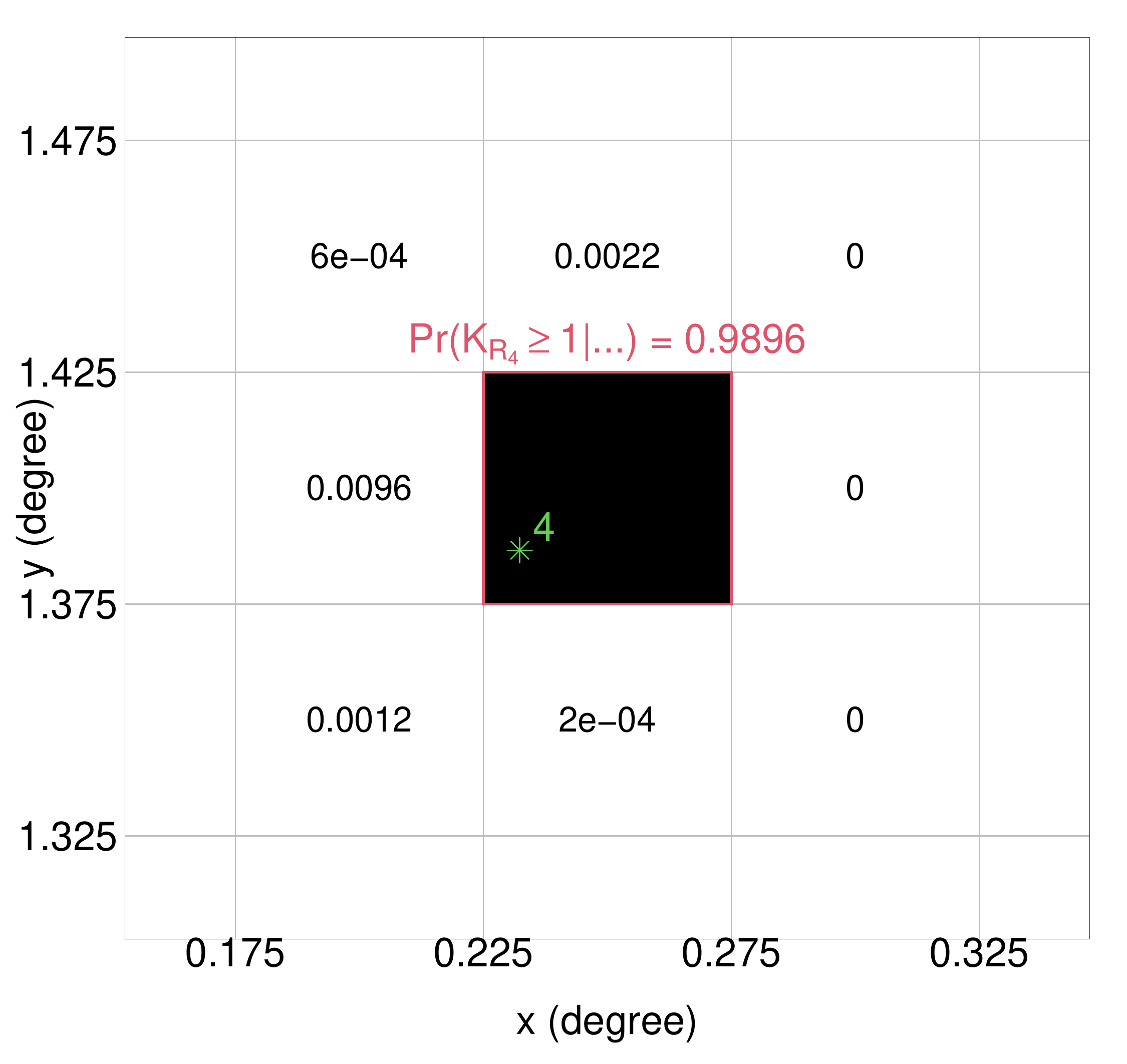}\hspace{-.2cm}
		\includegraphics[height=4.8cm, width=4.8cm]{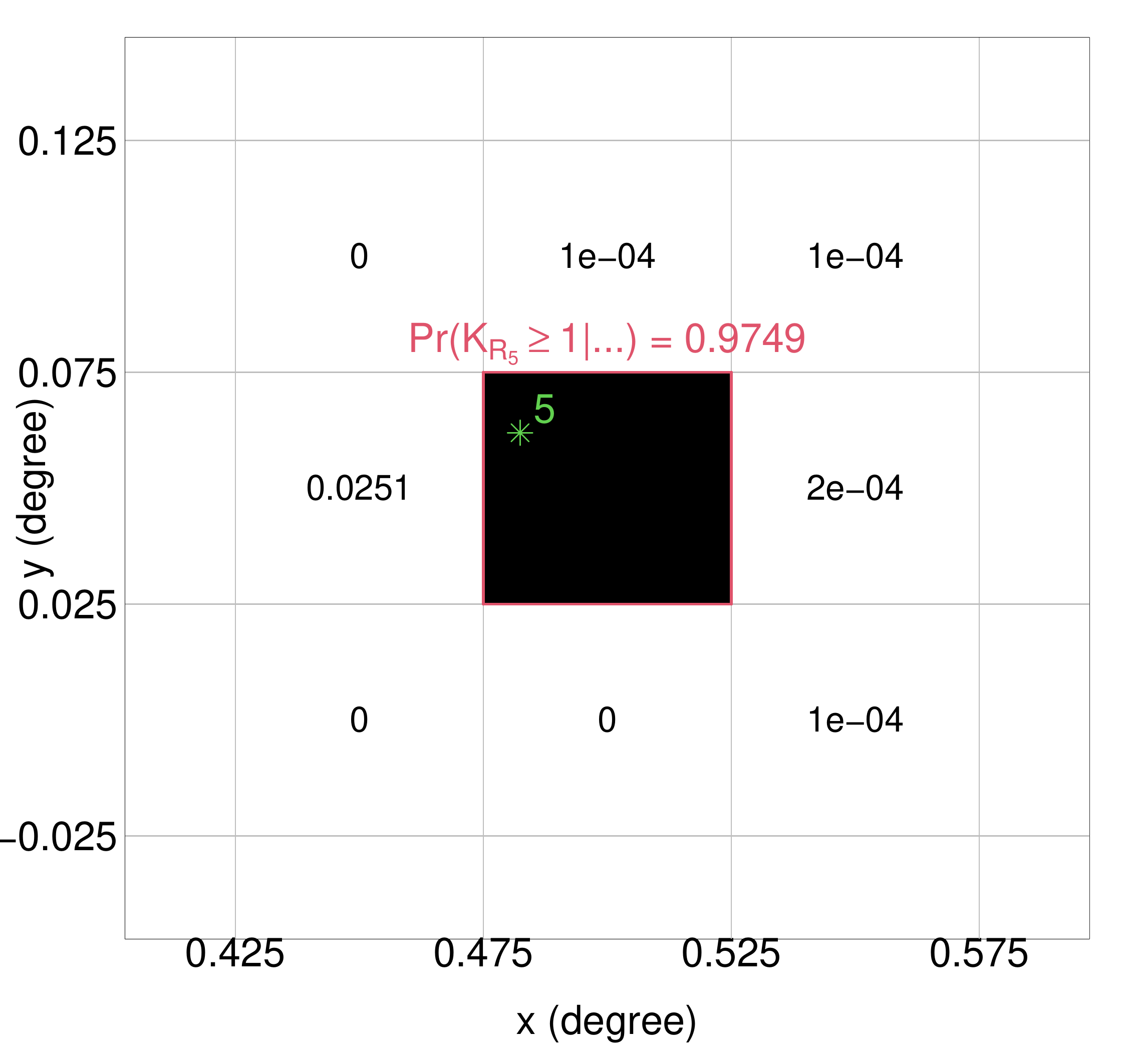}\hspace{-.2cm}
		\includegraphics[height=4.8cm, width=4.8cm]{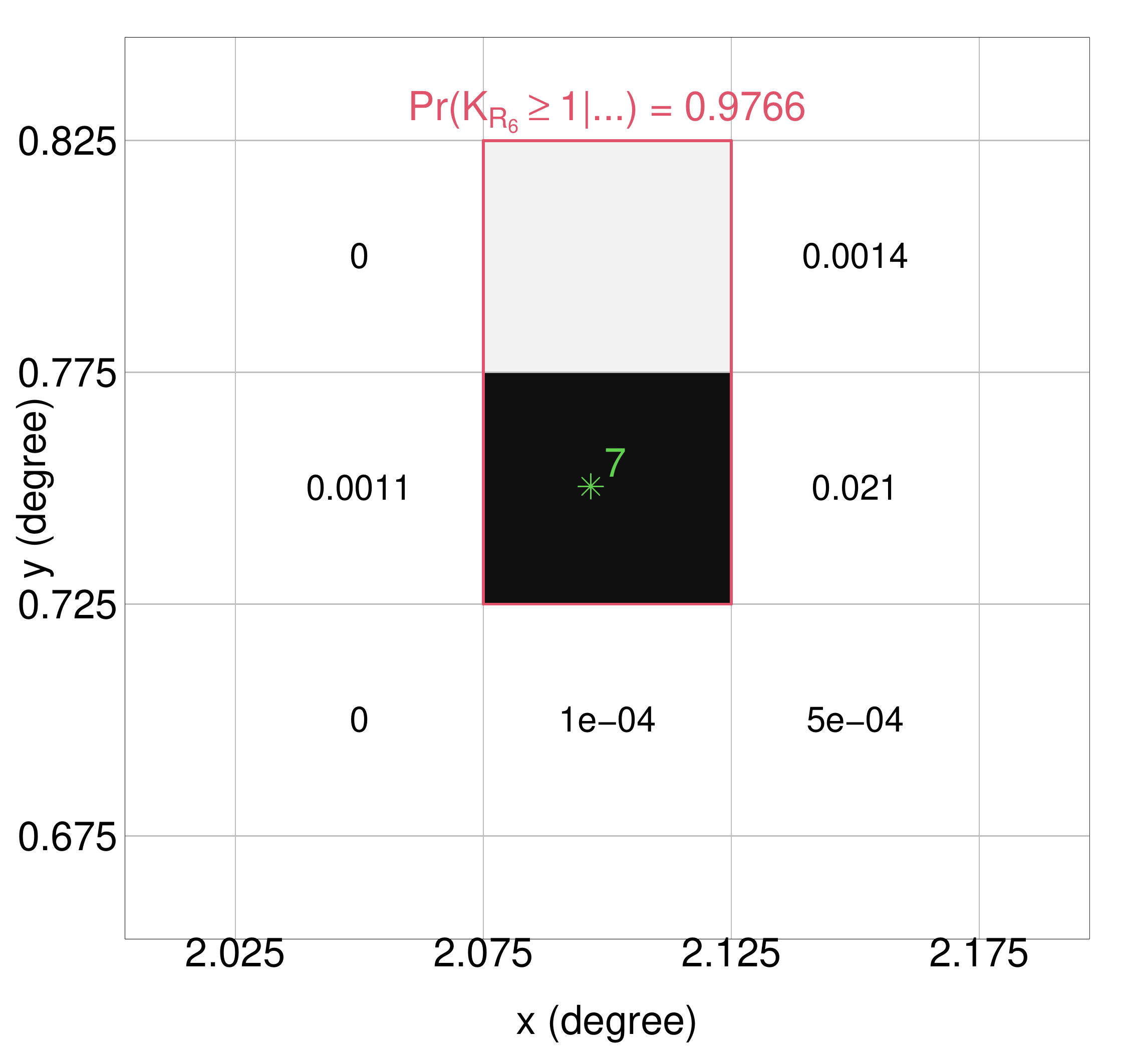}}\\
	\vspace{-.1cm}
	\scalebox{.95}{
		\hspace{-.4cm}
		\includegraphics[height=4.8cm, width=4.8cm]{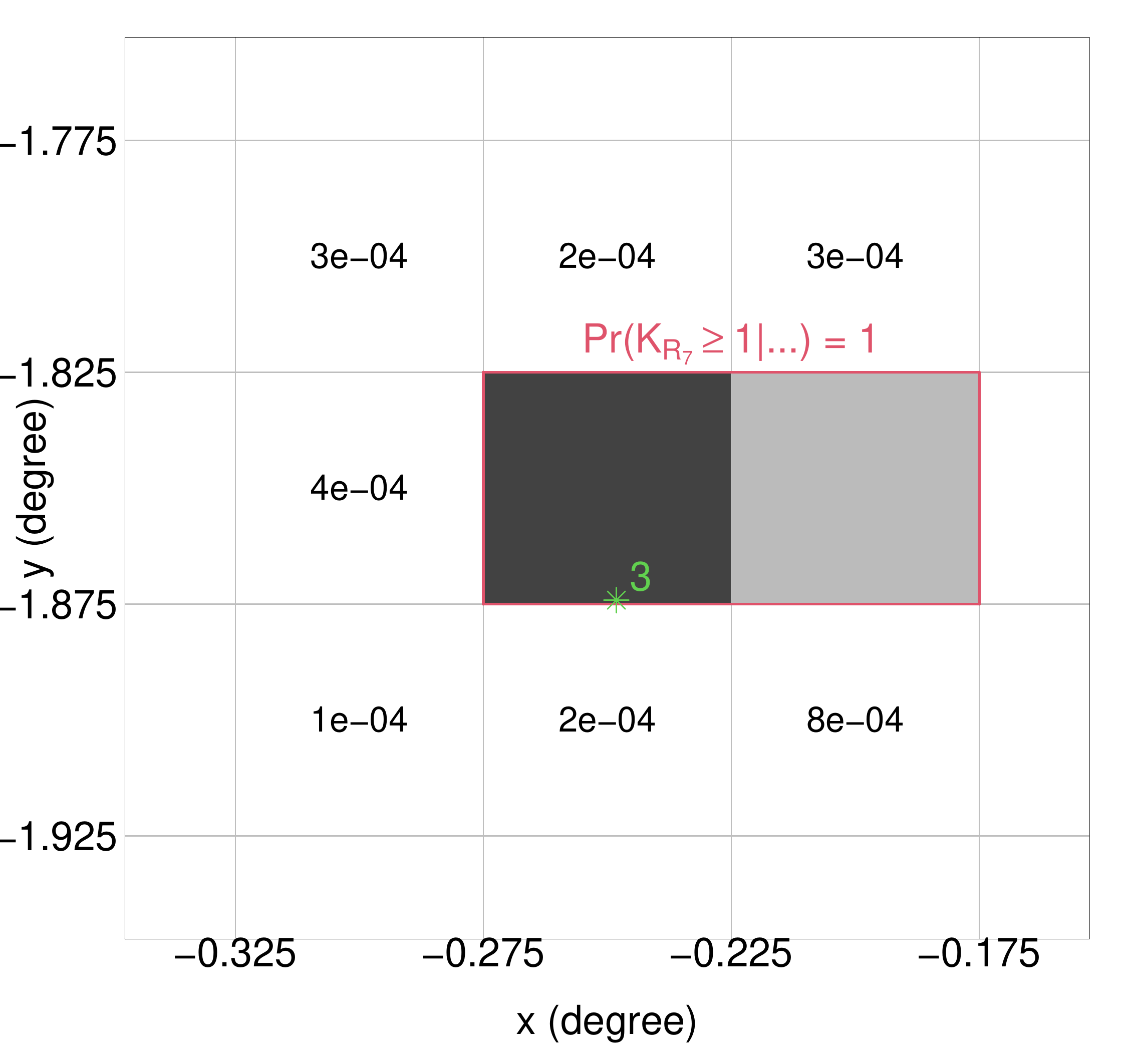}\hspace{-.2cm}
		\includegraphics[height=4.8cm, width=4.8cm]{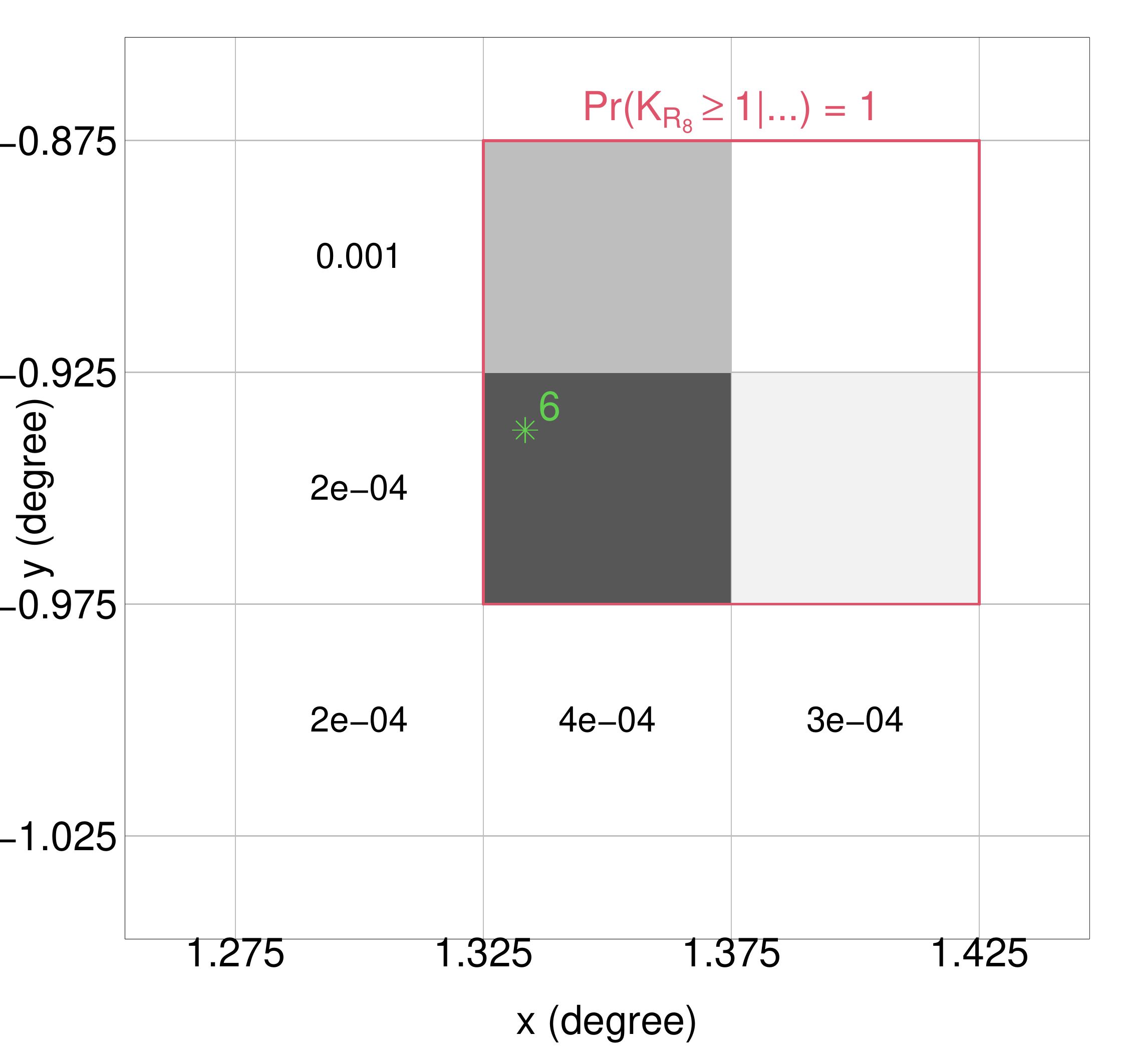}\hspace{-.2cm}
		\includegraphics[height=4.8cm, width=4.8cm]{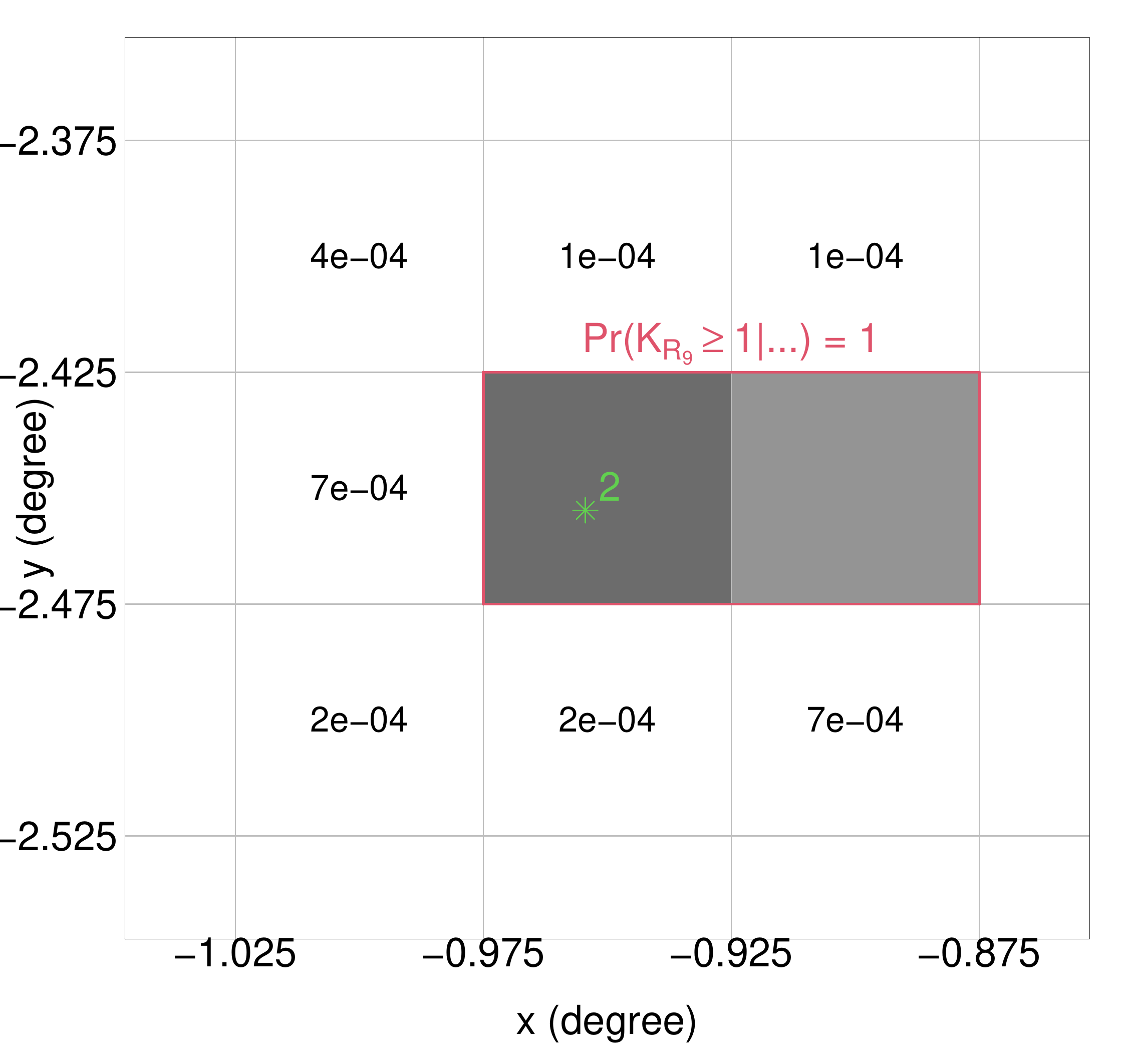}}\\
	\vspace{-.1cm}
	\scalebox{.95}{
		\hspace{-.4cm}
		\includegraphics[height=4.8cm, width=4.8cm]{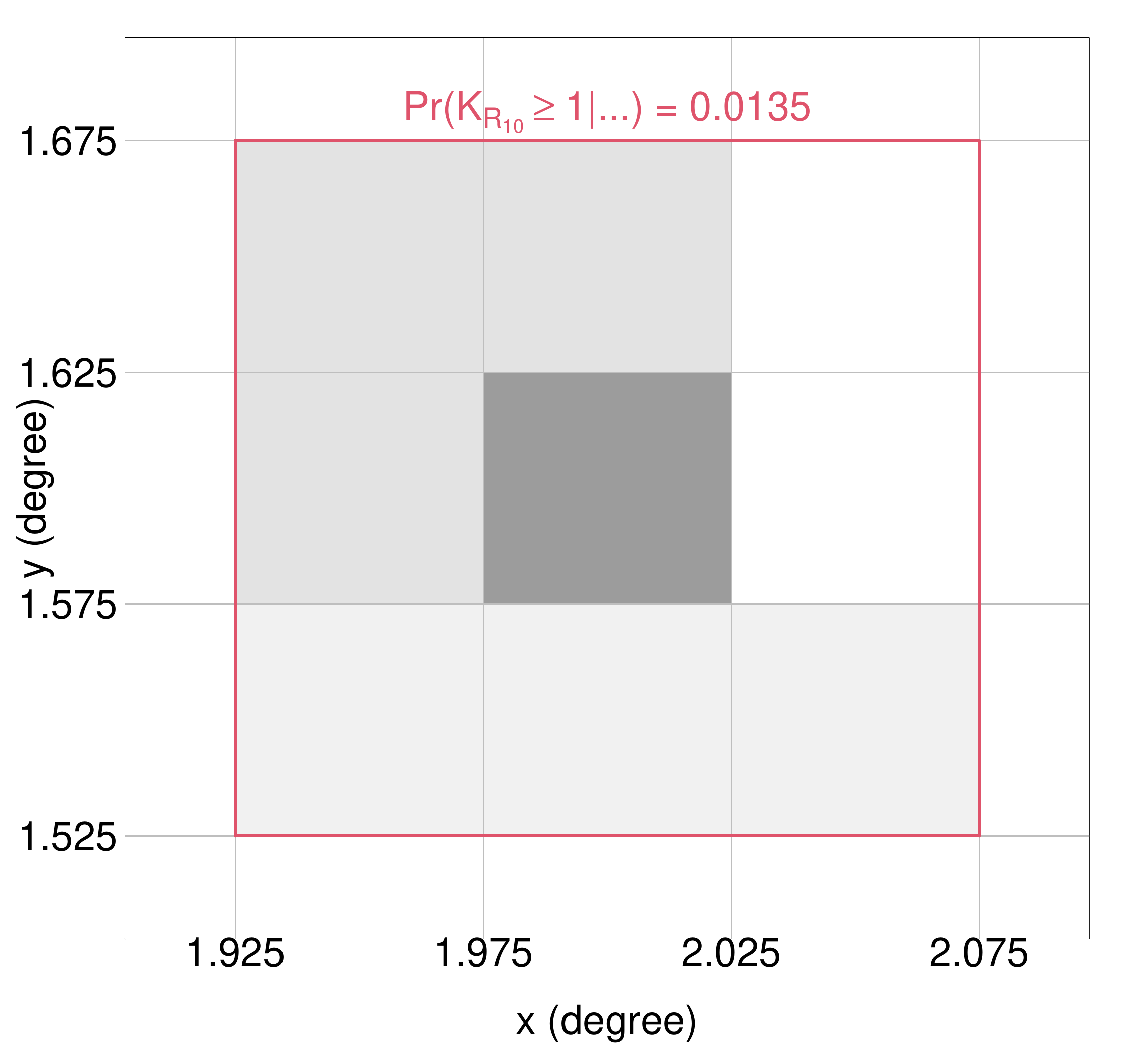}\hspace{-.2cm}
		\includegraphics[height=4.8cm, width=4.8cm]{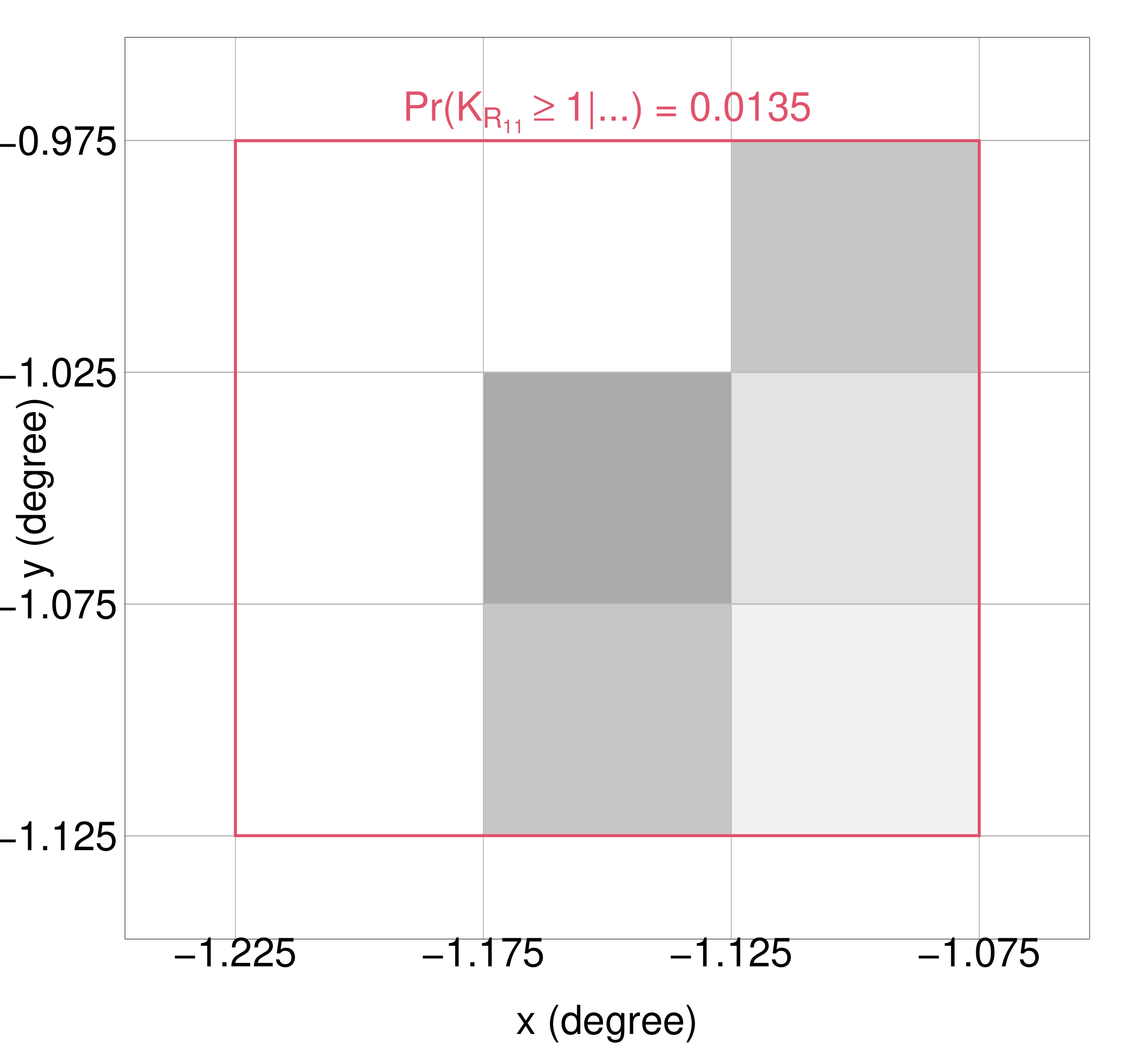}\hspace{-.2cm}
		\includegraphics[height=4.8cm, width=4.8cm]{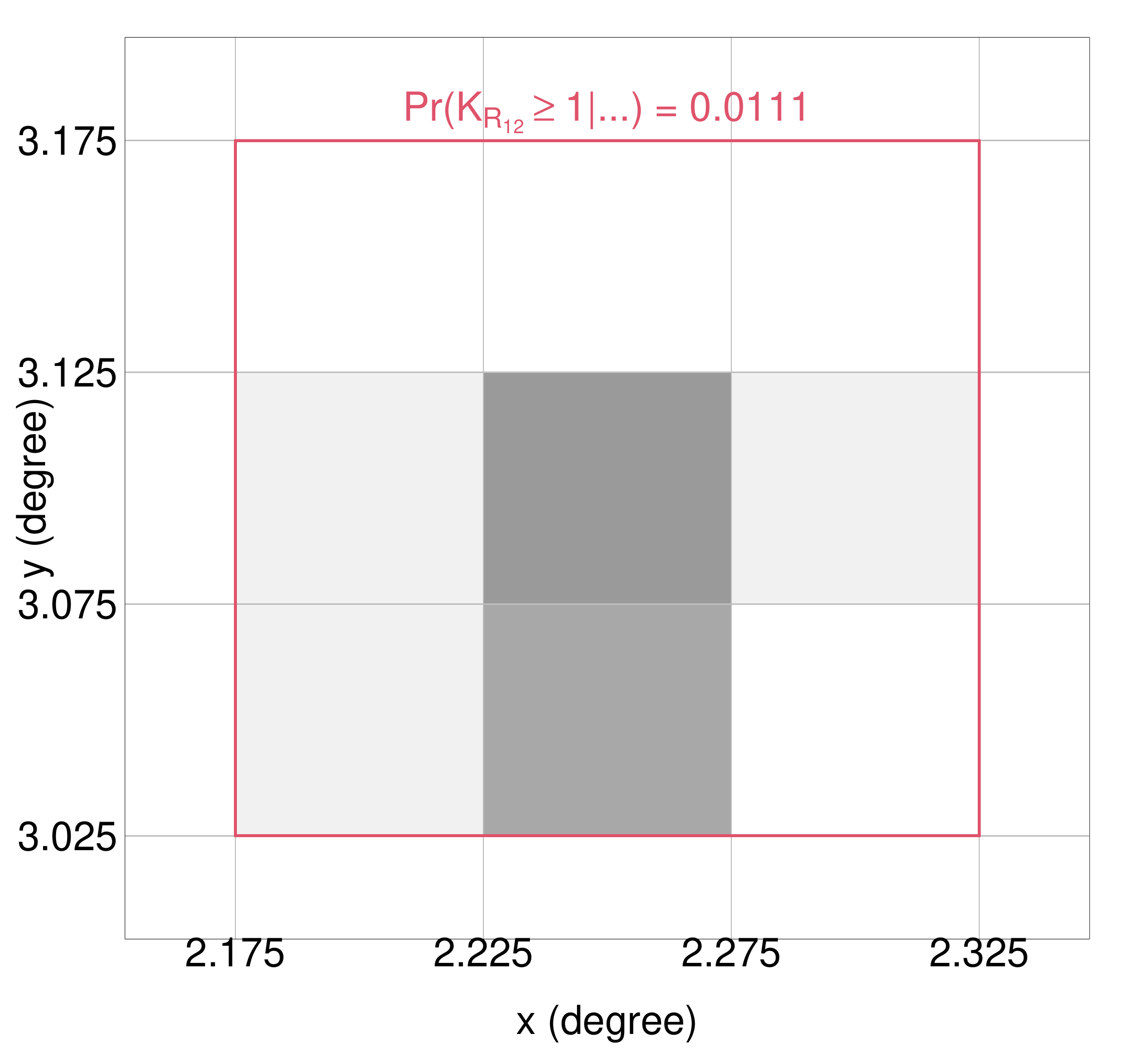}}\\
	\hspace{1.2cm}
	\includegraphics[height=1cm, width=5cm]{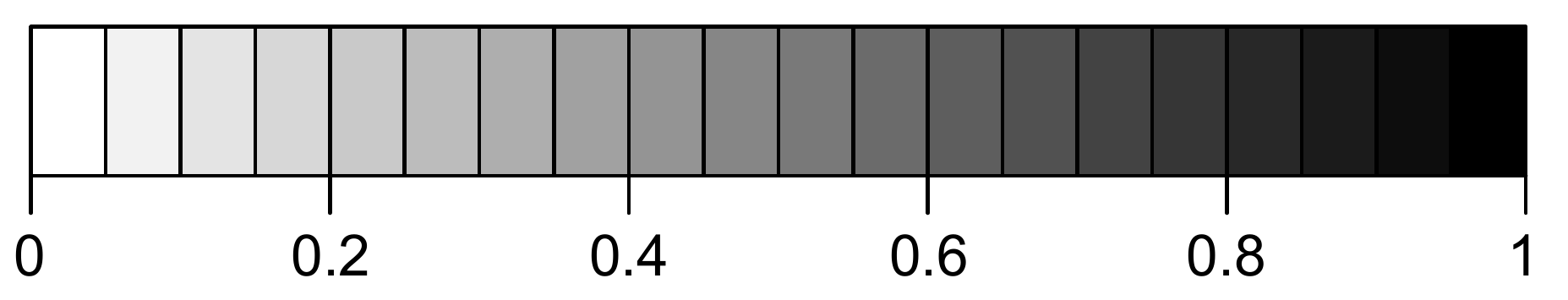}
	\caption{
		Results from Section \ref{subsection:illustrative_example}. Panels display the regions $\mathcal{R}_1,\dots,\mathcal{R}_{12}$, highlighted by red lines. The numbers give the probability that each pixel of size $0.05\degree\times 0.05\degree$ includes a point source. Pixels within the regions are coloured according to the posterior distribution of $\Mu$, conditional on $K_{\mathcal{R}_m} = 1$, so assuming that $\mathcal{R}_m$ includes exactly one source. Green stars represent the source locations for the simulated data and green numbers denote the source labels as in Figure \ref{fig:applied1}. The bottom row of locations does not include any sources.}
	\label{fig:posterior_regions}
\end{figure}

We run four separate Markov chains of length 10,000 in parallel using Algorithm \ref{algorithm:nestedPolya}, and discard the first three quarters of each as burn-in. The remaining draws are then combined to obtain a final posterior sample of the model parameters. The right panel of Figure \ref{fig:applied1} displays the posterior distribution of the number of sources, $k_s$, and the 95\% highest posterior density (HPD) interval (dashed vertical lines). The HPD interval indicates that there are between 9 and 17 sources, with a maximum \emph{a posteriori} estimate of 12. Deploying Algorithm \ref{algorithm:labelswitching} using the same grid of pixels of size $0.05\degree\times 0.05\degree$ used to simulate the data, we examine the 12 distinct regions of the map, $\mathcal{R}_1,\dots,\mathcal{R}_{12}$: Figure \ref{fig:posterior_regions} shows these regions, delimited by red lines, together with their surrounding pixels. For every region, $\Pr(K_{\mathcal{R}_m}>1|K_{\mathcal{R}_m}>0,\dots)<.01$, that is, the probability that the region contains overlapping sources is negligible. Pixels within the regions are colour-coded according to the posterior distribution of $\Mu$, given $K_{\mathcal{R}_m} = 1$ for all $m$. If follows that, if $\mathcal{R}_{m}$ consists of a single pixel, the probability distribution of $\Mu$ is concentrated within that pixel. 

The first nine regions of Figure \ref{fig:posterior_regions} contain the nine actual locations of the simulated sources (indicated by green stars); the probabilities under our model are $\Pr(K_{\mathcal{R}_m}\geq 1|\dots) > .95$, for $m = 1,\dots,9$. By contrast, the last three regions do not include any of the simulated point sources, and indeed the posterior probability that each contains a source is small, i.e.,  $\Pr(K_{\mathcal{R}_m}\geq 1|\dots) \approx .01$, for $m = 10,11,12$.

\begin{figure}[t]
	\centering
	\hspace{-.5cm}
	\includegraphics[width=8cm, height=6.5cm]{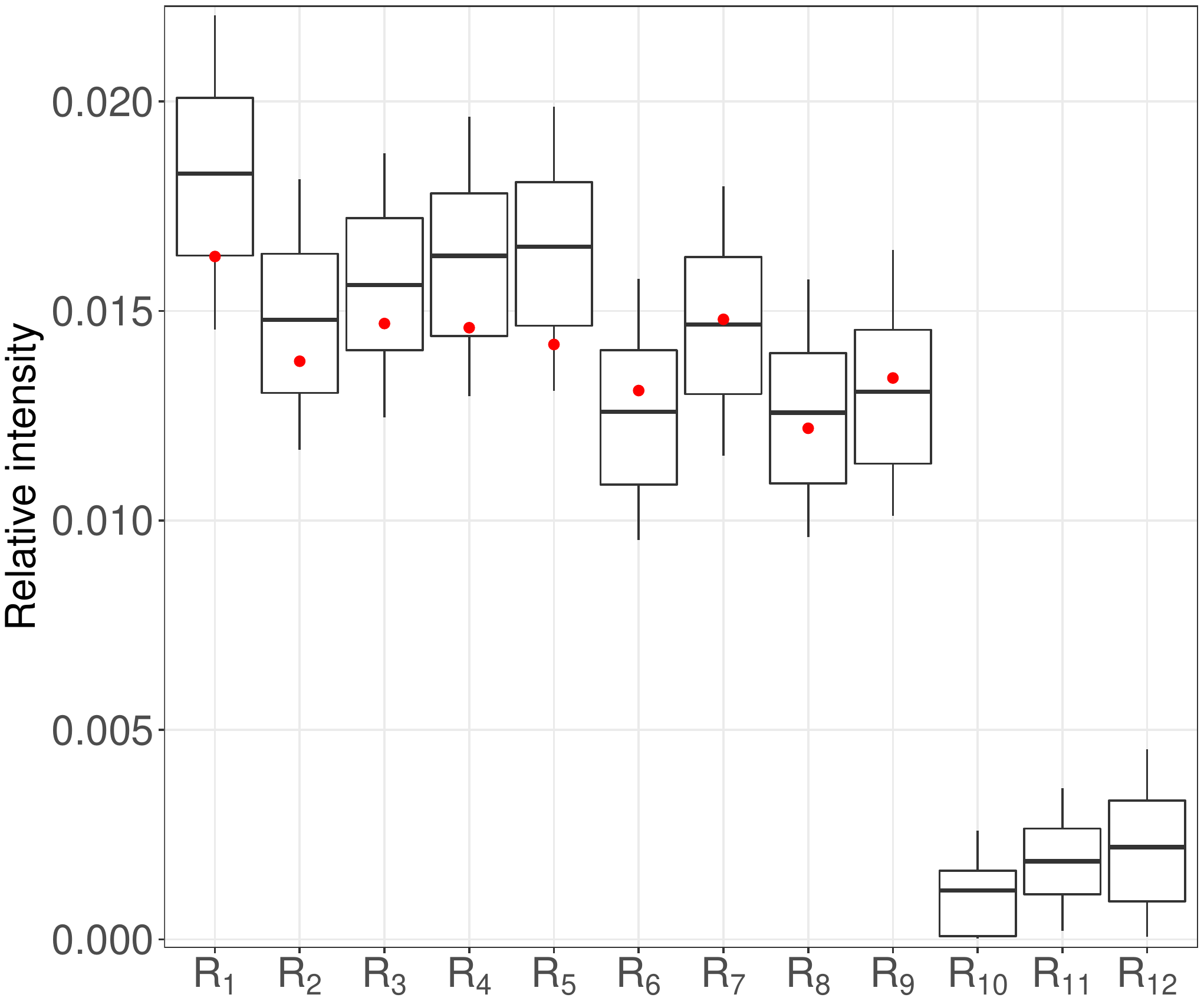}
	\caption{Posterior relative intensity of the sources in $\mathcal{R}_m$, assuming $K_{\mathcal{R}_m}=1$, for $m=1,\dots,12$. The center line of each box, the boxes themselves, and the whiskers represent the posterior means, the 68\% HPD intervals and the 95\% HPD intervals, respectively. Red points are the actual proportion of photons from the 9 sources in the dataset: 7 of them fall within the 68\% intervals, and the remaining two fall into the 95\%. Boxes without red points denote the spurious source regions.}
	\label{fig:applied2}
\end{figure}

Figure \ref{fig:applied2} displays the posterior distribution of the relative intensities of the sources, which equal the product of $\delta$ and DP weights of $\smixture$, within the 12 regions. Red points denote the simulated proportion of photons generated by the 9 sources. Seven of the nine true proportions fall within the 68\% HPD intervals, with the remaining two falling within the 95\% HPD intervals. These results  validate the capability of our method to identify sources in a relatively simple context.

\subsection{More challenging simulations and model comparison}
\label{subsec:model_comparison}

\begin{figure}[t]
	\centering
	\includegraphics[width=1\linewidth]{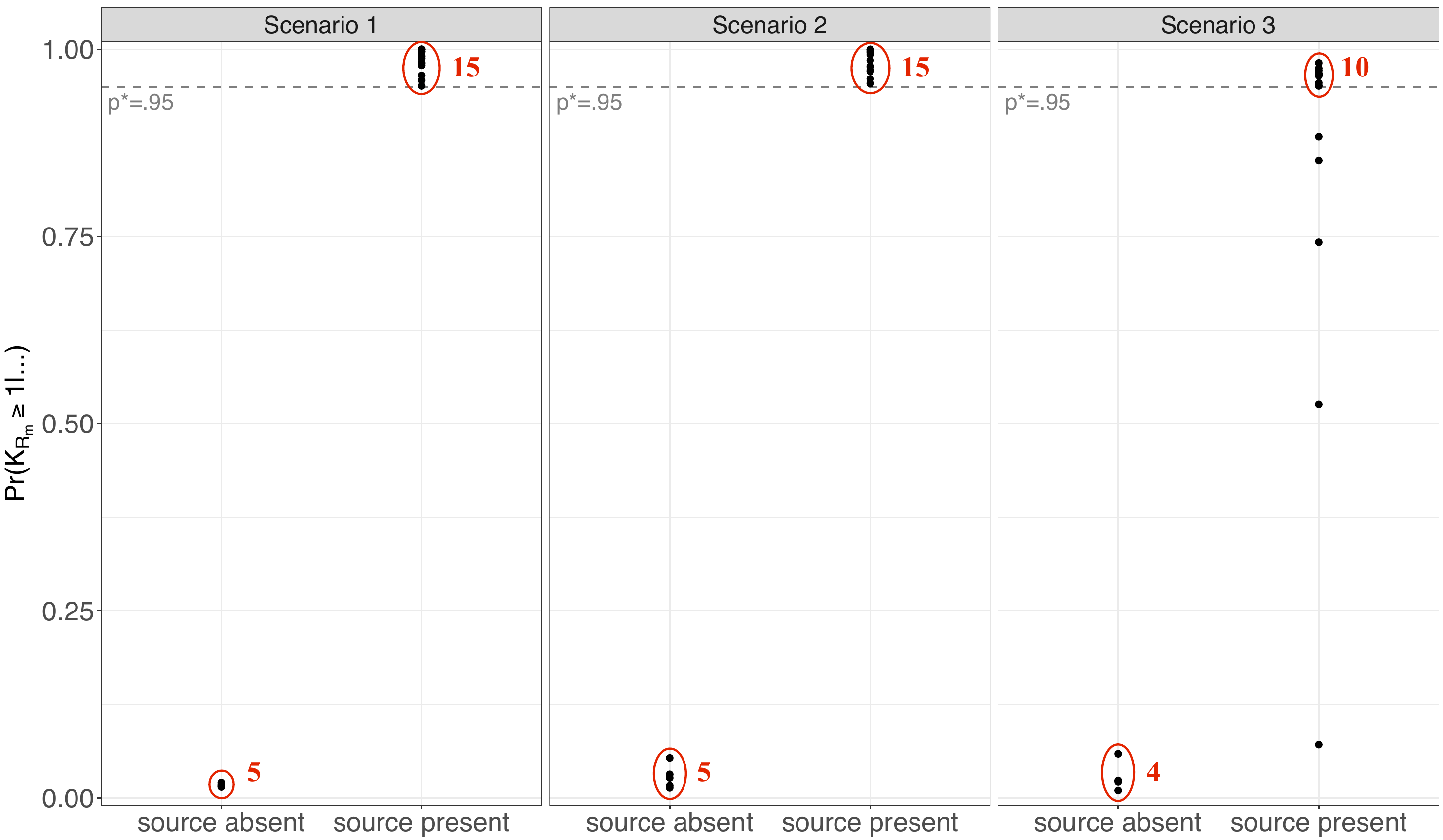}
	\caption{Comparison of results from the \imagemodel\ in the three simulation scenarios considered in the text. In each scenario, there are 15 sources with varying assumptions about their spectral properties. The x-axis splits the regions identified by Algorithm~\ref{algorithm:labelswitching} according to whether they do (``source present'') or do not (``source absent'') contain a real source; the y-axis shows the probability that the  regions contain a source. The red figures count the number of regions in each category.}
	\label{fig:scenarios}
\end{figure}

We carry out three additional simulation experiments to evaluate the ability of our method to identify sources in scenarios of growing complexity. In particular, we test whether including energy information substantially improves performance. We simulate 3 datasets (scenarios), each containing 15 point sources. The locations of the sources do not change across the datasets. Scenario 1 is simulated with $F_{0,\varsigma} = 10^{-9}$ and $\varrho_\varsigma = 2$ for $\varsigma=1,\dots,15$; this is thus identical to the illustrative example in Section \ref{subsection:illustrative_example} but with a larger number of sources. Scenario 2 is simulated with $F_{0,\varsigma} = 10^{-9}$, for each $\varsigma$,  but different values of $\varrho_\varsigma$, sampled at random from the third \emph{Fermi} LAT catalogue (3FGL) \citep{acero_etal.2015}. This scenario provides a check of the importance of inference model misspecification in the spectral domain; recall that, unlike this simulation model, our inference model assumes that all point sources share the same spectral shape parameter. Scenario 3 is simulated with varying source brightnesses by setting $F_{0,1},\dots,F_{0,15}$ to a sequence of 15 equispaced values between $2\cdot10^{-10}$ and $5\cdot10^{-10}$, while $\varrho_\varsigma = 2$, for each $\varsigma$. Scenario 3 is useful to assess the performance of the \jointmodel\ relative to the simpler \imagemodel. Moderate contamination is simulated from the same background model used in Section \ref{subsection:illustrative_example} and added to the three datasets. Finally, the simulated photon counts are again thinned to 10,000 by random sampling in order to reduce the computational cost. The three maps and the background model used to generate the contamination appear in Appendix \ref{app.B}, Figure \ref{fig:appendix_simulationdatasets}.

We fit both the \imagemodel~and the \jointmodel~to all three simulation scenarios using the same hyperparameters and MCMC settings as in Section~\ref{subsection:illustrative_example}. For inference, we set the prior distribution of $\eta_s$ to $\hbox{Gamma}(a_{\eta_s} =3.196,b_{\eta_s} =2.196)$, so that the \emph{a priori} mode of $\eta_s$ is one and its shortest 95\%-probability interval is (0, 3). This choice reflects the actual knowledge on sources with power-law spectra, which are known to have a shape parameter $\varrho\in (1,4)$, that corresponds to $\eta_s = \varrho-1$ in our Pareto model. We further set the prior of $\eta_b$ to $\hbox{Gamma}(a_{\eta_s} = 1.79,b_{\eta_s} =0.714)$, which has \emph{a priori} approximately the same mode of $\eta_s$, but twice the size of the 95\% HPD interval. This choice reflects the lack of knowledge that we have on the spectral shape parameter of the background. 

The results are shown in Figure \ref{fig:scenarios}. In Scenario 1, the 15 sources are identified with large probability ($\geq p^* = .95$, as in Algorithm \ref{algorithm:labelswitching}). Five 5 spurious regions also appear, but  with very small probability of containing a source. Similar results are obtained for Scenario 2. For Scenario 3, 10 of the 15 sources are identified with probability above the threshold $p^* = 0.95$; however, the remaining 5 sources are detected with probabilities less than $p^*$, meaning that the model struggles to detect faint signals. 

\begin{figure}[t]
	\centering
	\includegraphics[width=0.7\linewidth]{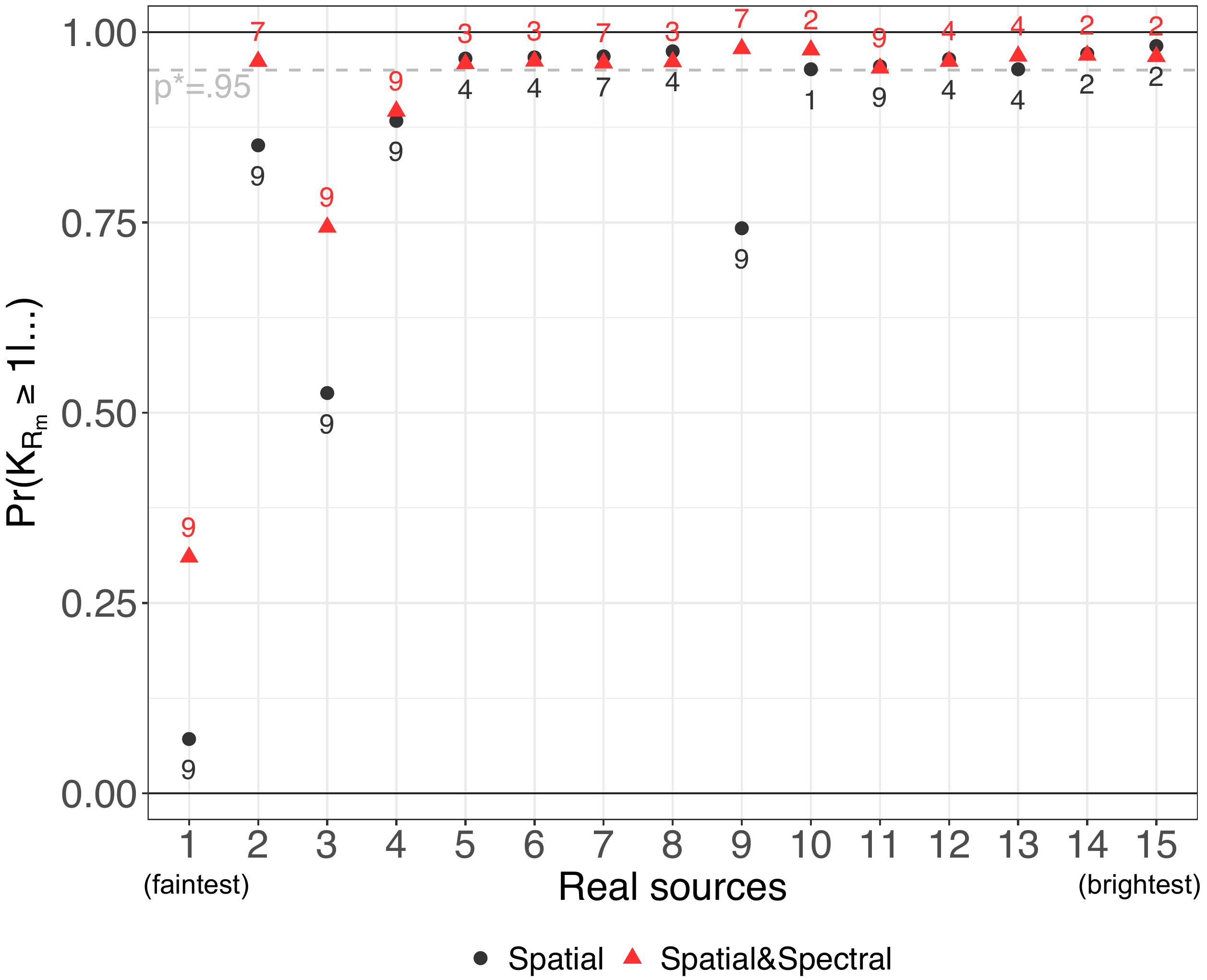}
	\caption{Results from Scenario 3.  The plot displays the regions  which include the simulated sources.
	The x-axis indexes the 15 sources, ordered from the faintest (Source 1) to the brightest (Source 15), and the y-axis is the probability that each region includes a source. Black circles are probabilities under the \imagemodel\ and red triangles are under the \jointmodel; the coloured labels correspond to the region sizes, expressed in number of pixels.}
	\label{fig:probabilitiesscenario5}
\end{figure}

We do not display results from Scenario 1 and 2 from the  \jointmodel~as they are similar to those from \imagemodel; however, results for Scenario 3 differ substantially and worth exploring in detail. For Scenario 3, both models identify actual sources from the simulation along with spurious sources; Figure \ref{fig:probabilitiesscenario5} displays only the actual sources. The x-axis sorts the regions in increasing order of source amplitude (so, $x=1$ corresponds to the source with the smallest amplitude, $F_0 = 2\cdot 10^{-10}$, and $x=15$ to the largest amplitude, $F_0 = 5\cdot 10^{-10}$), and the y-axis gives $\Pr(K_{\mathcal{R}_m}\geq 1|\dots)$, the probability of the region containing a source. Black circles are the probabilities under the \imagemodel, and red triangles under \jointmodel. Finally, the colored labels give the region sizes, expressed in number of pixels.

We notice that the \jointmodel~performs considerably better than the \imagemodel~in three different aspects. First, it recovers a higher proportion of the 15 sources and with high probability; 12 regions have  probability $\geq p^*$, against the 10 regions from the \imagemodel~with probability $\geq p^*$. Second, it returns smaller (that is, better localized) regions, allowing for a more precise evaluation of the source locations. Finally, it correctly extracts the true shape parameter of the sources $\varrho$, as demonstrated by the posterior traceplot of $\eta_s$ (see Figure \ref{fig:appendixspatspectr2} in Appendix \ref{app.B}). This result is particularly notable, showing that the fitted source spectrum is consistent with the spectra used to generate the data even though both the image and spectral models of the background are misspecified. In addition to the actual simulation sources, the \jointmodel~detects 4 spurious sources; however, their probability $\Pr(K_{\mathcal{R}_m}\geq 1|\dots)$ is low. (see left panel of Figure \ref{fig:appendixspatspectr2} in Appendix \ref{app.B}).

To conclude, we confirm the ability of our inferential models to identify sources under scenarios of various complexity. Scenario 3 confirms that, like in \cite{jones_etal.2015}, the inclusion of energy brings noteworthy improvements in the accuracy of the source detection. These results encourage future extensions of the \jointmodel, which would, for example, account for sources with different shape parameters or extend the energy range by explicitly including the energy-dependent effective area.

\section{Application to Fermi-LAT data\label{sec:Fermiapplication}}

\begin{figure}[th!]
	\centering
	\includegraphics[width=.55\linewidth]{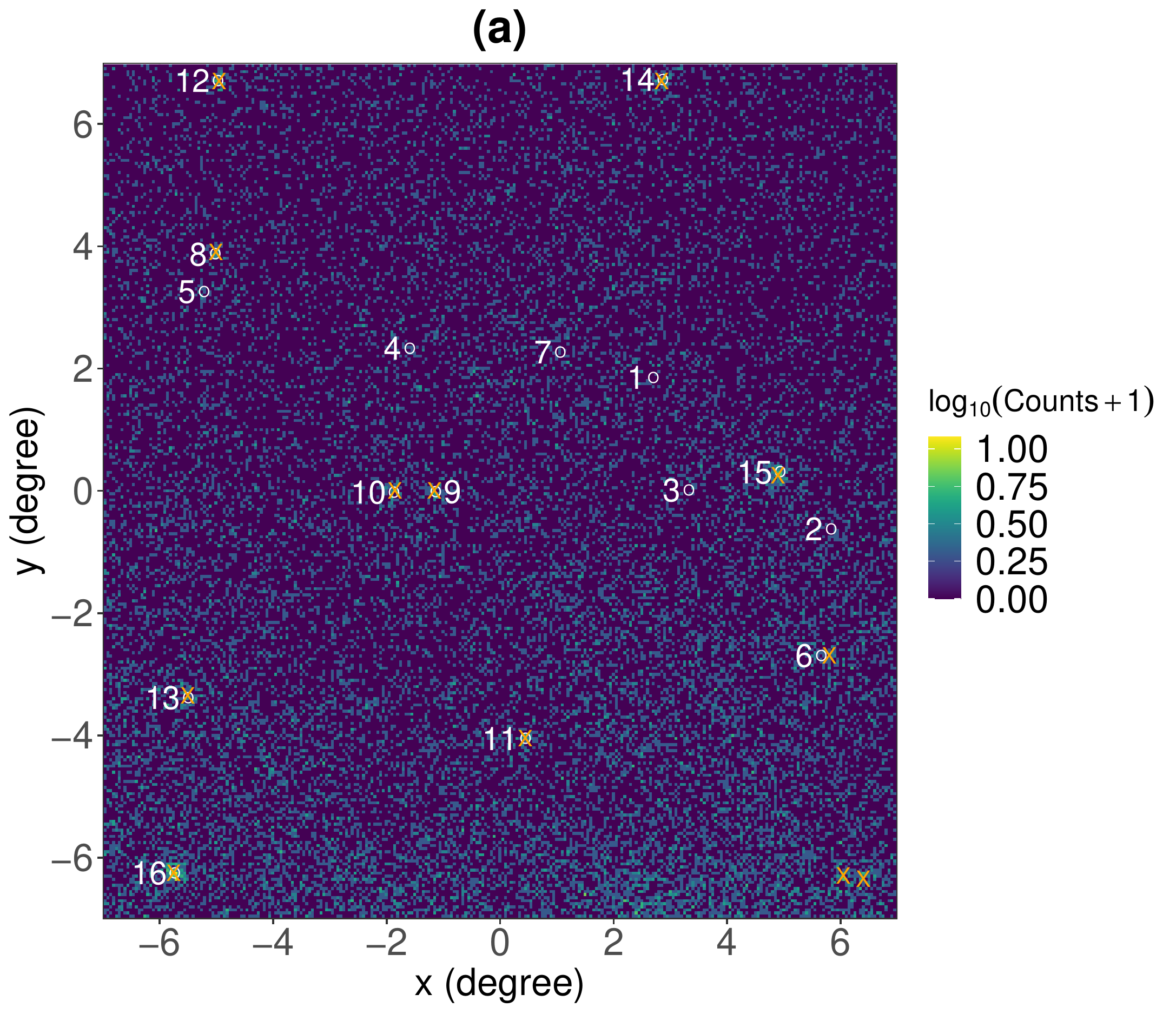}\\
	\includegraphics[width=.49\linewidth]{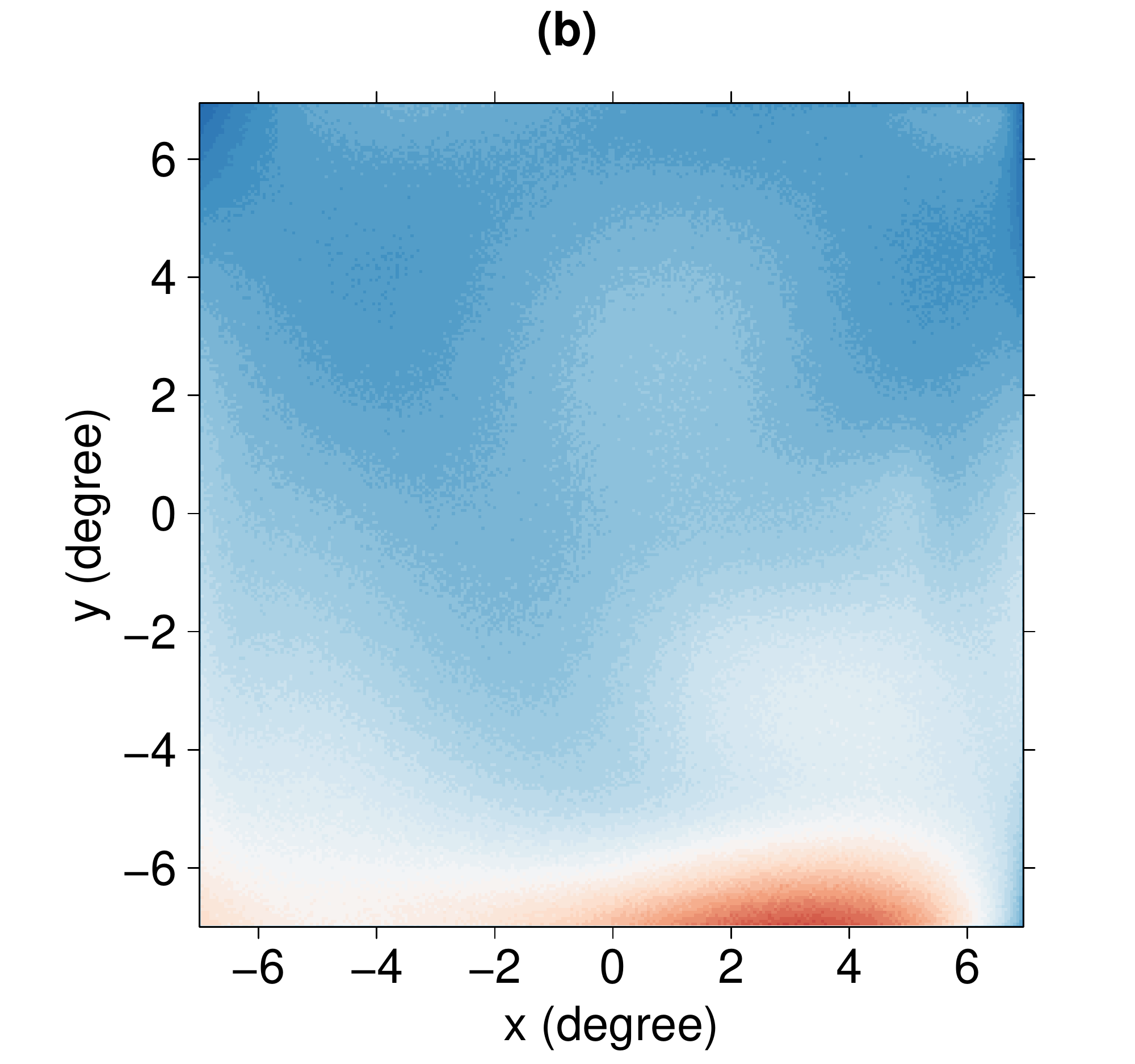}
	\includegraphics[width=.49\linewidth]{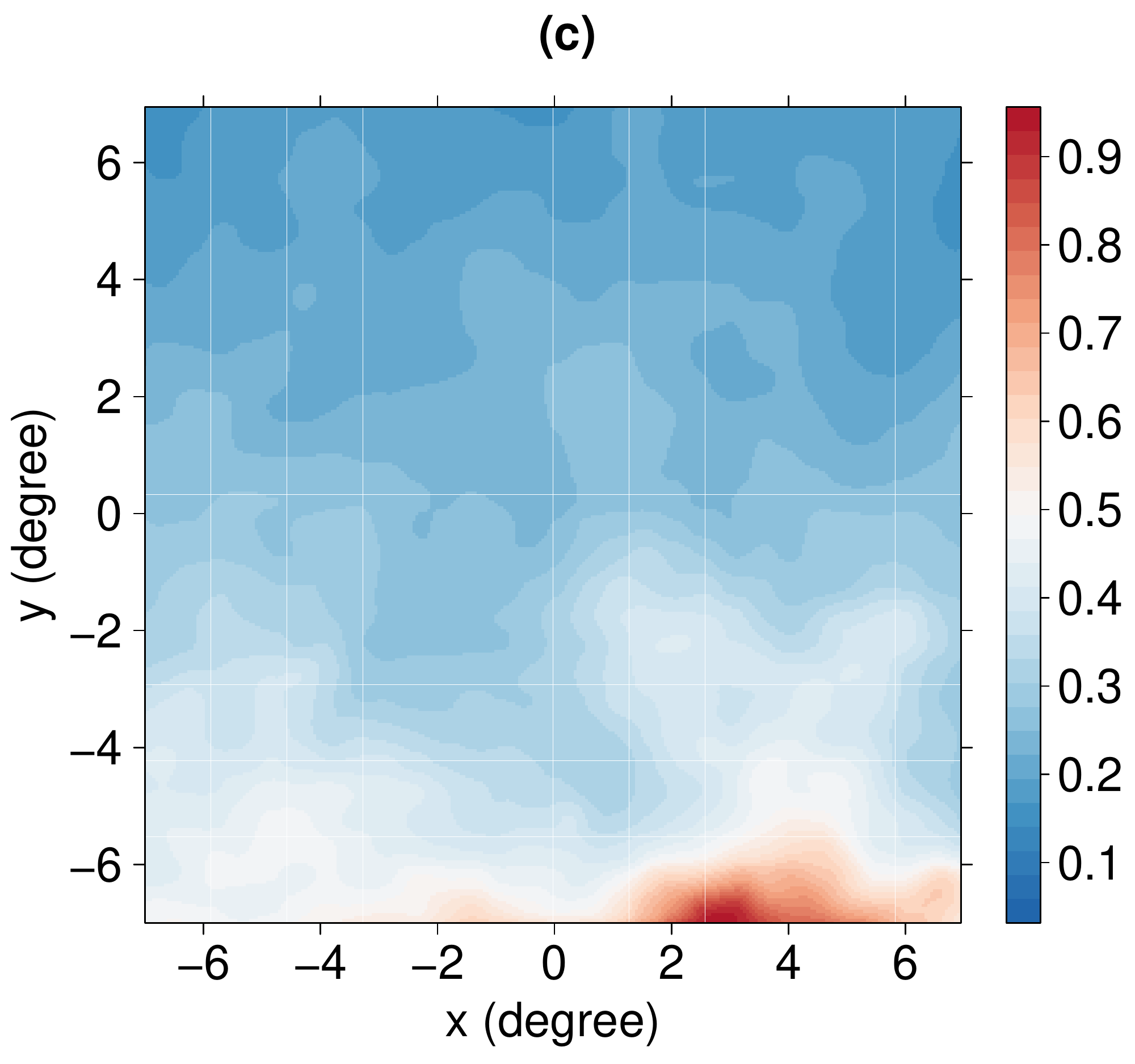}
	\caption{Panel {(a)}: Map of photons from a region surrounding the {Antlia-2} galaxy, gathered into a grid of $280\times280$ pixels of size $0.05\degree\times 0.05\degree$. The scale colour gives the $\log_{10}$-transformation of the counts per pixel observed. White circles denote 16 sources listed in the {Fermi}~LAT 4FGL source catalogue, numbered in increasing order by their expected flux in the energy range $(2.5~\mathrm{GeV} - 500~\mathrm{GeV})$. Orange crosses denote the centroids of the 12 regions identified by the \imagemodel. Panels {{(b-c)}}: Expected counts from our background model $\bmixture$ and from the model of \cite{acero_etal.2016}, respectively (using the same pixelization; color scale is linear in predicted pixel counts).}
	\label{figure:Antlia2_results}
\end{figure}

In this section, we apply our methods to $\gamma$-ray data collected over a 9.4 year period by the \emph{Fermi}~LAT from a region surrounding the newly discovered dwarf galaxy \emph{Antlia 2} \citep{torrealba_etal.2019}, a potentially interesting target for dark matter-induced $\gamma$-ray emission. This region is relatively close to the Galactic plane, and therefore subject to significant diffuse background processes that have strong gradients across the field of view. A first background model developed by the \emph{Fermi} Collaboration \citep{acero_etal.2016} is made up of an isotropic component plus a diffuse component representing diffuse galactic emission. The physical processes that give rise to the diffuse emission are difficult to model and we do not expect the \emph{Fermi} background model to capture all the detailed morphological and spectral characteristics of the true background. Therefore, a method independent from the \emph{Fermi} background model   would be more suitable for estimating the noise component of the data.

The original \emph{Fermi} dataset contains the photons from a region of $400\times 400$ spatial pixels of size $0.05\degree\times 0.05\degree$, across 30 energy pixels in the range $[0.5~{\rm GeV}-500~{\rm GeV})$. To reduce computational cost, we restricted our attention to an area at the centre of the image of size $280\times 280$ pixels, and we removed photons with energies less than $2.5~{\rm GeV}$. The resultant dataset consists of 23,897 photons and is illustrated in Figure~\ref{figure:Antlia2_results}(a). Strong emission from the Galactic plane is visible at the bottom of the map; this background emission progressively diminishes toward northern latitudes (top of the figure), away from the Galactic plane. The latest \emph{Fermi}~LAT source catalogue (4FGL) \citep{abdollahi_etal.2020} reports 16 sources in this area, denoted by red circles and labelled in ascending order according to their expected photon count in the energy range $(2.5~{\rm GeV} - 500~{\rm GeV})$.  Thus, we expect the smallest photon count from Source 1, and the largest from Source 16. 

We fit both the \imagemodel\ and the \jointmodel\ with the same prior set-up described in Section~\ref{subsec:model_comparison}. For both models, we run four separate Markov chains, each of length 10,000, and discard the first three quarters of each chain as burn-in. The remainder are combined to form a posterior sample of size 10,000. The simulations were implemented in the \textsc{R} language and environment for statistical computing \citep{RCoreTeam.2020}.  The four chains were run in parallel on the departmental computer cluster and took approximately one day to produce the results. 

\begin{figure}[t]
	\centering
	\includegraphics[width=.6\linewidth]{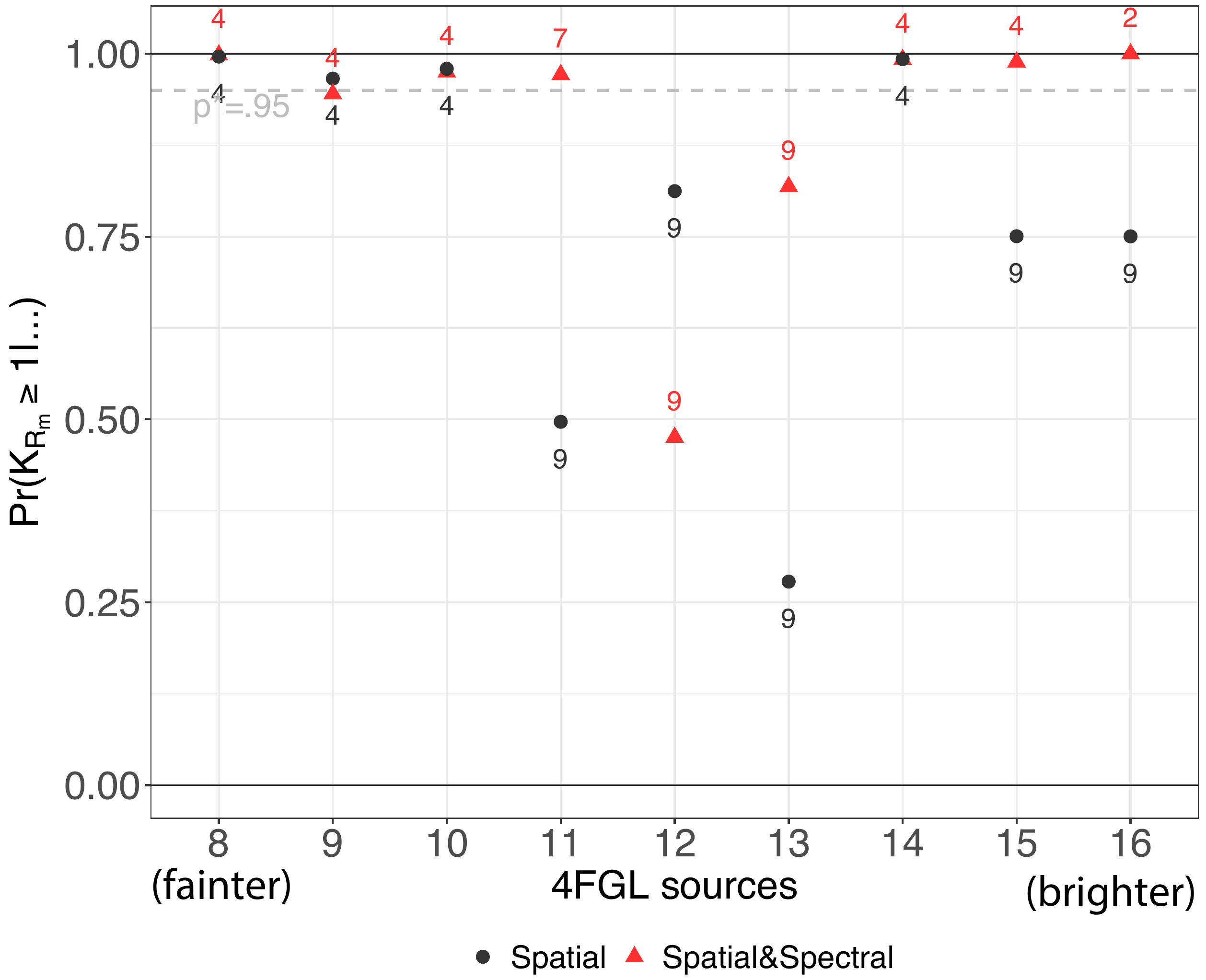}
	\caption{Results on the {Antlia-2} dataset. The plot displays only the regions inside which there are some sources listed in the catalogue of \cite{abdollahi_etal.2020}. The x-axis displays those sources in order of expected counts according to the catalog model. For every source, we give the probabilities of detecting it under both the \imagemodel~(black circles) and the \jointmodel~(red triangles).
	The coloured numbers state the number of pixels the regions are made of.}
	\label{fig:comparisonnoeneryesener2}
\end{figure}

Figure~\ref{figure:Antlia2_results}(b) shows the posterior expected counts per pixel from background under our B-spline DP mixture.  We generate this map by simulating, for every iteration $t$ of the posterior sample, $n^{(t)}_b$  photon counts from \eqref{formula:3.6_DPbackground} with parameters evaluated at iteration $t$, and then binning them into the $280\times 280$ grid of pixels. Averaging across the 10,000 iterations of the posterior sample gives a Monte Carlo estimate of the posterior expected pixel counts. In contrast, Figure~\ref{figure:Antlia2_results}(c) shows  the expected photon counts under the \emph{Fermi} collaboration model, which is a single ``best-fit'' model to the entire $\gamma$-ray sky. The two panels broadly agree in identifying (i) a prominent contamination from the nearby Galactic plane at southern latitudes (bottom of the images), (ii) some moderate and more uniform background areas in the middle of the maps, and (iii) a considerably reduced background contribution moving to northern latitudes (top of the images). Nevertheless, our method gives substantially lower expected counts in the bottom-right corner of the map than does the \emph{Fermi} model. This discrepancy occurs because our model fit indicates two potential sources in this region, whereas the {\it Fermi} background model subsumes these potential sources. We return to this point below. Our expected background, shown in Figure~\ref{figure:Antlia2_results}(b), also appears smoother than the \emph{Fermi} background of Figure~\ref{figure:Antlia2_results}(c).  This is because our  model adapts to the varying background morphology, unlike the fixed \emph{Fermi} model, which pixel-wise estimates the background without uncertainty. Under our mixture model we have access not just to the marginal expected pixel counts but to the joint distribution of counts in all pixels, allowing us, for example, to explore spatial correlations in the background structure. Additionally, the detailed, small-scale features in the \emph{Fermi} background map (Figure \ref{figure:Antlia2_results}(c)) are not detected in the $\gamma$-ray data. Rather, they correspond to patterns of gas clumps present in templates (see Section~\ref{sec:introsourceextraction}). By contrast, our B-spline background model only includes ``features'' that can be inferred from the $\gamma$-ray data. 

Both the \imagemodel\ and the \jointmodel\ detect nine of the 16 sources listed by the \emph{Fermi} catalogue (IDs 8-16), see Figure~\ref{fig:comparisonnoeneryesener2}. The \jointmodel~performs considerably better than the \imagemodel, providing higher posterior probabilities $\Pr(K_{\mathcal{R}_m}\geq 1|\dots)$ and smaller regions for eight sources. (The only exception is Source 12.) This result is encouraging, especially considering that, according to \cite{abdollahi_etal.2020}, the sources in questions have substantially different spectral shapes, which is something that our \jointmodel~does not account for.

\begin{figure}[th!]
	\centering
	\scalebox{.92}{
	    \hspace{-.5cm}
		\includegraphics[height=5cm, width=4.9cm]{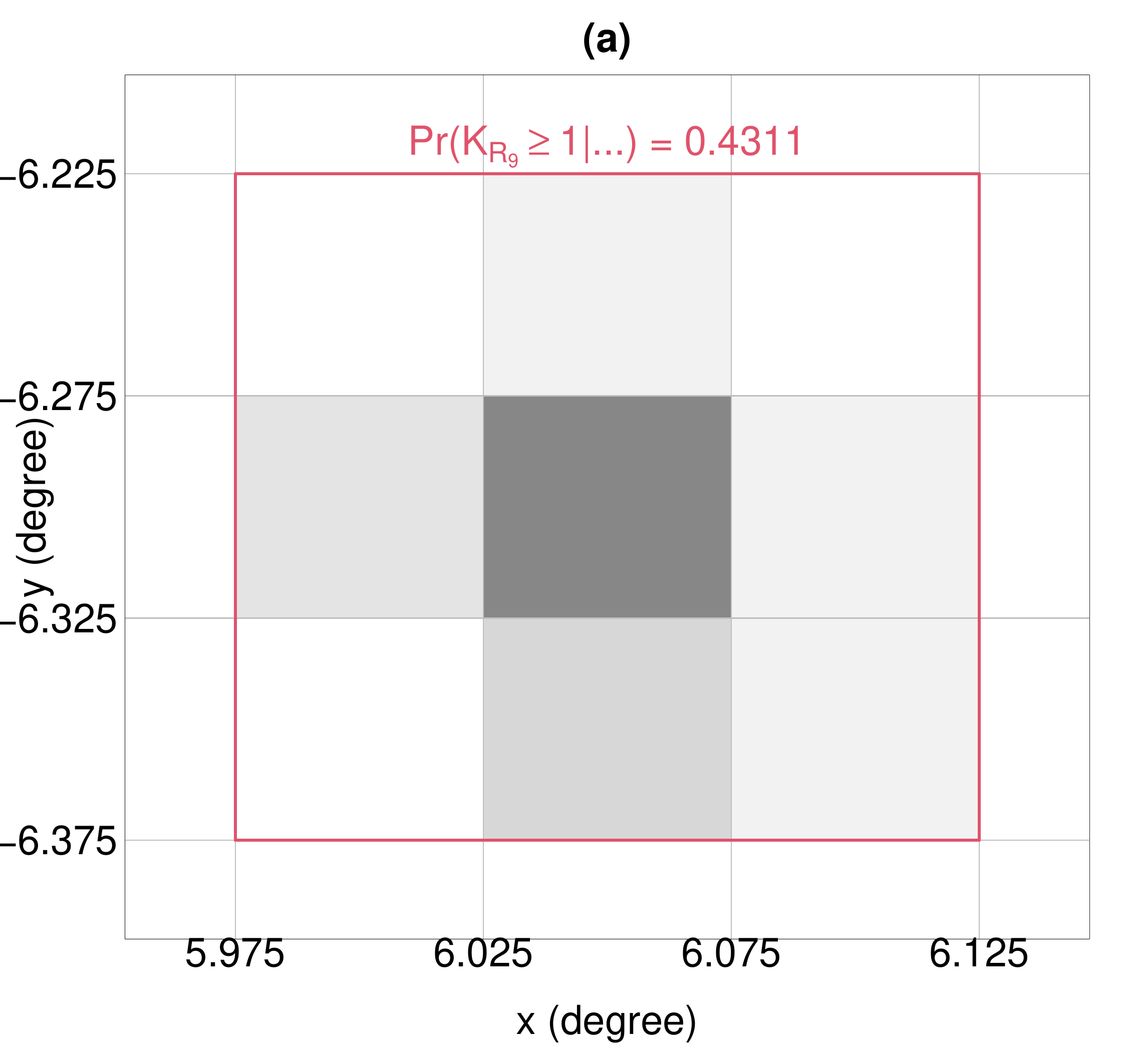}\hspace{-.3cm}
		\hspace{.1cm}
		\includegraphics[height=5cm, width=4.9cm]{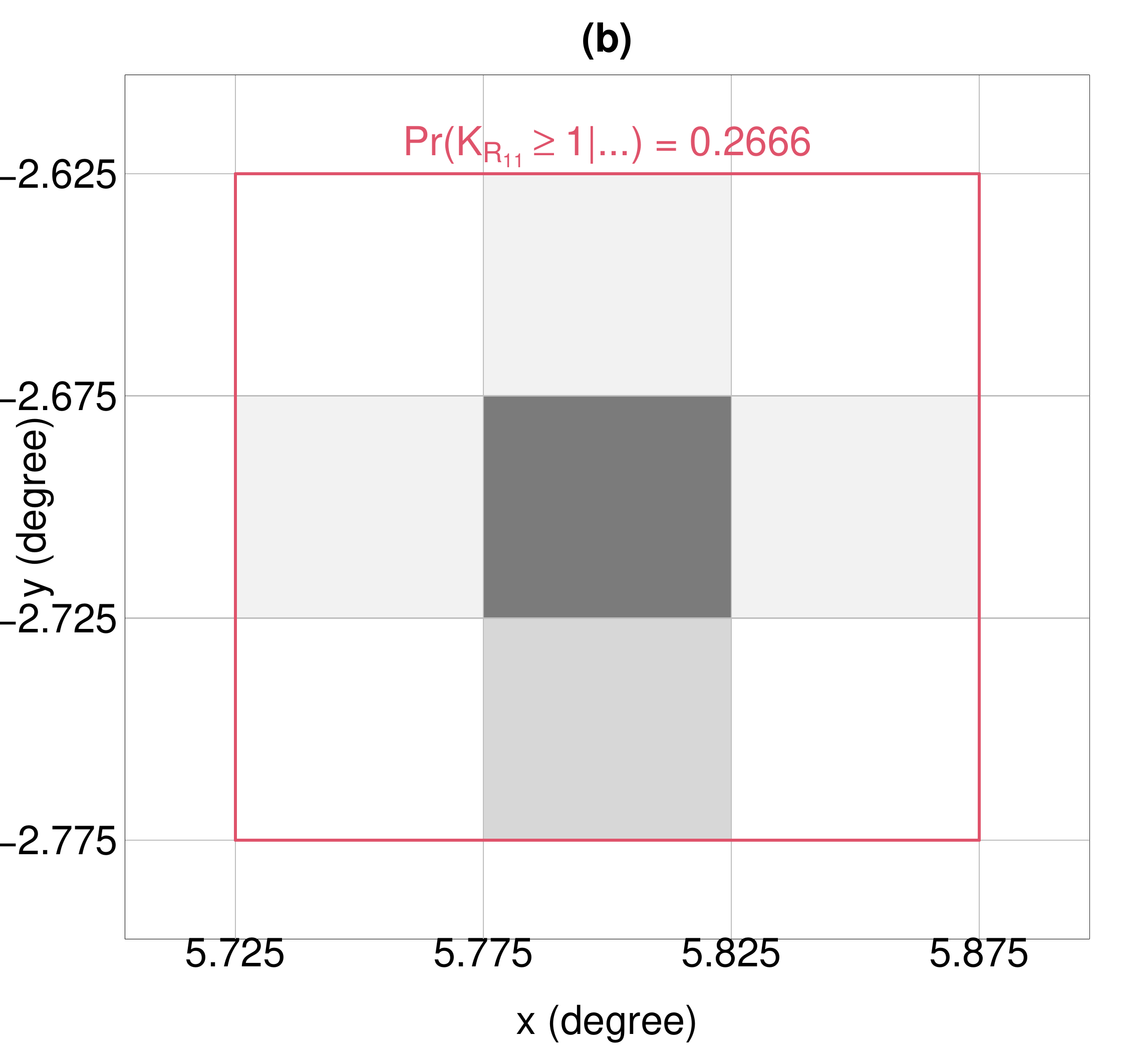}\hspace{-.3cm}
		\hspace{.1cm}
		\includegraphics[height=5cm, 
		width=4.9cm]{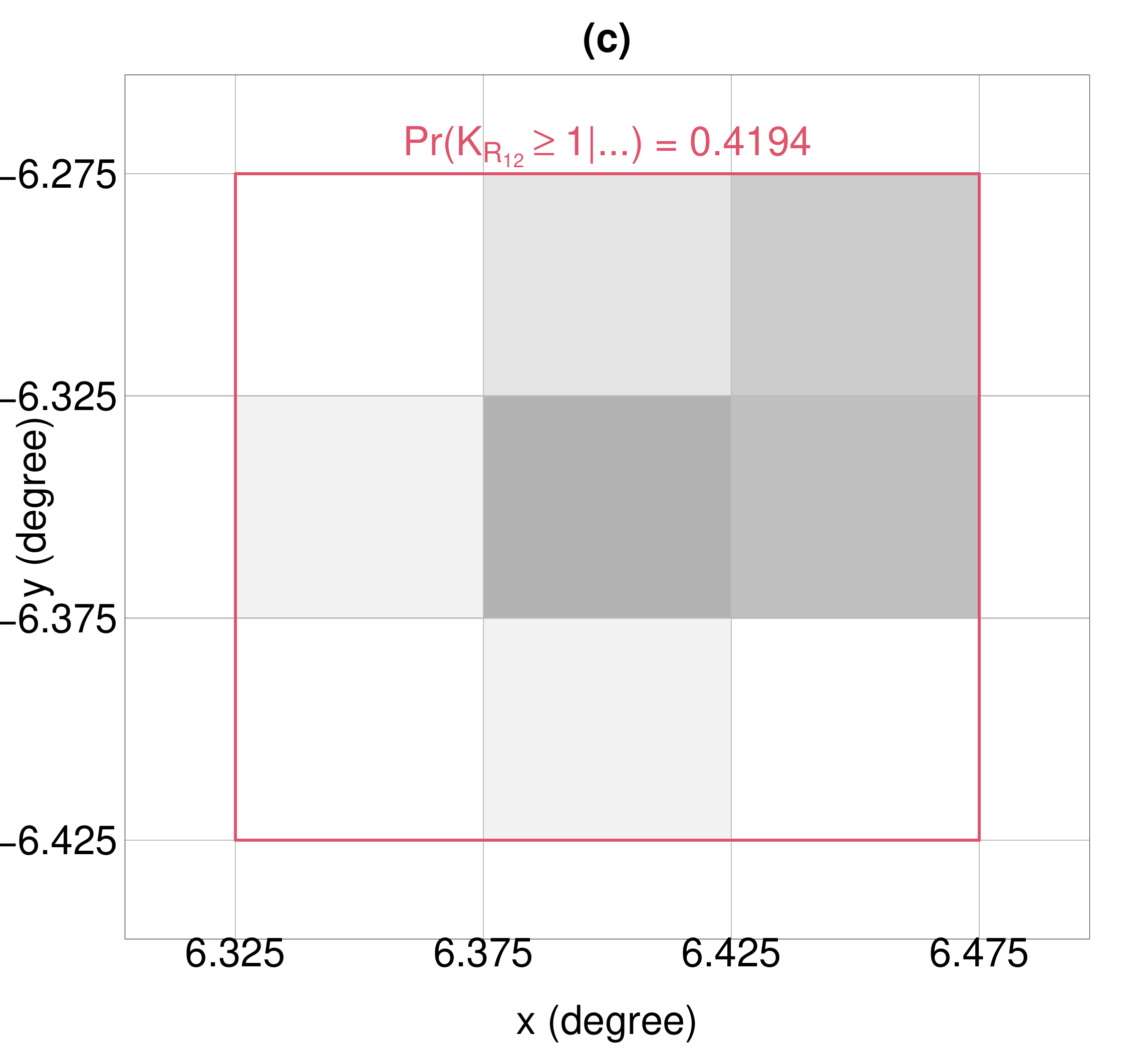}}\\
	\vspace{-.1cm}
	\scalebox{.92}{
		\hspace{-.5cm}
		\includegraphics[height=5cm, width=4.9cm]{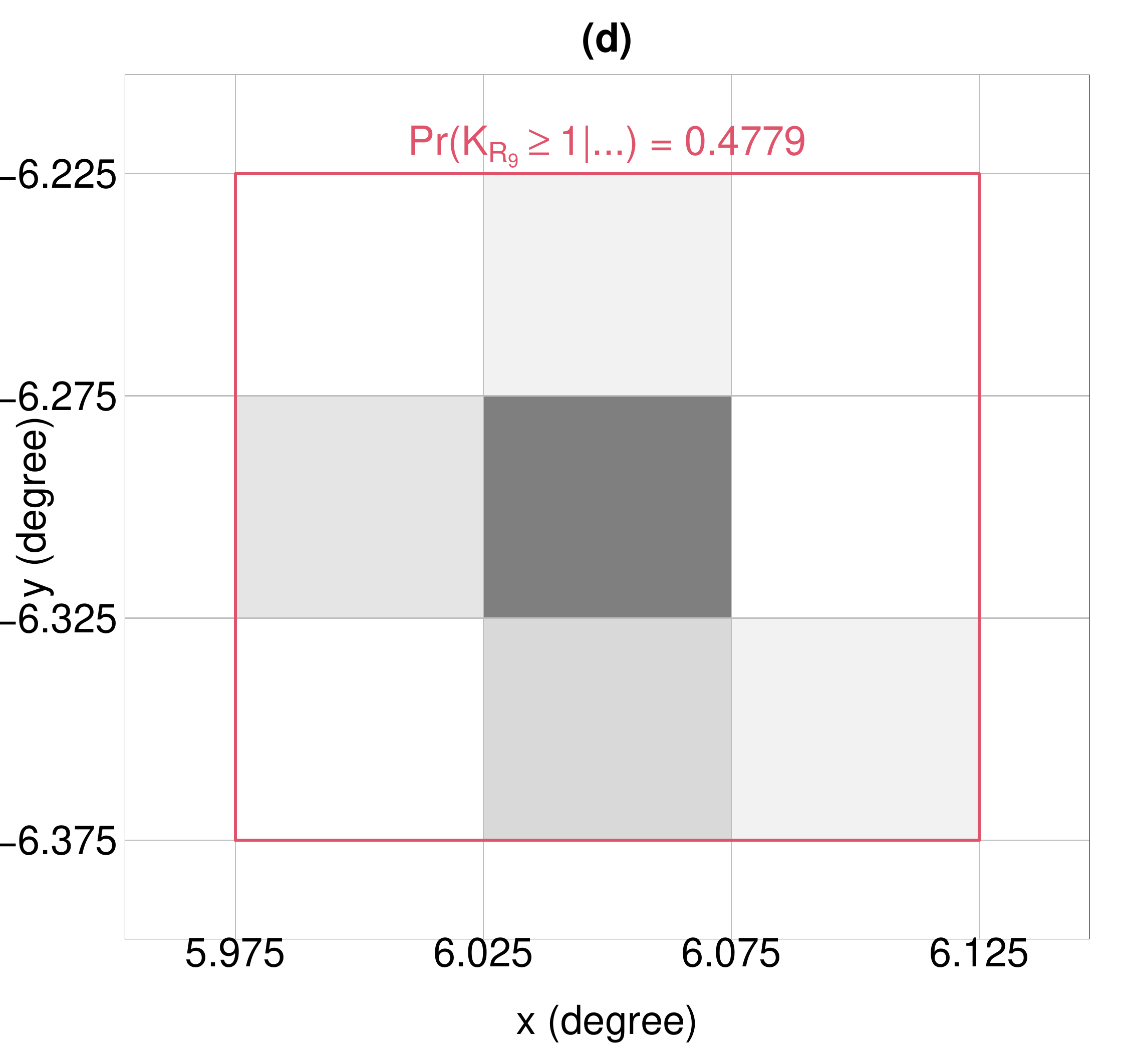}\hspace{-.3cm}
		\hspace{.1cm}
		\includegraphics[height=5cm, width=4.9cm]{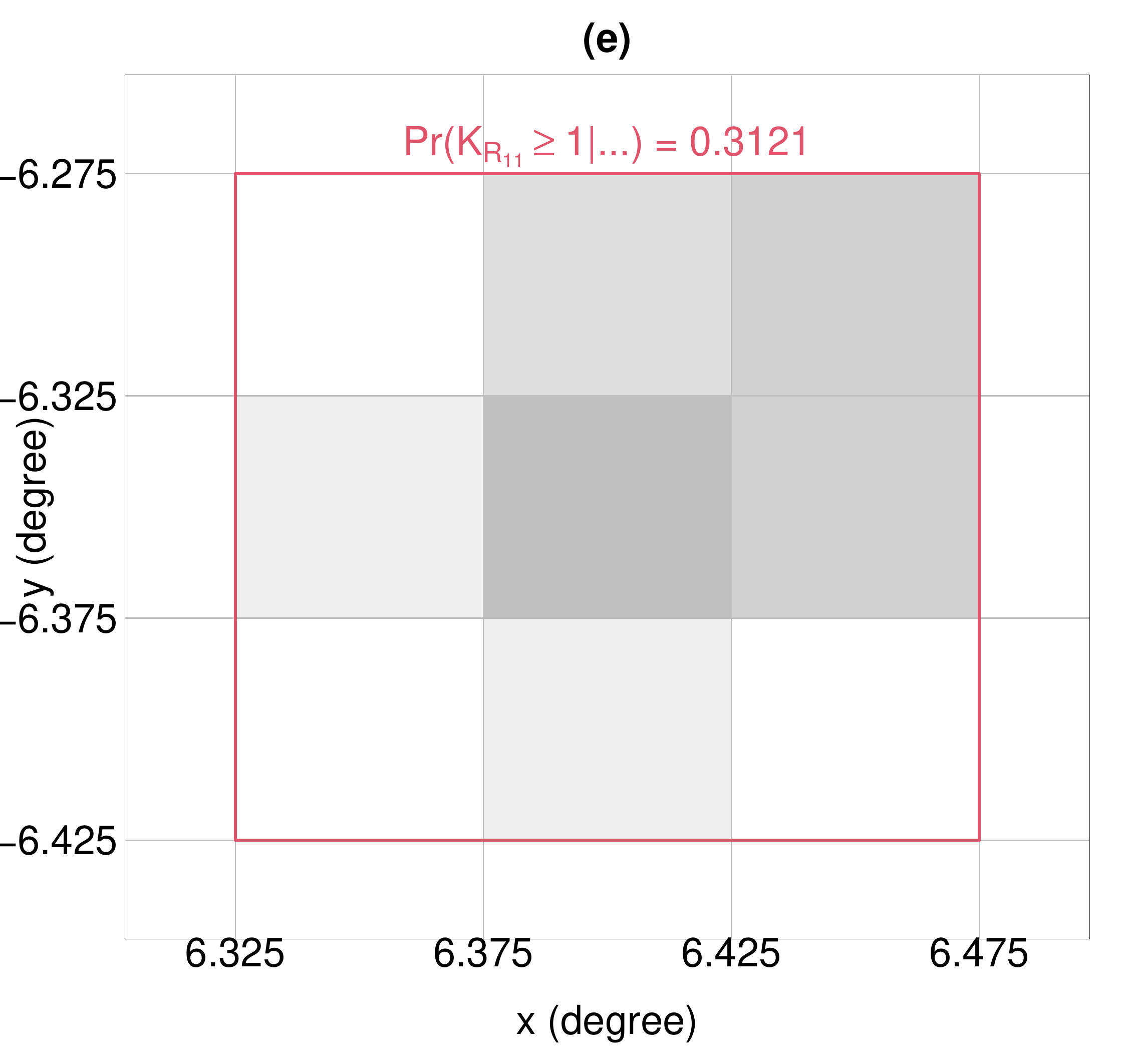}\hspace{-.3cm}
		\hspace{.1cm}
		\includegraphics[height=5cm, 
		width=4.9cm]{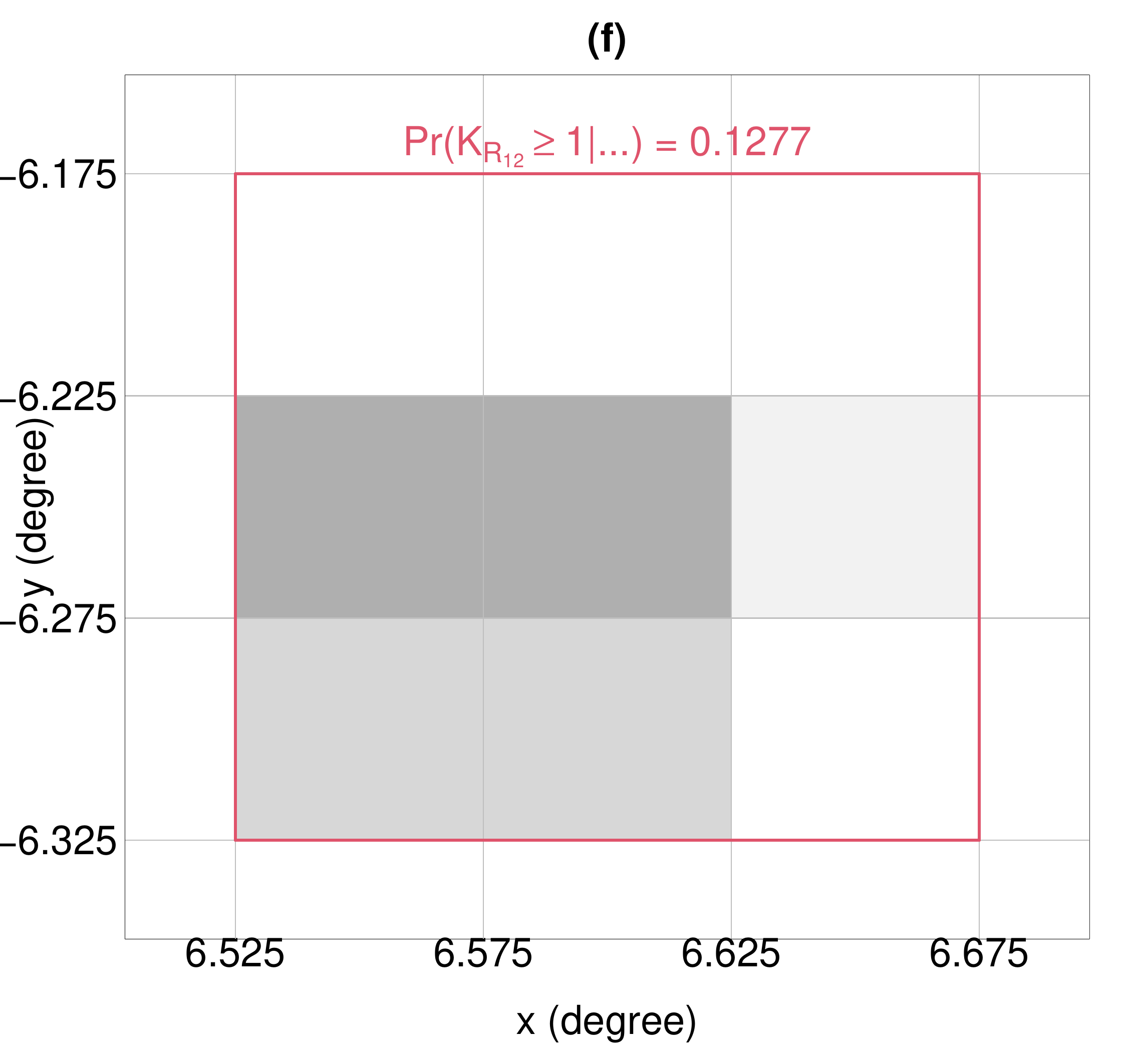}}\\
	\hspace{1.2cm}
	\includegraphics[height=1cm, width=5cm]{Figures/legend3.pdf}
	\caption{
		Panels (a-c): Regions identified by the \imagemodel~that do not contain any known catalogue source. Region $\mathcal{R}_{11}$ (Panel (b)) is about $0.1\degree$ away from Source 8, but it does not enclose it. Regions $\mathcal{R}_9$ and $\mathcal{R}_{12}$ (Panels (a) and (c)) are located instead in the bottom-right corner of the map (see the bottom-right corner of Figure \ref{figure:Antlia2_results}(a)). Panels (d-f): point sources identified by the \jointmodel\ which do not correspond to known sources; all of them are located in the bottom-right corner of the map. 
		Grey colours give the posterior distribution of the source location, assuming that there is only one in every region. Notice that Panels (a) and (d) are for the same region; and so are Panels (c) and (e).
	}
	\label{fig:new_posterior_regions_Ant2data}
\end{figure}

Although we do not detect Sources 1-7, \cite{abdollahi_etal.2020} reports that these sources are of low brightness ($F_0=5.7\cdot 10^{-10}$ for the brightest of them), and their spectra have a steep power-law ($\varrho > 2.4$). Thus, we expect a small number of photons from each, and mostly at low energies. We may detect these sources if we include photons with energies less than 1~GeV as done in \cite{abdollahi_etal.2020}.

Our analysis identified four additional point sources not recorded in the {\it Fermi} catalogue; details are given in Figure~\ref{fig:new_posterior_regions_Ant2data}. In particular, both the \imagemodel~and the \jointmodel~indicate the presence of points sources in the same regions in the bottom-right corner of the map. The first appears in Figure~\ref{fig:new_posterior_regions_Ant2data}(a) and Figure~\ref{fig:new_posterior_regions_Ant2data}(d),  with probabilities $\mathrm{Pr}(K_{\mathcal{R}_m}\geq 1|\dots)$ of 0.43 and 0.48 respectively. The second appears in Figure~\ref{fig:new_posterior_regions_Ant2data}(c) and Figure~\ref{fig:new_posterior_regions_Ant2data}(e) with probabilities $\mathrm{Pr}(K_{\mathcal{R}_m}\geq 1|\dots)$ of 0.42 and 0.31. The third region discovered by the \imagemodel~(Figure~\ref{fig:new_posterior_regions_Ant2data}(b)) has presumably recovered a faint signal from Source~6, as it is $0.1\degree$ away from its catalogue location (see Figure~\ref{figure:Antlia2_results}(a)). However, we conclude there is no source in this region as its probability is only 0.26. Lastly, there is a third region in the bottom-right corner of the map indicated by the \jointmodel~(Figure~\ref{fig:new_posterior_regions_Ant2data}(f)): since its probability of including a source is 0.13, we consider it spurious.

\section{Discussion}

We present an innovative approach to extract point sources in highly contaminated high-energy ($\gamma$-ray and X-ray) photon-count maps using both spatial and spectral information. The methods exploit advanced Bayesian nonparametric techniques in a way that simultaneously identifies point sources and fits the background contamination. A physics-based constraint on the background parameter space is imposed to reduce the possibility that the signal of some  sources is attributed to the background. Our mixture of DP mixture models is fitted to the data via Markov chain Monte Carlo simulation. In addition, we developed a post-processing algorithm that conducts inference on component-specific parameters, even when some posterior distributions are multimodal, and handles the well-known label switching problem in mixture models. We further illustrated its working with examples drawn from real astrophysical applications.

We proposed two models: the first exploits only the spatial information of the photons, the second extends it in a simple manner by including the energy, as well, under the simplifying assumptions that all sources share the same power-law spectrum. The two models exhibited good detection performance on several simulation experiments of growing complexity, and revealed only few spurious clusters which could be easily recognised using the posterior probability of corresponding to real sources computed with the proposed  algorithm for posterior analysis. Simulation experiments have also shown that the extended model is better suited to extracting point sources when they all share a similar spectral shape parameter. 

Finally, we carried out the analysis of a real-case dataset, a map of photons collected by the \emph{Fermi} LAT space observatory around the newly discovered galaxy \emph{Antlia-2}. As revealed by the comparison with the \emph{Fermi} collaboration background model \citep{acero_etal.2016}, our B-spline DP mixture is able to reconstruct the background morphology, thus demonstrating its ability to detect and separate point sources from background emission in highly contaminated maps. We successfully extracted the signal of 9 sources, whose presence was already known from the last \emph{Fermi} catalogue, and we identified additional regions where point sources may be present, and which do not coincide with any known source. The \jointmodel~ model performed considerably better than the \imagemodel~even if the discovered sources are known to have different spectral parameters. Our results from both simulated experiments and real-case data definitely encourage future extensions of the \jointmodel~ model in two different directions: first, allowing the source DP mixture to model the spectrum of multiple sources with different spectral parameters, and second, providing ad adequate model for the background energy.

With respect to the finite mixture approach of \cite{jones_etal.2015}, DP mixtures have both conceptual and computational advantages. Considering that the number of components in Bayesian nonparametric mixtures is potentially infinite, a larger number of records usually implies a larger size of the mixture. Thus, a DP-based model looks more appropriate than a finite mixture as, with sky surveys, new data become available and may guide to the identification of new sources that could not be recognized before. Moreover, the available algorithms for nonparametric Bayes models look more efficient for posterior inference and scale the mixture size faster than the reversible jump algorithm.

\appendix

\section{Gibbs sampler for B-spline knots}\label{app}

\label{app.A}

\noindent
Let $\{ x_i\}_{i=1}^{n}$, with $x_i\in (x_{\min},x_{\max})$, be the coordinates of the longitude for $n$ photons which are bounded into the interval $(x_{\min},x_{\max})$. Let us assume they distribute according to the density  $\tilde{\mathscr{B}}_4(\cdot|\boldsymbol{\ell})$, with $\boldsymbol{\ell}$ a five-dimensional vector of longitude knots. At the $t$-th iteration of the MCMC algorithm, the full-conditional density of $\ell_k$ is
$$
p(\ell|\tilde{\boldsymbol{\ell}}_{-k},\{x_i  \})\propto\prod_{i=1}^{m} \tilde{\mathscr{B}}_4(x_i|\tilde{\boldsymbol{\ell}})\mathcal{G}_0(\tilde{\boldsymbol{\ell}}),
$$
where $\tilde{\boldsymbol{\ell}}=(\boldsymbol{\ell}^{(t)}_{1:(k-1)},\ell,\boldsymbol{\ell}^{(t-1)}_{(k+1):5})$ if $k = 2,3,4$, $\tilde{\boldsymbol{\ell}}=(\ell,\boldsymbol{\ell}^{(t-1)}_{2:5})$ if $k=1$ and $\tilde{\boldsymbol{\ell}}=(\boldsymbol{\ell}^{(t)}_{1:4},\ell)$ if $k=5$. Furthermore, $\tilde{\boldsymbol{\ell}}_{-k}$ is the vector without its $k$-th element. Recall the base measure $\mathcal{G}_0$ given in Section~\ref{subsec:model_back}. Then, for $k=1,\dots,5$:
\begin{enumerate}
	\item let $(\ell_{k,\mathrm{left}},\ell_{k,\mathrm{right}})$ be the lower and the upper bounds of $\ell_k$ given by Table~\ref{table:spline.appendix}, and 
	$$
	\tilde{b} = \max_{\ell_{k,\mathrm{left}}<\ell<\ell_{k,\mathrm{right}}} p(\ell|\tilde{\boldsymbol{\ell}}_{-k},\{x_i  \})(\ell_{k,\mathrm{right}}-\ell_{k,\mathrm{\mathrm{left}}});
	$$
	\item draw $\ell^*\sim \mathcal{U}(\ell_{k,\mathrm{left}},\ell_{k,\mathrm{right}})$ and $u\sim \mathcal{U}(0,1)$.  If
	$$
	p(\ell^*|\tilde{\boldsymbol{\ell}}_{-k},\{x_i  \})> \frac{u\tilde{b}}{(\ell_{k,\mathrm{right}}-\ell_{k,\mathrm{left}})}
	$$
	and $\ell^*$ satisfies the constraint on the B-spline variance \eqref{formula:constraint}, then $\ell^{(t)}_k = \ell^*$. Otherwise, reject $\ell^*$ and repeat Step 2.
\end{enumerate}
The same sampling scheme can be adopted to draw also from the posterior distribution of the latitude knots $\lat$ by replacing $\{ x_i\}$ with $\{ y_i\}$.

\begin{table}[t]
	\centering
	\caption{Left and right bounds of the support for the full-conditional distributions of the B-spline knots.  Here, $x_{(1)}$ and $x_{(n)}$ denote the smallest and the largest values in $\{x_i \}_{i=1}^{n}$, respectively.}
	\label{table:spline.appendix}
	\begin{tabular}{lrr}
		\hline 
		\emph{Knot} & \emph{Left bound} & \emph{Right bound} \\ 
		\hline 
		\hline
		$\ell_1$ & $x_{\min}$& $\min(x_{(1)}, \ell^{(t-1)}_{2})$\\ 
		$\ell_2$ & $\ell_1^{(t)}$ & $\ell^{(t-1)}_{3}$ \\ 
		$\ell_3$ & $\ell_2^{(t)}$ & $\ell^{(t-1)}_{4}$ \\ 
		$\ell_4$ & $\ell_3^{(t)}$ & $\ell^{(t-1)}_{5}$  \\ 
		$\ell_5$ & $\max(x_{(n)},\ell_4^{(t)})$ & $x_{\max}$ \\ 
		\hline 
	\end{tabular} 
\end{table}

\section{Additional results from Section 4.2}\label{app.B}

\begin{figure}[th!]
	\centering
	\includegraphics[width=.47\linewidth]{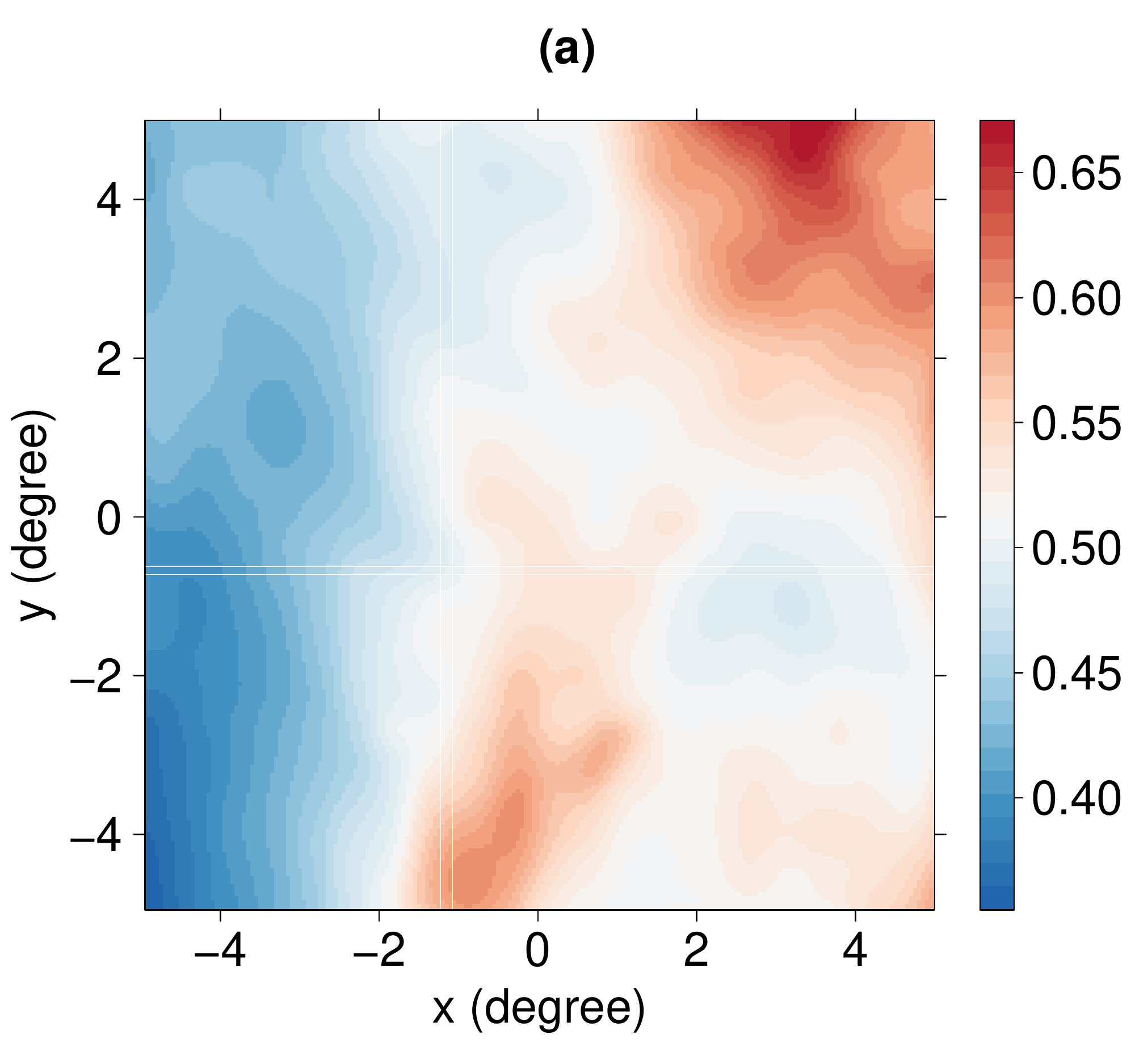}
	\includegraphics[width=.48\linewidth]{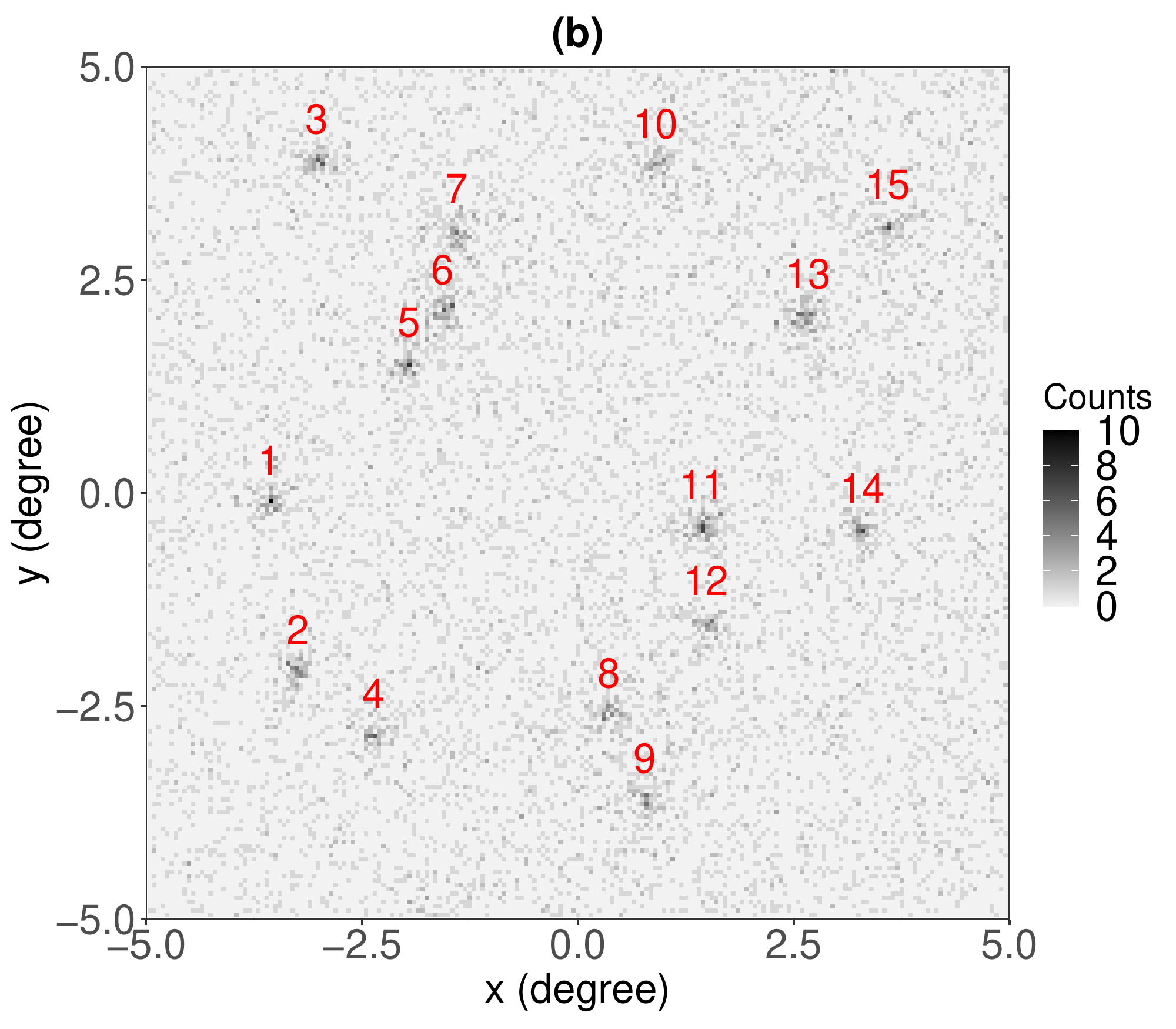}\\
	\includegraphics[width=.48\linewidth]{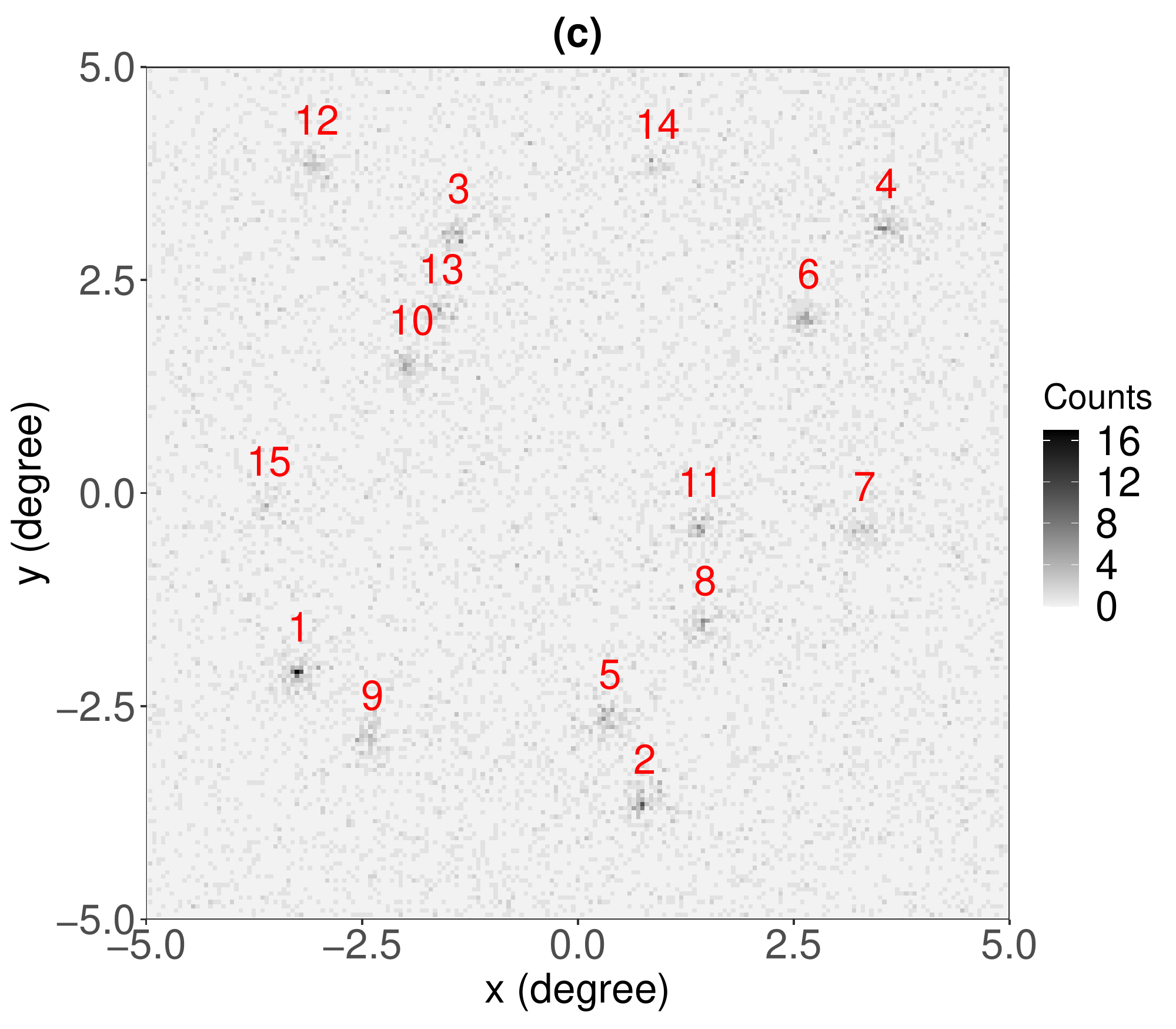}
	\includegraphics[width=.48\linewidth]{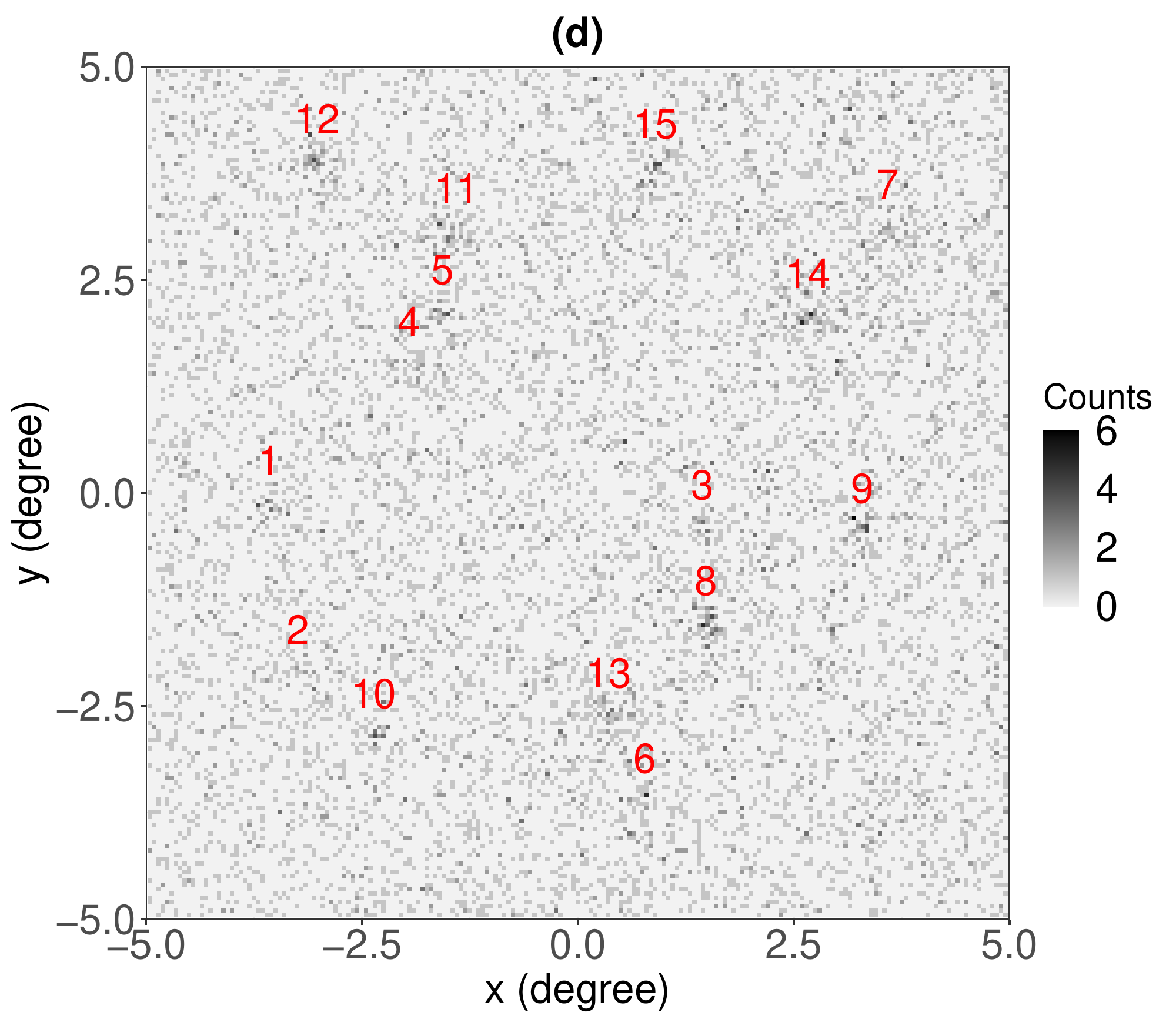}
	\caption{Panel {(a)}: Background region extracted from the model of \cite{acero_etal.2016} and used to generate the background for the experiments in Sections~\ref{subsection:illustrative_example} and \ref{subsec:model_comparison}. Panels (b)-(d): photon counts maps of Scenarios~1, 2 and 3 proposed in  Section~\ref{subsec:model_comparison}. Red numbers label the 15 sources.
	}
	\label{fig:appendix_simulationdatasets}
\end{figure}

This section contains some additional graphs related to the simulation experiments conducted in Section~\ref{subsec:model_comparison}. Figure~\ref{fig:appendix_simulationdatasets}(a) shows the expected photon counts of the background model by \cite{acero_etal.2016} which we used to simulate the background data for the experiments of Sections~\ref{subsection:illustrative_example} and \ref{subsec:model_comparison}.  Panels (b-d) show the three simulated datasets of Section~\ref{subsec:model_comparison}. The sources in the three scenario are labelled with different criteria.  In Scenario~1, they are labelled from 1 to 15 according to increasing longitude.  In Scenario~2, they are labelled according to the spectral shape parameter $\varrho_\varsigma$: Source~1 has the shallowest spectrum ($\varrho_1=1.6514$) and Source~15 has the steepest spectrum ($\varrho_{15}=2.4926$). Last, in Scenario~3, Source~1 is the faintest source ($F_{0,1}=2\cdot10^{-10}$) and Source~15 the brightest ($F_{0,15}=5\cdot10^{-10}$) one.

\begin{figure}[t]
	\centering
	\includegraphics[width=0.48\textwidth]{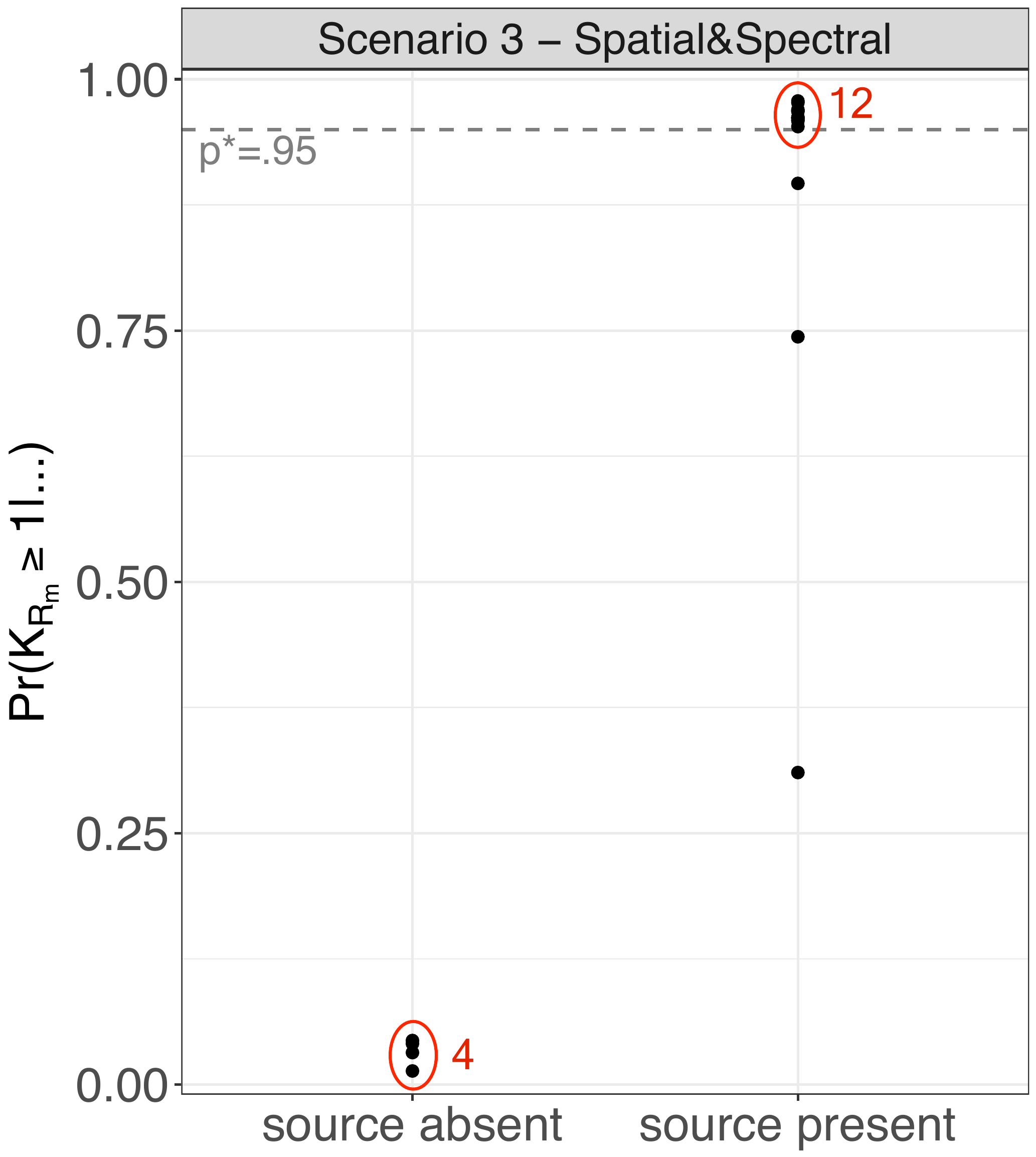}
	\raisebox{0.25\height}{\includegraphics[width=0.5\textwidth]{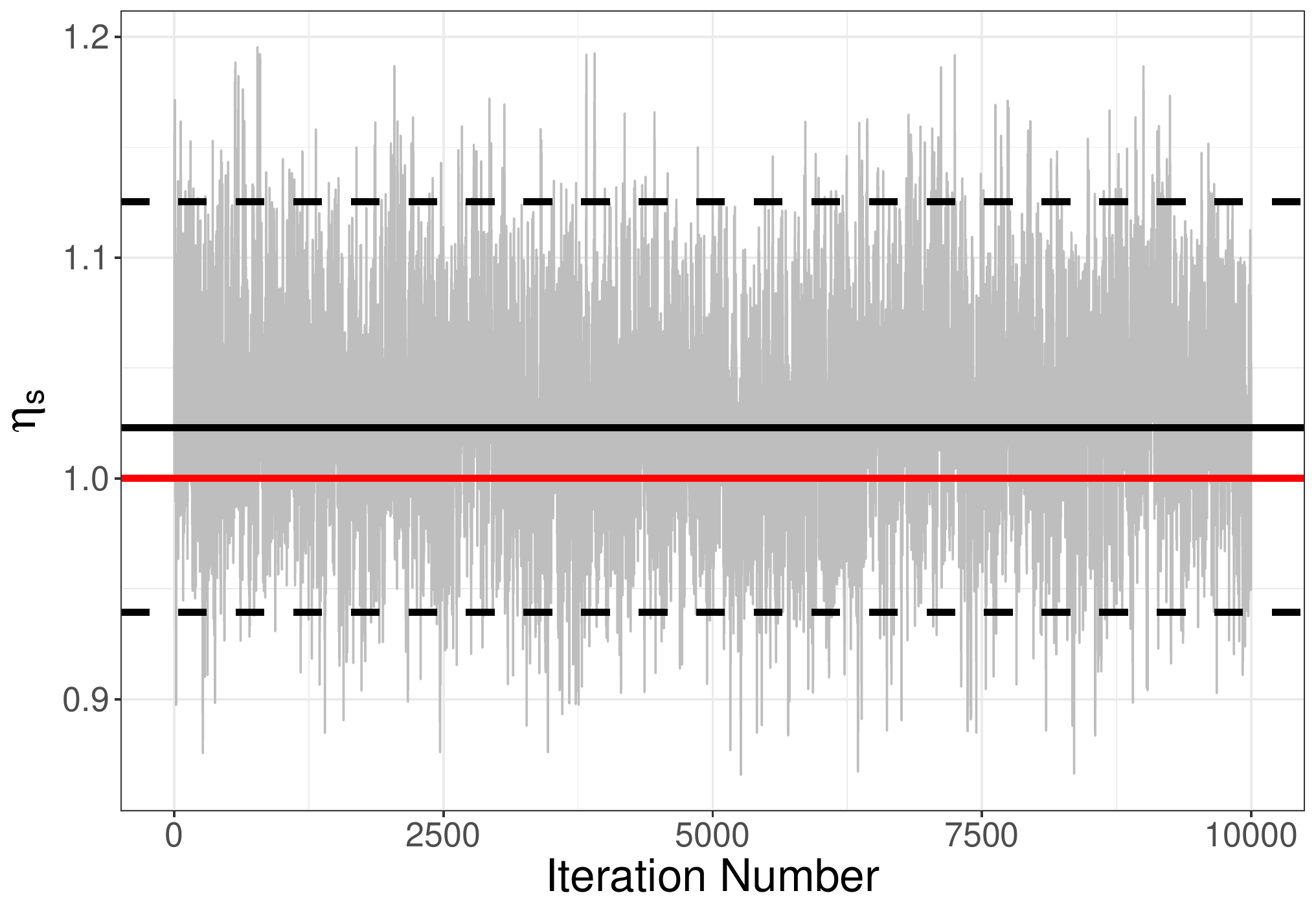}}
	\caption{Results on Scenario 3. Left: probability that the regions discovered under the \jointmodel\ contain a source, differentiated according to whether they do (``source present") or they do not (``source absent") contain a real source. Right: posterior traceplot of $\eta_s$ across 10,000 iterations. The red line is the true shape parameter used to generate the signal of the 15 sources rescaled by 1; the solid and dashed black lines give the {maximum a posteriori} estimate and the 95\% HPD interval, respectively.}
	\label{fig:appendixspatspectr2}
\end{figure}

The left panel of Figure~\ref{fig:appendixspatspectr2} displays the regions discovered under the \jointmodel\ for Scenario~3, differentiated according to whether they do (``source present") or do not (``source absent") include a real source. The right panel displays the traceplot of $\eta_s$ obtained from a MCMC run of length 10,000.  The red line is the spectral shape parameter of the 15 simulated sources, rescaled by 1 to match the parametrization of the Pareto density function (in fact, $\varrho = \eta_s+1$). The posterior mode and the 95\% HPD interval are also displayed using solid and dashed black lines, respectively.

\section*{Acknowledgements}

The authors thank Antonio Canale and Michele Guindani for their many useful discussions and constructive comments. This project was supported by SID grant \textquotedblleft Advanced statistical modelling for indexing celestial objects" (BIRD185983) awarded by the Department of Statistical Sciences of the University of Padova. This work was conducted in collaboration with the CHASC International Astrostatistics Center. CHASC is supported by NSF DMS-18-11308, DMS-18-11083, and DMS-18-11661. We thank CHASC members for many helpful discussions. DvD's and RT's work was supported in part by a Marie-Skodowska-Curie RISE (H2020-MSCA-RISE-2015-691164, H2020-MSCA-RISE-2019-873089) Grants provided by the European Commission. RT's work was partially supported by STFC under grant number ST/T000791/1.


\end{document}